\documentclass[usegraphicx,usedcolumn,usenatbib]{mn2e}
\usepackage{color,epsfig,rotating,amssymb,longtable}
\usepackage{array,colortbl}
\usepackage[colorlinks=true,linkcolor=magenta,citecolor=blue,breaklinks=true]{hyperref}

\usepackage{ifpdf}

\ifpdf
\usepackage{epstopdf}
\else
\usepackage[dvips]{graphicx}
\usepackage{lscape}
\fi

\def\kms{~km\,s$^{-1}$}

\def\ergsc{~erg\,s$^{-1}$cm$^{-2}$}

\def\HII{H{\sc ii}}
\def\SII{S{\sc ii}}

\def\NII{N{\sc ii}}

\def\OIII{O{\sc iii}}

\usepackage{color}

\definecolor{dgreen}{rgb}{0,.5,.1} 
\definecolor{pink}{rgb}{.9,.4,.7}

\usepackage{graphicx}

\title[GHIIRs in NGC\,7479 and NGC\,6070]{Giant H{\sc ii} Regions in NGC\,7479 \& NGC\,6070} 
\author[V. Firpo et al.]{V. Firpo$^{1}$\thanks{vfirpo@fcaglp.unlp.edu.ar}, 
G. Bosch$^{1}$\thanks{IALP-CONICET, Argentina.}, G. F. H\"agele$^{1,2}$, 
N. Morrell$^{3}$ 
\\  
$^{1}$ Facultad de Ciencias Astron\'omicas y Geof\'{\i}sicas, Universidad Nacional  
de la La Plata, Paseo del Bosque s/n, 1900 La Plata, Argentina.\\ 
$^{2}$ Departamento de F\'{\i}sica Te\'orica, C-XI, Univerdidad Aut\'onoma de
Madrid, 28049 Madrid, Spain.\\
$^{3}$ Las Campanas Observatory, Carnegie Observatories, Casilla 601, La Serena, Chile.} 

\begin{document} 

\date{Accepted . Received ; in original form } 


\maketitle

\begin{abstract} 
We present new results
from our search for Giant H\,{\sc ii} Regions in galaxies visible from the southern hemisphere. In
this work we study two galaxies: NGC\,7479 and NGC\,6070. Using
high-resolution spectra, obtained with different instruments at Las Campanas
Observatory, we are able to resolve the emission-line profile widths
and determine the intrinsic velocity dispersion of the ionised gas. We
detect profile widths corresponding to supersonic velocity dispersions in the six observed H\,{\sc ii} regions.     
We find that all of them show at least two distinct kinematical components: a relatively narrow feature (between ~11 and ~22\kms) and a broader (between
~31 and ~77\kms) component. Two of the regions show a complex narrow profile in all ion lines, which can
be further split into two components with different radial velocities. Whereas the wing broadening of the overall profile can
be fitted with a low-intensity broad component for almost all profiles, in one region it was better reproduced by two separate shell-like wings. We have analysed the impact that the presence of multiple components has on the location of the H{\sc ii} regions in the $\log(L) - \log(\sigma)$ plane. Although the overall distribution confirms the presence of a regression, the precise location of the regions in the plane is strongly dependent on the components derived from the profile fitting. 
\end{abstract}


\section{Introduction} 

The most extended and luminous objects which are observed on the disks of
spirals, in irregulars and in starburst galaxies are called Giant Extragalactic
H{\sc ii} Regions (GH{\sc ii}Rs). GH{\sc ii}Rs  are places of very active star
formation and they are characterised by the large emission of ultraviolet
photons from a large number of young and massive stars. Owing to the large
number of ionising photons most of the surrounding gas is ionised, and the
recombination process provides a strong signature of recent or ongoing massive
star formation. 

The emission-line profile widths of the giant H{\sc ii} regions imply the
existence of supersonic motions in the gas \citep{SW70}. Furthermore, \cite{TM81} found a correlation between the
gas velocity dispersion ($\sigma$) and the total luminosity (L) emitted
in the respective line. \cite{H86} and \cite{1988A&A...201..199A} confirmed the existence of such a regression, but there was no
agreement on the values derived for its slope so no conclusion could
be reached regarding the origins of this motion. \cite{2000AJ....120..752F}
and \cite{B02} using different techniques on different samples
agreed on a slope close to a value of 4, which seemed to favour the
gravitational source of energy for the observed supersonic motion. Even if the nature of this behaviour is not fully understood, it still
allows us to distinguish giant H{\sc ii}
regions from an agglomeration of classical H{\sc ii} regions by means of high
resolution spectroscopy. GH{\sc ii}Rs  provide the link between small-scale
star forming regions, such as Orion in our Galaxy and violent star formation
processes taking place in starburst galaxies. 

In \cite{Firpo05} we have confirmed the giant nature of three candidates to
giant H{\sc ii} regions in the southern sky which were identified as very
luminous H{\sc ii} regions by \cite{F97}. 

Here we continue our search and detailed analysis of GH{\sc ii}Rs  in
local Universe galaxies. We have selected from the Feinstein catalogue, a
sample of the brightest H{\sc ii} regions from two spiral galaxies: NGC\,7479
and NGC\,6070. 

NGC\,7479 is a barred spiral galaxy, at a distance of 31.92
Mpc \citep{1999A&AS..135..145R}, classified as SB(s)c by \cite{dV91}. \cite{1989ApJ...346..126D} and
\cite{1997A&A...323..363M} have classified it also as a starburst galaxy,
whereas \cite{1983ApJ...269..466K} and \cite{1997ApJS..112..315H} have
identified it as a LINER and a Seyfert 1.9 galaxy respectively. Many studies have been performed on this 
galaxy owing to the presence of a certain asymmetry in the spiral structure
\citep{1995ApJ...441..549Q} that leads several authors to suggest this
might be owed to a recent merger event \citep{1996PhDT........82L,1999MNRAS.308..557L}. This scenario can also
account for perturbations in the velocity field \citep{1998MNRAS.297.1041L}
and the latest minor merger model indicates that the remnant may be situated
within the bar \citep{2001Ap&SS.276..667L} as no remnant 
of the merger could be identified in optical images \citep{2003A&A...409..899S}.

NGC\,6070, at a distance of 29.8 Mpc \citep{2002A&A...389...68G}, is classified as SA(s)cd by
\cite{dV91} although \cite{2004A&A...423..849G} state that a bar with two 
main arms can be identified. \cite{2006A&A...454..759P} further studied the
bar structure classifying it as Type II.o.CT according to
\cite{2008AJ....135...20E}. The galaxy shows a normal arm distribution, with
two arms that break into several on each side of the central region (see
Figure \ref{fig6070chart}). The disk dynamics was studied by
\cite{2002A&A...393..389M}, and \cite{2002A&A...389...68G} studied the
distribution of \HII\ regions in the galaxy. 

In order to verify the nature of our giant H{\sc ii} region candidates, we obtained high resolution spectra to measure the emission-line profile
widths and to estimate if the velocity dispersion is indeed supersonic. To
determine the optimum number of Gaussian components to fit each line profile
in order to minimise the uncertainty in the velocity dispersion of the ionised
gas, we have considered information from other studies available in the
literature.  Previous papers have dealt with the fact that a single Gaussian
profile might not be realistic enough when fitting the observed emission
lines. Several studies favour the presence of an omnipresent broad component which
explains the integral profile wings. They include \cite{M99} who identified a broad
component in the recombination lines in the central region of the 30 Doradus
nebula, \cite{1987MNRAS.226...19D} and \cite{1996MNRAS.279.1219T} in the M33 giant \HII\
region NGC\,604, \cite{1997ApJ...488..652M} in four Wolf-Rayet galaxies, and
\cite{1999ApJ...522..199H} in the starburst galaxy NGC\,7673.
More recently, \cite{H07,2008PhDT........35H,2009MNRAS.396.2295H} and \cite{Hagele+10}
found that the best Gaussian fits in circumnuclear 
star-forming regions involved the presence of broad and narrow components for the
emission lines of the ionised gas.

A variety of line broadening mechanisms have been proposed to interpret the existence of the broad supersonic component measured in the emission line profile of GH{\sc ii}Rs. Dynamics of virialised systems (Tenorio-Tagle et al.~\citeyear{TT93}), superposition 
of multiple gas bubbles in expansion (Chu \& Kennicutt~\citeyear{CHK94}), or turbulence of the same 
interstellar gas (Medina Tanco et al.~\citeyear{MT97}). More recently, Westmoquette et al.~\citeyear{2007MNRAS.381..894W,2007MNRAS.381..913W} concluded that the narrow component represents the general disturbed ionised interstellar medium (ISM), arising through a convolution of the stirring effects of the starburst and gravitational virial motions. On the other hand, the broad component results from the highly turbulent velocity field associated with the interaction of the hot phase of the ISM with cooler gas knots, setting up turbulent mixing layers.

Other studies support the existence of shell-like profiles on the blue and red wings
of the main feature as found by \cite{CHK94}, who state that the integrated
line profiles of the entire 30 Dor region may be fitted with a broad Gaussian
with low-intensity wings. \cite{Relano05} claim that a large fraction of the
H{\sc ii} regions in three barred spiral galaxies shows high-velocity and low-intensity features in its integrated line profiles and
\cite{2006A&A...455..539R} find that the great majority of their sampled H{\sc
  ii} regions show that the brightest principal component is accompanied by
low-intensity components symmetrically shifted in velocity. 

In this paper we present echelle data obtained at Las Campanas Observatory (LCO) of six GH{\sc ii}Rs candidates in the two mentioned galaxies. We have determined the velocity dispersion and we have evaluated the possible presence of a broad component or two symmetric low-intensity components in the fit with the observed emission line profile widths. In section \ref{sec:observations} we present the observations and the data reduction. Section \ref{sec:analysis} details the analysis performed on the emission-line profiles and discusses each conspicuous case individually. The summary and conclusions of this work are included in Section 4.

\section{Observations and Reductions} 
\label{sec:observations} 

\begin{figure}
\begin{center}
\caption[NGC7479 finding chart]{H$\alpha$ image with the GH{\sc ii}R
  candidates observed in NGC\,7479, identified by circles. H$\alpha$ images
  were kindly made available by C.\ Feinstein.}
\label{fig7479chart}
\includegraphics[angle=270,width=.47\textwidth]{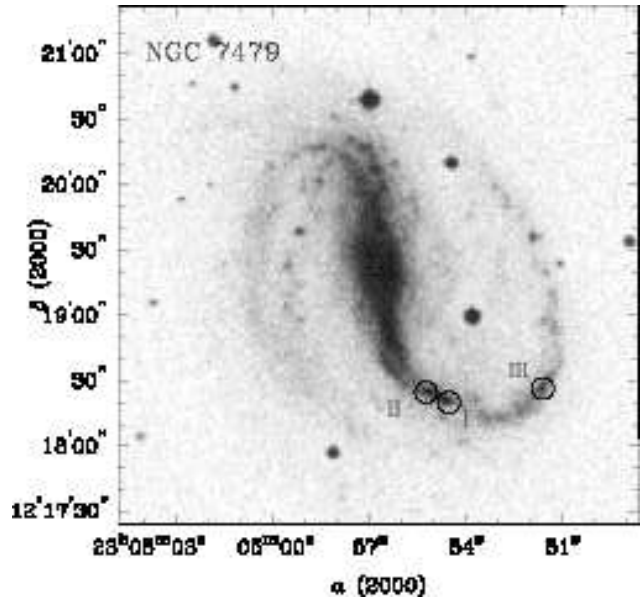}
\end{center}
\end{figure}

\begin{figure}
\begin{center}
\caption[NGC6070 finding chart]{Idem as Figure \ref{fig7479chart} for NGC\,6070}\label{fig6070chart}
\includegraphics[angle=270,width=.46\textwidth]{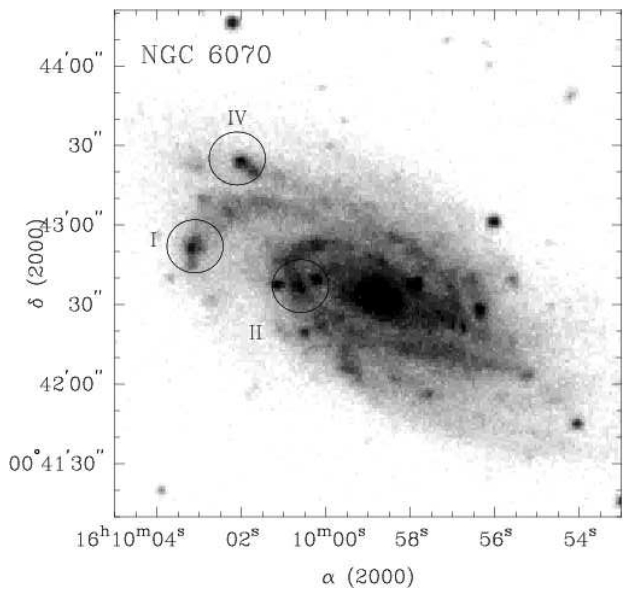}
\end{center}
\end{figure}

According to canonical denomination for GH{\sc ii}Rs we have labelled our candidates following their brightness ranking from Feinstein catalogue. NGC\,7479 I is therefore the brightest region in NGC\,7479, NGC\,7479 II is the second brightest and so on. Observed regions from NGC\,7479 and NGC\,6070 are shown in Figures \ref{fig7479chart} and \ref{fig6070chart} respectively. High resolution 
spectra were obtained with an echelle
spectrograph at the 100-inch du Pont Telescope, Las Campanas
Observatory (LCO) between 19 and 22, July 2006. The spectral range covered by the 
observations was from 3800 to 9500\AA. Observing conditions were good
with 1 arcsec seeing and photometric sky.  A 2$\times$2 binning was applied to the CCD in order to minimise readout contribution to the final spectrum noise. The spectral resolution achieved in our du Pont Echelle data (obtained with 1 arcsec effective slit width) is R$\simeq$25000:  $\Delta\lambda$=0.25\AA\ at $\lambda$ 6000\AA, as measured from the FWHM of the ThAr comparison lines. This translates in a resolution of $\sim$12 kms$^{-1}$.

The standard stars, HR\,7950, and HR\,4468 and the CALSPEC spectrophotometric
standard star, Feige\,110 (Bohlin et al.~\citeyear{Bohlin01}), were also
observed for flux calibration purposes. In addition, Th-Ar comparison spectra,
milky flats (sky flats obtained with a diffuser , during the afternoon) and bias
frames were taken every night. The exposure time for standard stars was five
seconds for the bright HR stars and 1200 seconds for Feige\,110. 

We also obtained high resolution spectra for the three H{\sc ii} regions
in the NGC\, 6070 galaxy using the echelle
double spectrograph Magellan Inamori Kyocera Echelle (MIKE) at the 6.5-m,
Magellan II (Clay) Telescope, LCO, in July 2004. No binning was applied to the 2K$\times$4K CCD detector and a 1 arcsec slit was used. The spectral resolution measured on our MIKE spectra is 11 km$s^{-1}$, which is quite similar than that obtained at du Pont. The exposure
time for CALSPEC spectrophotometric standard star BD+28D4211 was 900
seconds. 
Th-Ar comparison lamps, milky flats (internal flat field lamp with 
a diffuser glass slide) and bias frames were used to calibrate these data.
Table \ref{Regions} lists the journal of observations, together with the exposure times, air masses and the standard stars
that were used for flux calibrating each region. \\

\begin{table}
\caption[Observed dates]{Journal of observations for the candidate GH{\sc ii}Rs and flux standard stars. The candidate regions have labelled following their brightness ranking from Feinstein catalogue (see \cite{F97}).}
\label{Regions}
\begin{center}
\begin{tabular}{@{}llccccl@{}}
\hline 
Telescope & Region & Date & Exp. & sec z & Standard \\
\hline
du Pont  &7479\,I  & 2006 Jul 19 & 1800  & 1.6  & Feige 110\\
         &7479\,II & 2006 Jul 20 & 1800  & 1.5  & HR\,7950 \\  
         &7479\,III& 2006 Jul 20 & 1800  & 1.4  & HR\,7950 \\
         &6070\,I  & 2006 Jul 21 & 1800  & 1.2  & HR\,4468 \\
         &6070\,II & 2006 Jul 21 & 1800  & 1.2  & HR\,4468 \\
         &6070\,IV & 2006 Jul 21 & 1800  & 1.2  & HR\,4468 \\
\\
Clay &6070\,I  & 2004 Jul 12 &1800  & 1.3   & BD+28D4211 \\
            &6070\,II & 2004 Jul 12 &1800  & 1.2   & BD+28D4211 \\
            &6070\,IV & 2004 Jul 12 &1200  & 1.7   & BD+28D4211 \\
\hline
\end{tabular}
\end{center}
\end{table}

The data analysis was carried out with {\sc IRAF}\footnote{Image Reduction and 
Analysis Facility, distributed by NOAO, operated by AURA, Inc., under
agreement with NSF.} software. After bias subtraction and flat field corrections with Milky
Flats, the bidimensional images were corrected for cosmic rays with the task
{\tt cosmicrays} which detects and removes cosmic rays using a flux
ratio algorithm. The corrected data were reduced by {\sc IRAF} routines
following similar procedures to those described in \cite{Firpo05}. 

We also compared the red end of the wavelength calibrated spectra with the night-sky
spectrum by \cite{O96}. This turned out to be a very reliable confirmation of the
goodness of the wavelength solution, and we were able to check that differences
between our wavelengths and the sky line wavelengths were below 0.05\AA.

We performed flux calibrations using the observed
spectrophotometric standard stars as described in \cite{Firpo05}. The observed standard stars had their fluxes 
tabulated every 16\AA, and the amount of defined intervals within an echelle order ranged
from four to twelve, depending on the quality of the spectrum. 

For the Magellan spectra, and in the case of the NGC 7479\,I region du Pont data, we were able to obtain
CALSPEC spectrophotometric standard stars (Bohlin et al.~\citeyear{Bohlin01}), whose fluxes were tabulated every 2\AA. 
These stars are ideal for calibrating high resolution echelle spectra,
although their relatively low brightness (Feige 110, V=11.83 and BD+28D4211, V=10.51)
makes them time-consuming targets for two metre class telescopes.

\section{Analysis of line profiles} 
\label{sec:analysis} 

Making use of the known redshifts z=0.006685 for NGC\,6070
\citep{2005ApJS..160..149S} and z=0.007942 for NGC\,7479
\citep{1998AJ....115...62H} we identify the hydrogen
recombination lines, such as H$\alpha$ and H$\beta$, and collisionally excited
lines, such as [N{\sc ii}]$\lambda$6548,6584\AA, [S{\sc ii}]$\lambda\lambda$6717,6731\AA, present in the spectra. These strong lines are used to analyse the structure of the profiles as they allow us to verify the existence of more than one component.
The adopted laboratory wavelengths are taken from work by
\cite{G-R05}. 

To determine the radial velocities and the velocity dispersions of the ionised
gas, we measure the central wavelength and width of several emission lines.  
The radial velocity of each emission line is obtained from its central
wavelength determined from the Gaussian profile fitting, and their errors are
given by the fitting error provided by the resampling done within the \texttt
{\sc ngauss} task of {\sc IRAF} and taking into account the rms of the wavelength calibration. The true velocity
dispersion ($\sigma$), of each emission line is calculated as described in
\cite{Firpo05}. The width of the observed profile
($\sigma_o$) is affected by the contribution of thermal random motions
($\sigma_t$) and the instrumental profile ($\sigma_{i}$). Then, we can obtain
the true velocity dispersion as
$\sigma^2$\,=\,$\sigma_o^2$\,-\,$\sigma_{i}^2$\,-\,$\sigma_t^2$. 
As in \cite{Firpo05}, a typical kinetic temperature $T= 10^{4}$K is assumed,
and the instrumental profile ($\sigma_{i}$) is very well approximated by a
single Gaussian function. The velocity dispersion errors have been calculated
using the observational errors in the observed profile and assuming negligible
errors in $\sigma_{i}$ and $\sigma_t$. Owing to the generally
high metallicity of these kinds of objects as a class, the temperatures are
probably lower than $10^{4}$K \citep[see][and references
  therein]{2007MNRAS.382..251D}. The low excitation of these objects estimated
from their spectra makes any temperature sensitive line too weak to be
observed. In fact, in most cases, the [\OIII]$\lambda$5007\AA\ line,
which is typically one hundred times more intense than the auroral
[\OIII]$\lambda$4363\AA\ one, can barely be seen.
In this way, we have also calculated the velocity dispersions
assuming a kinetic temperature of 5000K and we do not find great
differences with the previous calculations derived using $10^{4}$K. These
differences are between 0.02 and 3\kms\ depending on the measured velocity
dispersion since $\sigma_t$ is constant for a given temperature and subtracted in quadrature to
$\sigma_o$. Obviously, the estimated $\sigma$ are always higher for the lower assumed
kinetic temperatures. Then, for consistency with \cite{Firpo05} 
we use a kinetic temperature of $10^{4}$K. Therefore, this effect owing to the
kinetic temperature uncertainty, would produce an underestimation in the
derived true velocity dispersions.

When fitting single Gaussian profiles to the emission lines we notice that,
although it works well for most of the observed profiles, there is a
residual present in the wings of several lines. This has been observed
in GH{\sc ii}Rs before. Some authors \citep[][among
others]{M99,H07,2009MNRAS.396.2295H,Hagele+10} have shown that the
profile wings are well fitted by single (broad) Gaussian of supersonic widths, 
whereas other authors \citep{CHK94,Relano05,2006A&A...455..539R}
proposed that the profiles 
appear to be dominated by a few, well-resolved, expanding shells. Wherever
possible, we have considered and applied both suggestions.

In order to fit multiple components to the observed profiles we make use of the
{\sc ngaussfit} task within {\sc IRAF}. {\sc Ngaussfit} performs iterative fitting of multiple
Gaussian profiles to spectral lines. The task allows us to select individual parameters to
be fitted and delivers an estimate for the uncertainty of derived parameters.
The task needs initial  guesses  for  the  function coefficients, and these 
are specified through previously generated tables. 

In those cases when more than one component is evident, we follow an
iterative procedure in which we only allow the task to fit a limited subset
of parameters at one time. This is done to constrain the universe of possible
solutions to the fit by making use of the different information available from
each emission line. Basically, after obtaining an initial guess for the second
component parameters, we fit the amplitude, centre and width of each
component at one time, leaving the parameters of the other one untouched. Once
the central wavelength of all components is relatively well known, we
set them as fixed parameters and allow the task to perform a final fit of amplitude and width of
all components at the same time. 

The profile fittings show the presence of a residual in the
emission line wings for all cases, becoming more evident 
in the strong H$\alpha$ emission line. Therefore, we
evaluate the presence of either a broad component or shell
features.  For the former assumption, we introduce a broad component with an initial width which
is three times the width of the narrow component(s). This guess comes from an
overall view of the values found by \cite{H07} when performing a similar
analysis. We fit the peak values for the narrow components and then
their profile widths. The broad component parameters are then fitted before
we start the iterative procedure again. In the latter scenario, a similar
iterative procedure takes place, but in this case we add a blue
and a red narrow component to account for the additional emission.
The validity of the profile multiplicity and broadening is checked over the
different emission lines available for each region, although the most reliable
profile analysis comes from the strongest lines, such as H$\alpha$.  

Out of our sample of six GH{\sc ii}Rs we find that all of them show evidence
of wing broadening. For these regions we attempt to fit an overall broad
component or two narrow components symmetrically shifted in velocity with respect
to the intense component. In the following subsections we will discuss our
findings grouping them by the outcome profiles.

\subsection{NGC7479\,I} 

\cite{1999A&AS..135..145R} undertook a complete study to date of the H{\sc ii}
regions in NGC\,7479 galaxy, determining positions, angular sizes, and
absolute fluxes of over 1000 H{\sc ii} regions and they constructed the
luminosity function for the regions over the whole galaxy. We have
cross-correlated these results with our sample and find that the
three NGC\, 7479 regions 
selected in our work can be identified in the Rozas catalogue: NGC7479\,I as
RZH99=1, NGC7479\,II as RZH99=7 and NGC7479\,III as RZH99=74 \citep[RZH99 is
  the number identification used by][]{1999A&AS..135..145R}. 

We identify and fit the Gaussian profiles to the H$\alpha$,
[\NII]$\lambda$6584\AA\ and [\SII]$\lambda\lambda$6717,6731\AA\
lines in NGC 7479\,I region with {\tt splot} routines. The Gaussian fits
reveal the presence of two distinctly separated kinematic components
(labelled A and B) in all analysed profiles. Component A shows a much broader
profile than component B, and both components show little spread in their
individual radial velocities among different emission lines present in the
spectrum. The reliability of these values is confirmed when we improve the
profile fitting using {\sc ngaussfit} which in turn yields values for the
profile width of each component.  

\begin{figure*}
\begin{center}
\caption[A+BbroadFig]{{\sc Ngauss} fits with three Gaussian components in
  the NGC7479\,I emission line profiles, two narrow principal
  components and a broad one. In order: H$\alpha$, [\NII]6584\AA, and
  [\SII]6731\AA.}
\label{figAB_broad}
\includegraphics[trim=0cm 0cm 0cm 0cm,clip,angle=0,width=8cm,height=5cm]{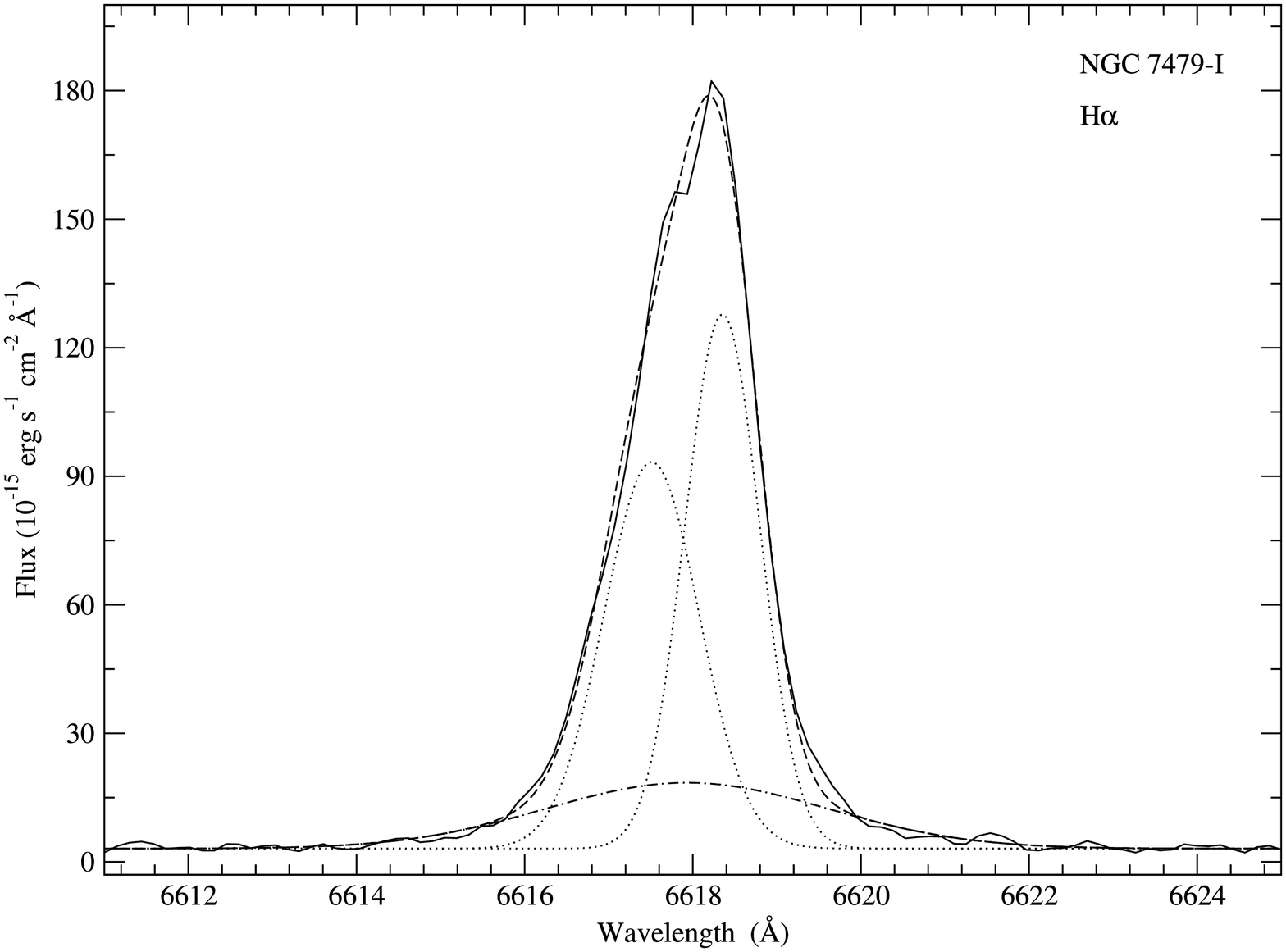}
\includegraphics[trim=0cm 0cm 0cm 0cm,clip,angle=0,width=8cm,height=5cm]{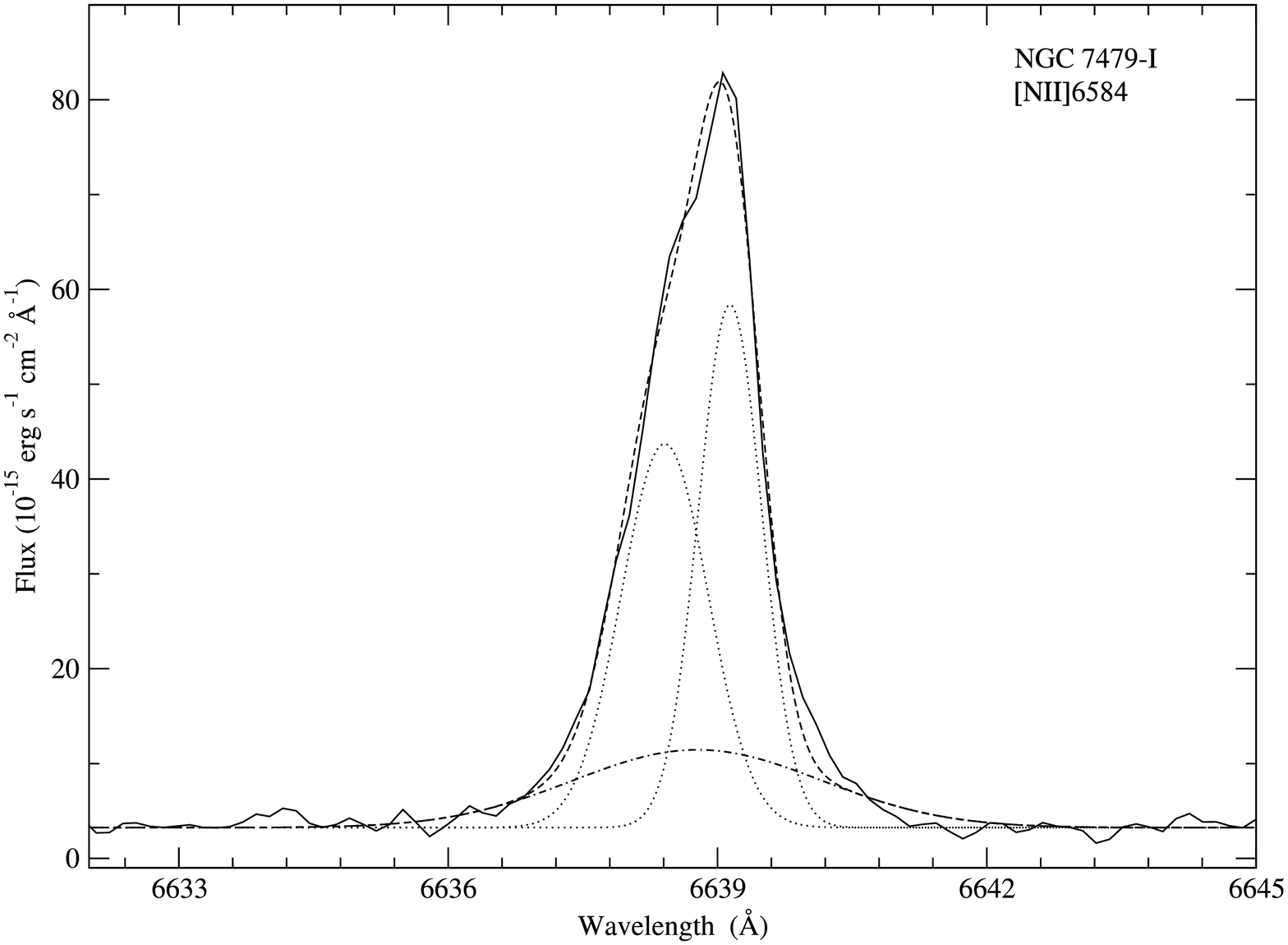}
\includegraphics[trim=0cm 0cm 0cm 0cm,clip,angle=0,width=8cm,height=5cm]{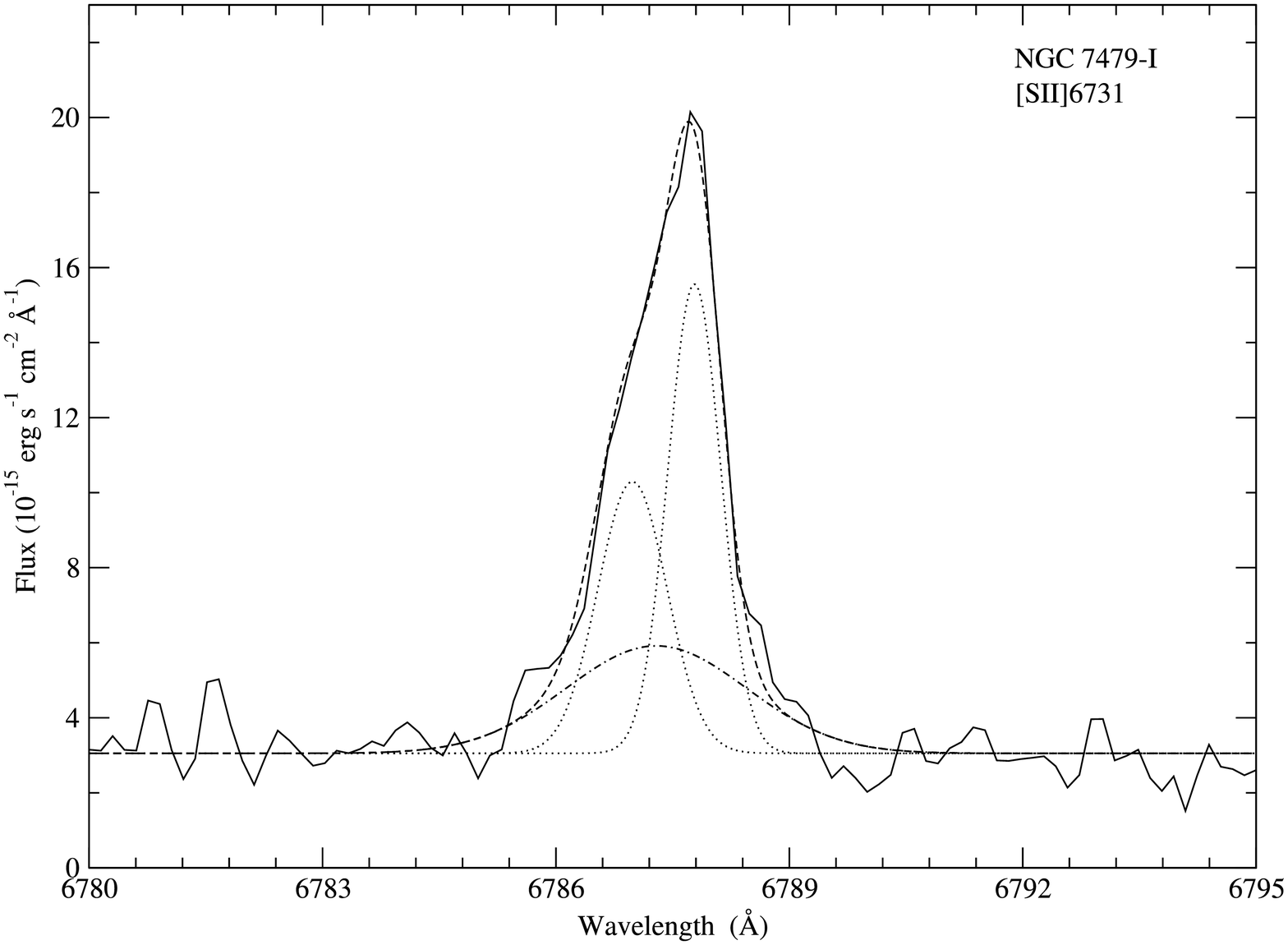}
\end{center}
\end{figure*}

Even considering the presence of two distinct components with different radial
velocities, the overall fit cannot account for a residual emission in the
profile wings. Following the procedures outlined in the previous section, we
are able to fit a broad component, with a velocity dispersion of
about 75\kms\ from the H$\alpha$ emission line 
and about 60\kms\ from the other ones. An attempt to fit twin narrow profiles did
not converge to a reliable solution. Table \ref{tab7479_ngauss} shows
parameters for the three components that fit the global profile. Individual Gaussian component fluxes are listed as emission measure (EM) relative to the total line flux following the work by \cite{Relano05}. The overall H$\alpha$ flux, uncorrected for reddening, is found to be 3.37 x 10$^{-14}$\ergsc. In Figure \ref{figAB_broad} we show the {\sc ngauss} fitting done with three different Gaussian components in the emission lines which have enough signal to provide an accurate fit. The emission lines with a low signal-to-noise ratio for which the fits do not provide accurate results for the broad component, are not listed in the table. The velocity dispersion for the ionised gas is derived by taking $\sigma_{i}$ = 5.1 kms$^{-1}$ as instrumental width,
and a kinetic temperature of $T \simeq 10^{4}K$ as mentioned above.

\begin{table*}
\caption[NGC7479_comp]{Results of Gaussian profiles fitting to the observed emission lines in NGC 7479. Each emission line is identified by its ion laboratory wavelength and ion name in columns 1 and 2. According to the different fits performed on each line, column 3 identifies each "narrow component" (A and B, where applicable), a broad component and/or blue and red wings (b wing and r wing respectively). Radial velocities (V$_r$) and intrinsic velocity dispersions ($\sigma_{int}$) together with their respective errors are expressed in \kms. The intrinsic velocity dispersions are corrected for the instrumental and thermal widths. Emission measures (EM) are included and shown as a percentage of the component flux relative to the total EM of the region.}
\label{tab7479_ngauss}
\begin{center}
\begin{tabular}{llllllllllllllllll}
\hline
\multicolumn{3}{c}{} & \multicolumn{5}{c}{\em NGC\,7479 I} &  \multicolumn{5}{c}{\em NGC\,7479 II} & \multicolumn{5}{c}{\em NGC\,7479 III}\\
$\lambda_{0}$&ion&comp.&V$_r$&error&$\sigma_{int}$&error&EM&V$_r$&error&$\sigma_{int}$&error&EM&V$_r$&error&$\sigma_{int}$&error&EM\\
\hline
6563	&	H$\alpha$	&	A	&	2500.4	&	0.7	&	22.3	&	0.9	&	39	&	2492.4	&	0.6	&	14.8	&	0.8	&	66	&	2481.1	&	0.7	&	11.3	&	2.9	&	67	\\
	&		&	B	&	2538.9	&	0.6	&	15.3	&	0.8	&	42	&	2532.2	&	0.7	&	11.1	&	0.9	&	23	&	\ldots	&	\ldots	&	\ldots	&	\ldots	&	\ldots	\\
	&		&	broad	&	2519.6	&	3.9	&	76.5	&	4.8	&	19	&	\ldots	&	\ldots	&	\ldots	&	\ldots	&	\ldots	&	2472.5	&	2.1	&	37.8	&	0.9	&	33	\\
	&		&	b wing	&	\ldots	&	\ldots	&	\ldots	&	\ldots	&	\ldots	&	2442.7	&	1.7	&	13.5	&	2.0	&	5	&	\ldots	&	\ldots	&	\ldots	&	\ldots	&	\ldots	\\
	&		&	r wing	&	\ldots	&	\ldots	&	\ldots	&	\ldots	&	\ldots	&	2573.8	&	1.9	&	18.0	&	1.7	&	6	&	\ldots	&	\ldots	&	\ldots	&	\ldots	&	\ldots	\\
\\																																			
6584	&	[NII]	&	A	&	2506.8	&	0.7	&	21.6	&	0.9	&	39	&	2494.2	&	1.1	&	16.4	&	1.0	&	53	&	2482.9	&	0.8	&	12.8	&	2.3	&	52	\\
	&		&	B	&	2540.1	&	0.7	&	15.2	&	0.8	&	39	&	2532.5	&	2.0	&	16.7	&	1.3	&	30	&	\ldots	&	\ldots	&	\ldots	&	\ldots	&	\ldots	\\
	&		&	broad	&	2523.4	&	4.4	&	63.6	&	3.6	&	22	&	\ldots	&	\ldots	&	\ldots	&	\ldots	&	\ldots	&	2478.4	&	4.0	&	40.4	&	1.1	&	48	\\
	&		&	b wing	&	\ldots	&	\ldots	&	\ldots	&	\ldots	&	\ldots	&	2459.8	&	9.1	&	24.8	&	4.3	&	13	&	\ldots	&	\ldots	&	\ldots	&	\ldots	&	\ldots	\\
	&		&	r wing	&	\ldots	&	\ldots	&	\ldots	&	\ldots	&	\ldots	&	2581.5	&	6.5	&	14.7	&	2.5	&	4	&	\ldots	&	\ldots	&	\ldots	&	\ldots	&	\ldots	\\
\\																																			
6717	&	[SII]	&	A	&	\ldots	&	\ldots	&	\ldots	&	\ldots	&	\ldots	&	\ldots	&	\ldots	&	\ldots	&	\ldots	&	\ldots	&	2480.9	&	0.8	&	10.2	&	5.3	&	33	\\
	&		&	B	&	\ldots	&	\ldots	&	\ldots	&	\ldots	&	\ldots	&	\ldots	&	\ldots	&	\ldots	&	\ldots	&	\ldots	&	\ldots	&	\ldots	&	\ldots	&	\ldots	&	\ldots	\\
	&		&	broad	&	\ldots	&	\ldots	&	\ldots	&	\ldots	&	\ldots	&	\ldots	&	\ldots	&	\ldots	&	\ldots	&	\ldots	&	2471.1	&	4.1	&	52.2	&	1.4	&	67	\\
	&		&	b wing	&	\ldots	&	\ldots	&	\ldots	&	\ldots	&	\ldots	&	\ldots	&	\ldots	&	\ldots	&	\ldots	&	\ldots	&	\ldots	&	\ldots	&	\ldots	&	\ldots	&	\ldots	\\
	&		&	r wing	&	\ldots	&	\ldots	&	\ldots	&	\ldots	&	\ldots	&	\ldots	&	\ldots	&	\ldots	&	\ldots	&	\ldots	&	\ldots	&	\ldots	&	\ldots	&	\ldots	&	\ldots	\\
\\																																			
6731	&	[SII]	&	A	&	2502.4	&	1.1	&	19.3	&	1.2	&	30	&	\ldots	&	\ldots	&	\ldots	&	\ldots	&	\ldots	&	2481.0	&	0.8	&	15.4	&	4.3	&	50	\\
	&		&	B	&	2538.0	&	0.8	&	14.3	&	0.9	&	39	&	\ldots	&	\ldots	&	\ldots	&	\ldots	&	\ldots	&	\ldots		\ldots	&	\ldots	&	\ldots	&	\ldots	\\
	&		&	broad	&	2516.2	&	4.2	&	52.1	&	5.1	&	31	&	\ldots	&	\ldots	&	\ldots	&	\ldots	&	\ldots	&	2454.7	&	6.4	&	51.8	&	1.2	&	50	\\
	&		&	b wing	&	\ldots	&	\ldots	&	\ldots	&	\ldots	&	\ldots	&	\ldots	&	\ldots	&	\ldots	&	\ldots	&	\ldots	&	\ldots	&	\ldots	&	\ldots	&	\ldots	&	\ldots	\\
	&		&	r wing	&	\ldots	&	\ldots	&	\ldots	&	\ldots	&	\ldots	&	\ldots	&	\ldots	&	\ldots	&	\ldots	&	\ldots	&	\ldots	&	\ldots	&	\ldots	&	\ldots	&	\ldots	\\
\hline 
\end{tabular}
\end{center}
\end{table*}

\begin{figure}
\begin{center}
\caption[2DimageFig]{2D H$\alpha$ velocity image of NGC7479\,I region with the [\NII]$\lambda$6584\AA\ (red) contours overlapped. Three sections are numbered in the 2D image.}
\label{fig2D_image} 
\includegraphics[trim=0cm 0cm 0cm 0cm,clip,angle=0,width=8cm,height=3cm]{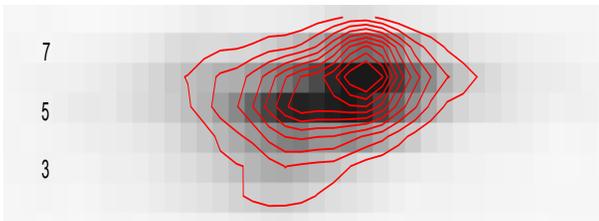}
\end{center}
\end{figure}

Since part of the spatial information along the slit is lost through the
aperture extraction process we generate 2D velocity images in
H$\alpha$, [\NII]$\lambda$6584\AA\ and [\SII]$\lambda$6717\AA\ lines. 
Taking into account that the slit size is 1''x 4'', the data are binned by two pixels in both
directions and the seeing was 1'', we take aperture sizes of 0.5\,arcsec
$\simeq$ 1px, therefore dividing
the slit into eight sections. We extract traces of the spectrum at eight
representative cuts along the slit, preserving the same spatial
resolution. This is done for each echelle order where a line of
interest is present and 
each spectral section is rebinned to velocity space. In this reference frame,
individual slices are stacked
together using the {\tt imstack} task, which allows us to generate 2D velocity
images. Figure \ref{fig2D_image} shows the 2D H$\alpha$ velocity image with the
[\NII]$\lambda$6584\AA\ contours overlapped. It can be readily seen that individual knots can only be identified when the kinematical information is considered in this $V_r$ vs. slit section plane. This suggests that 3D spectroscopy will be able to provide a powerful insight on the structure of this type of star forming knots.

\begin{table}
\caption[AB+broad_cortesTab]{Gaussian fitting results for individual sections of the H$\alpha$ emission line profile in NGC\,7479 I. Kinematical parameters (radial velocity (V$_r$) with errors (\kms), the intrinsic velocity dispersions ($\sigma_{int}$) corrected for the instrumental and thermal widths (\kms), and emission measure (EM) of each component as a fraction of the total EM of the sampled section (\%) data are shown.}
\label{tab7479I_ngaussx3_cortes}
\begin{center}
\begin{tabular}{llclllll}
\hline 
$\lambda_{0}$ & section & comp &  V$_r$&  error& $\sigma_{int}$&error& EM\\
\hline
6563	&	H$\alpha$-1	&	A	&	2501.8	&	0.9	&	12.6	&	1.2	&	51	\\
	&		&	broad	&	2500.7	&	4.8	&	59.3	&	4.3	&	49	\\
\\															
6563	&	H$\alpha$-2	&	A	&	2503.0	&	0.9	&	11.6	&	1.5	&	63	\\
	&		&	broad	&	2500.3	&	1.6	&	37.7	&	2.5	&	37	\\
\\															
6563	&	H$\alpha$-3 	&	A	&	2504.7	&	0.6	&	16.4	&	0.8	&	43	\\
	&		&	broad	&	2506.8	&	1.4	&	47.0	&	1.7	&	57	\\
\\															
6563	&	H$\alpha$-4	&	A	&	2500.8	&	0.7	&	15.8	&	0.9	&	43	\\
	&		&	broad	&	2519.6	&	2.9	&	55.0	&	3.6	&	30	\\
	&		&	B	&	2536.8	&	0.7	&	10.7	&	0.9	&	27	\\
\\															
6563	&	H$\alpha$-5 	&	A	&	2500.9	&	0.6	&	10.5	&	0.8	&	25	\\
	&		&	broad	&	2515.6	&	1.5	&	55.9	&	1.9	&	35	\\
	&		&	B	&	2535.6	&	0.6	&	13.1	&	0.8	&	40	\\
\\															
6563	&	H$\alpha$-6	&	broad	&	2502.6	&	1.5	&	45.5	&	1.7	&	39	\\
	&		&	B	&	2538.6	&	0.6	&	17.4	&	0.8	&	61	\\
\\															
6563	&	H$\alpha$-7	&	broad	&	2519.4	&	1.2	&	46.0	&	1.3	&	56	\\
	&		&	B	&	2540.7	&	0.6	&	6.0	&	0.8	&	44	\\
\\															
6563	&	H$\alpha$-8	&	broad	&	2523.9	&	2.1	&	44.9	&	1.9	&	80	\\
	&		&	B	&	2532.6	&	0.6	&	\ldots	&	1.2	&	20	\\
\hline
\end{tabular}
\end{center}
\end{table}

\begin{figure*}
\begin{center}
\caption[A+Bbroad_cortesFig]{{\sc Ngauss} fits with three (two) Gaussian
  components in the NGC7479\,I sections, two (one) narrow principal components and a broad one. In order: H$\alpha$ (sections 2, 4, 5 and 7), [\NII]6584\AA\ (section 4 and 5).}\label{figAB_broadcortes}
\includegraphics[trim=0cm 0cm 0cm 0cm,clip,angle=0,width=8cm,height=5cm]{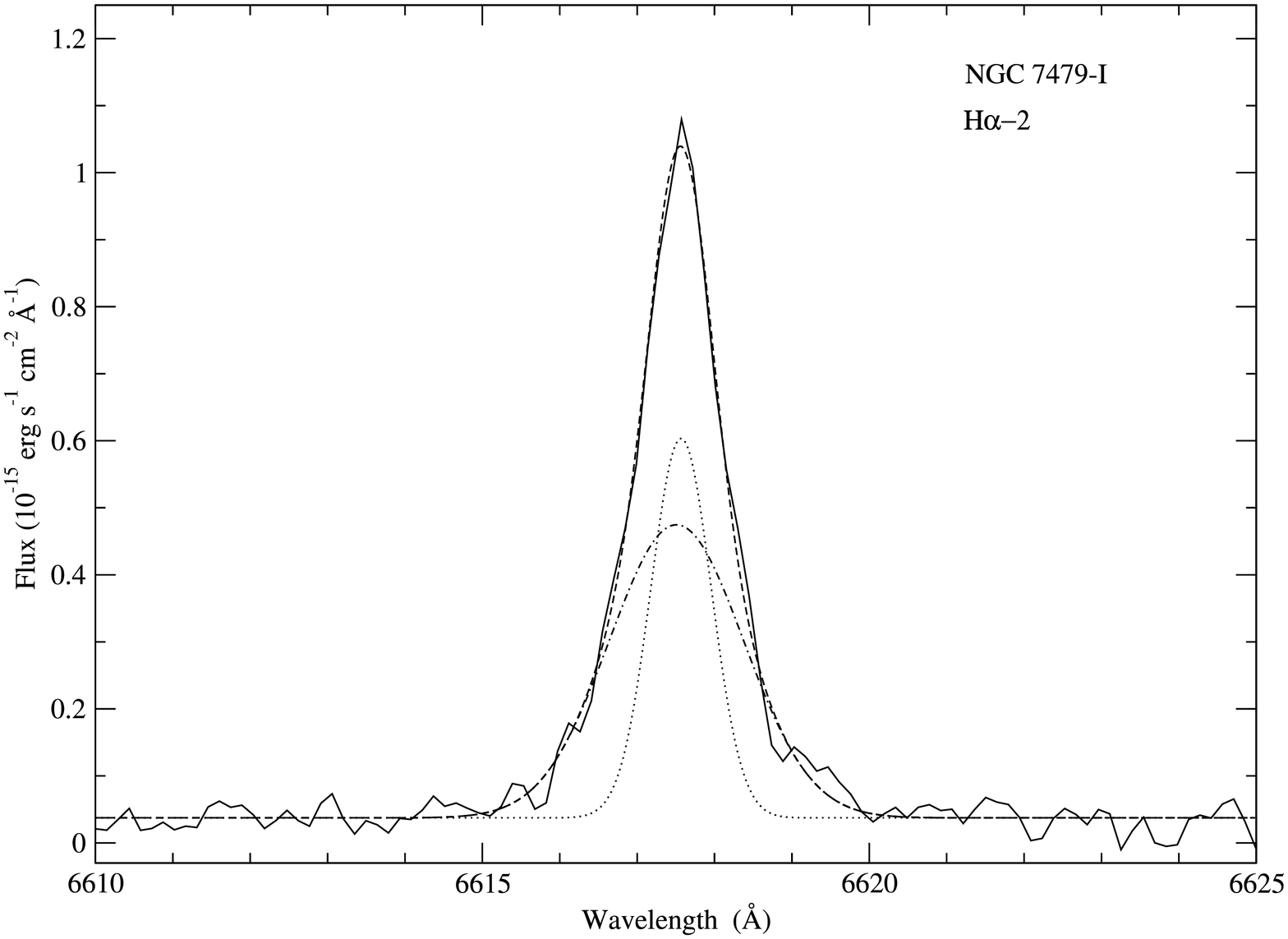}
\includegraphics[trim=0cm 0cm 0cm 0cm,clip,angle=0,width=8cm,height=5cm]{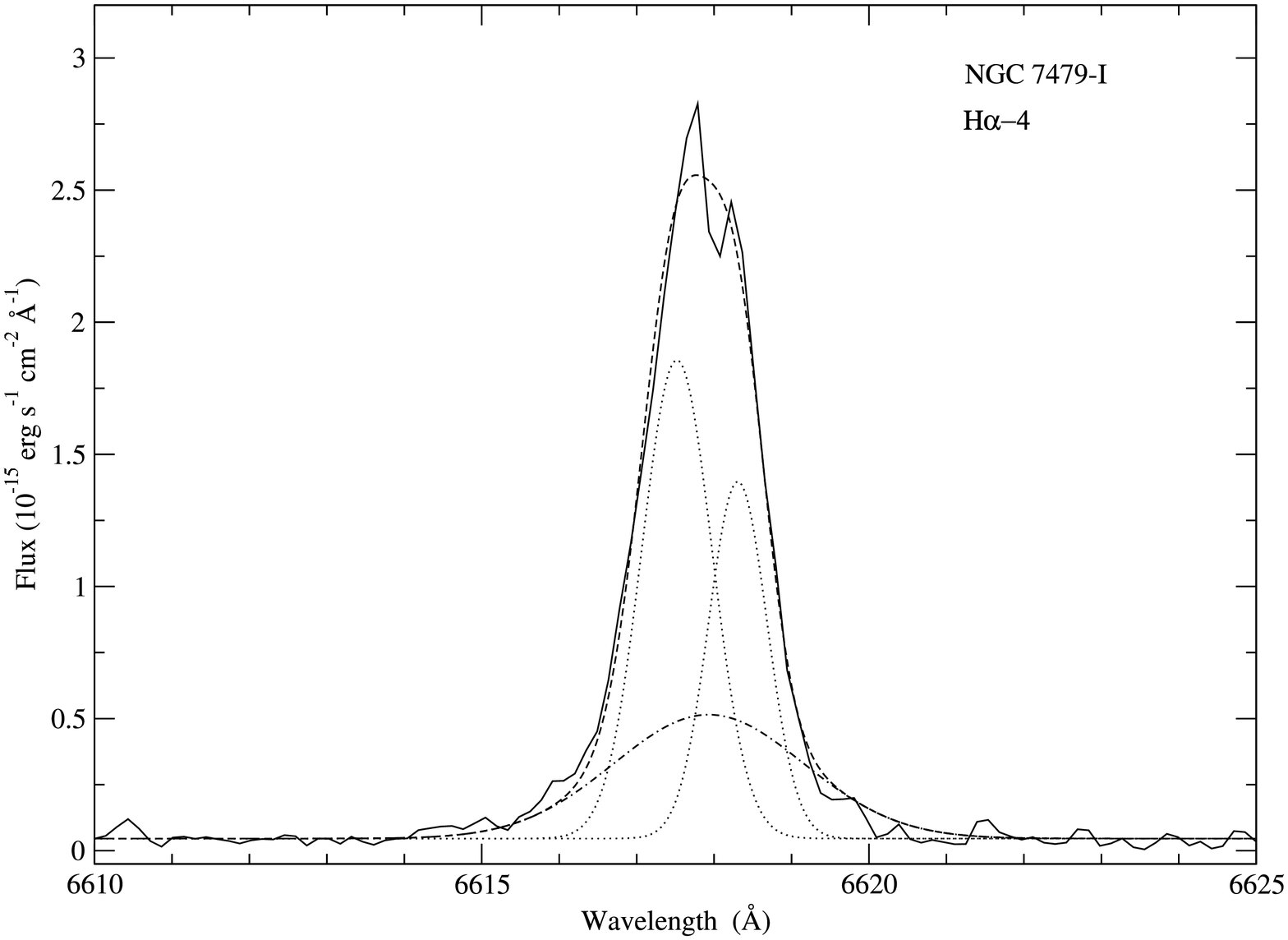}
\includegraphics[trim=0cm 0cm 0cm 0cm,clip,angle=0,width=8cm,height=5cm]{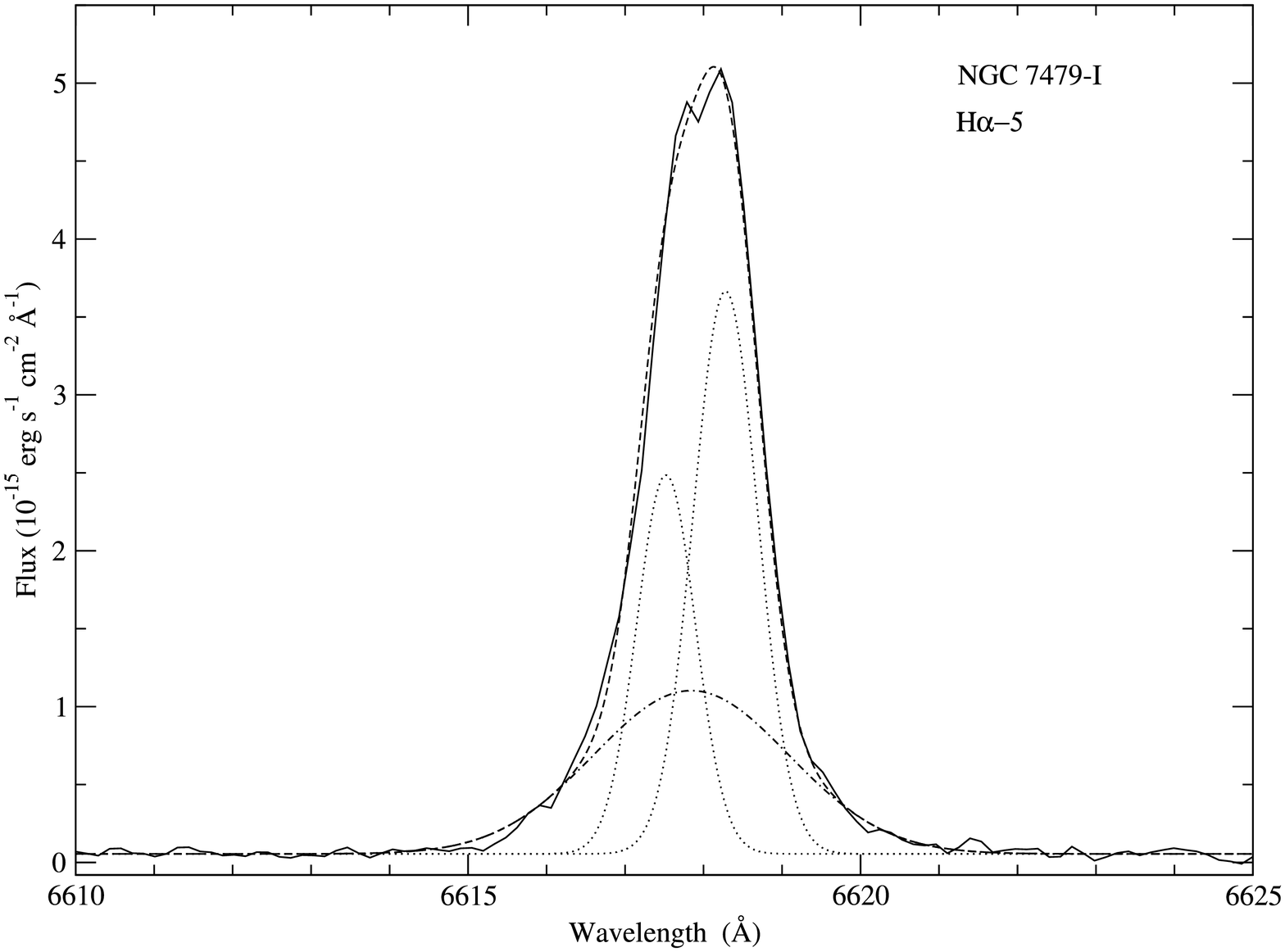}
\includegraphics[trim=0cm 0cm 0cm 0cm,clip,angle=0,width=8cm,height=5cm]{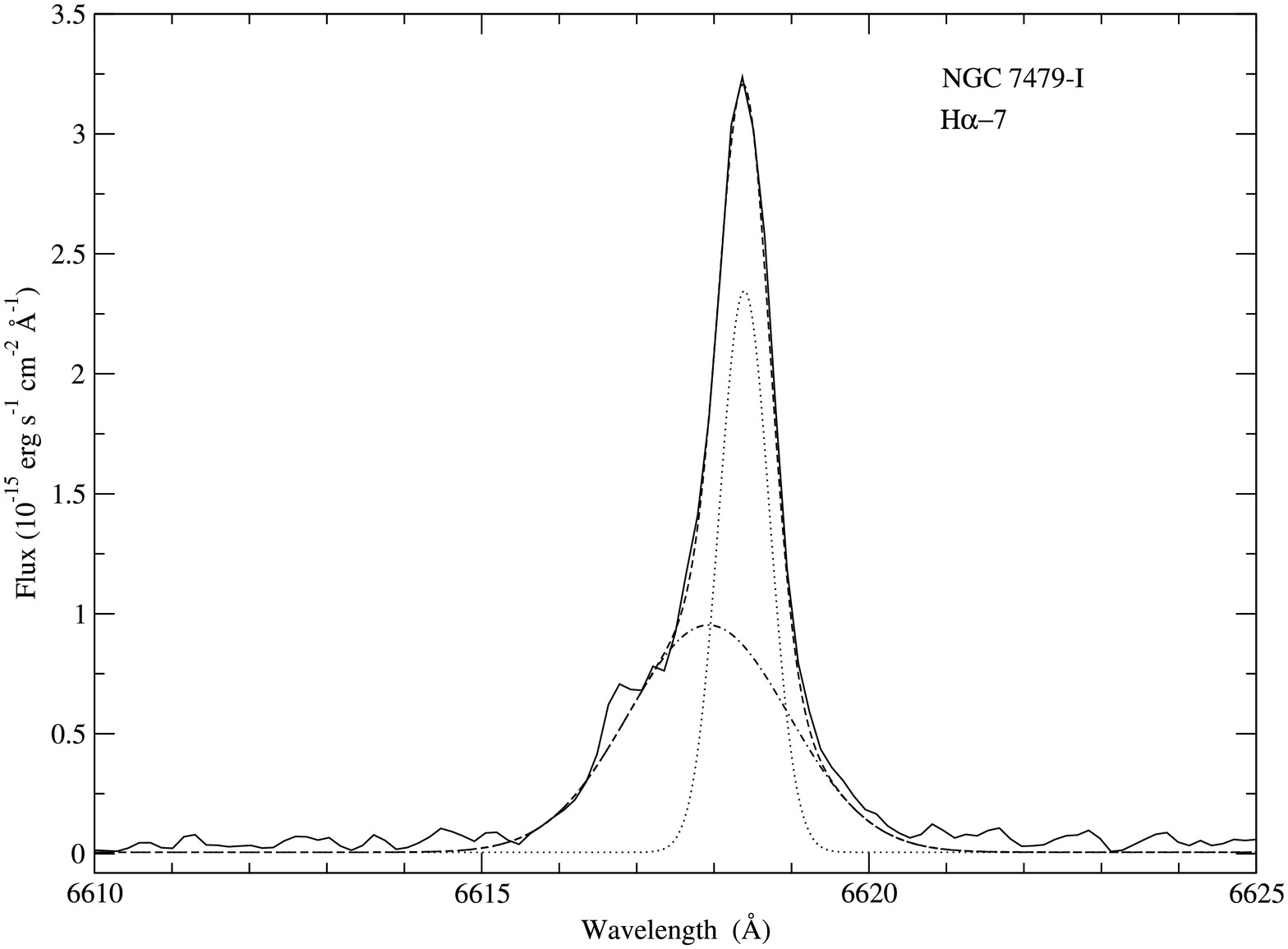}
\includegraphics[trim=0cm 0cm 0cm 0cm,clip,angle=0,width=8cm,height=5cm]{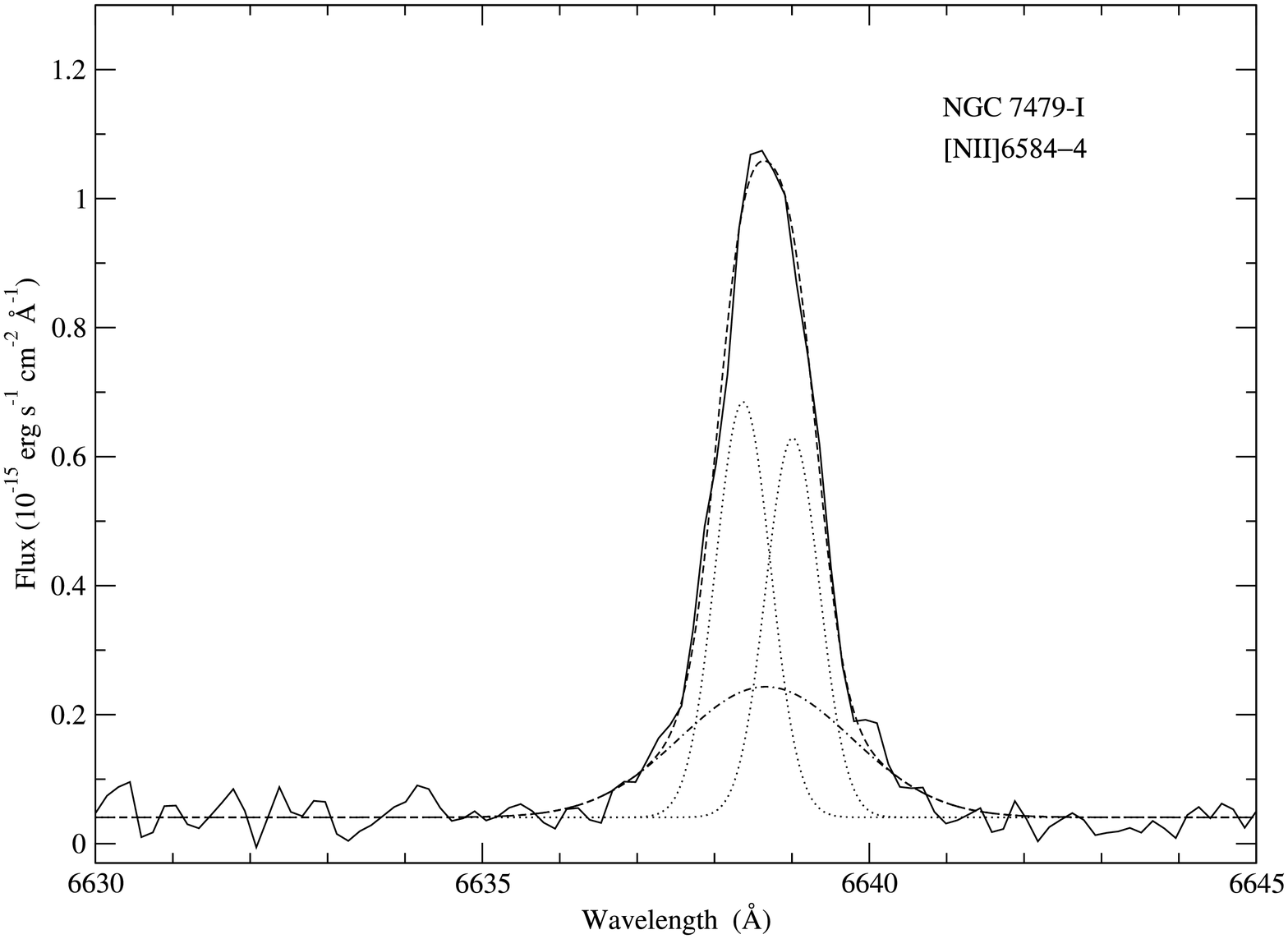}
\includegraphics[trim=0cm 0cm 0cm 0cm,clip,angle=0,width=8cm,height=5cm]{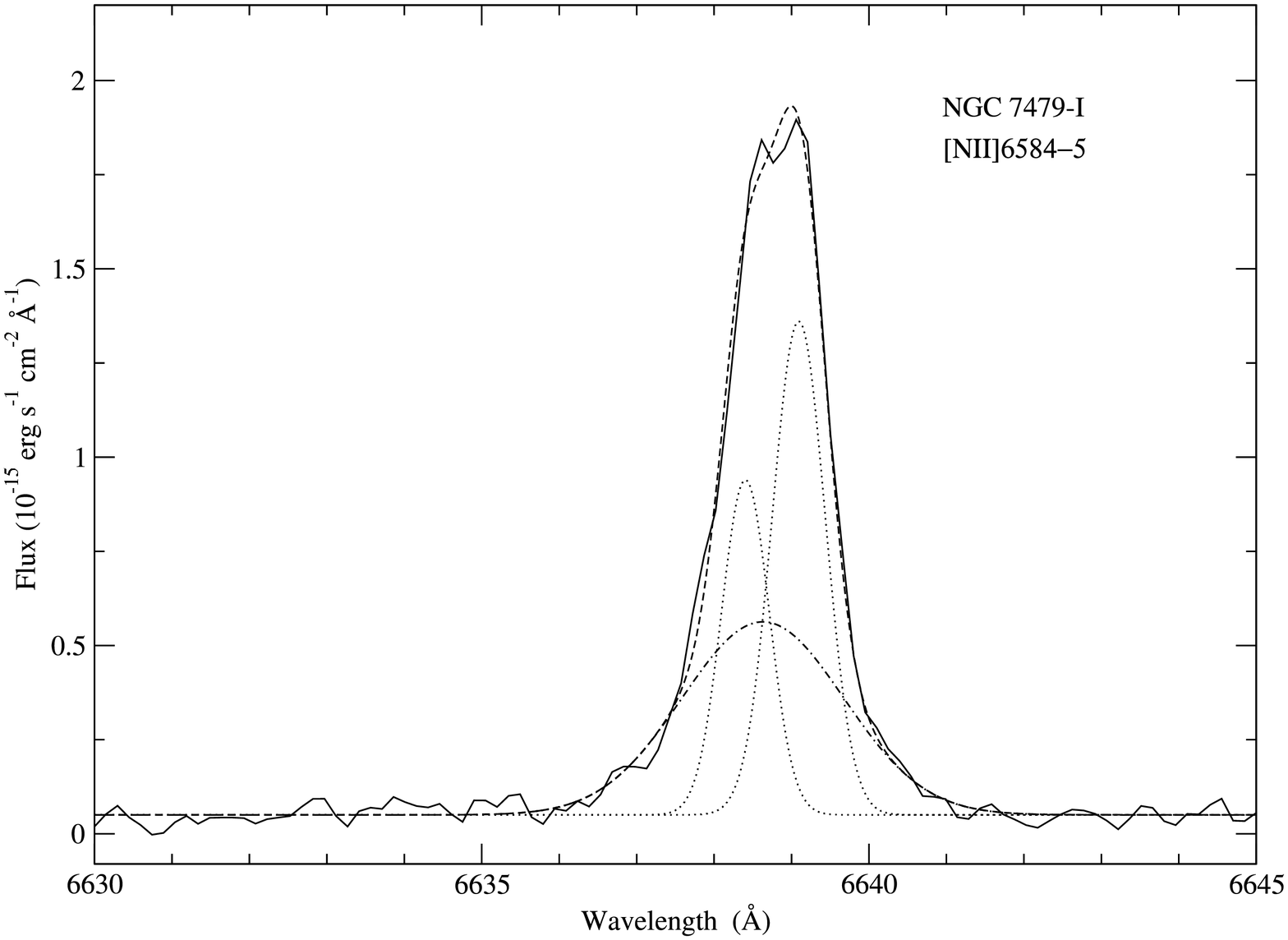}
\end{center}
\end{figure*}

We have performed a detailed analysis on 
the slit stack, by fitting Gaussian profiles at each slit section. The resulting parameters are listed in Table \ref{tab7479I_ngaussx3_cortes} and the graphics output can be seen in Figure \ref{figAB_broadcortes} where is shown 2, 4, 5, and 7 H$\alpha$ sections, and 4 and 5 [\NII]$\lambda$6584\AA\ sections.
Focusing on the slit edges first (sections 1 to 3 and 6 to 8) we identify two distinct kinematical components 
at 2500 \kms\ and 2540 \kms\ labelled A and B respectively. Each of these relatively narrow features has a broader
component associated with it. Although the narrow component can still be deblended in the central region, we are not able
to disentangle the broad ones, which results in a single broad profile with a velocity similar to the average between components A and B. Regarding the velocity dispersion of each
component, we have listed them in Table \ref{tab7479I_ngaussx3_cortes} as a reference, but they should not be
taken as a definite value as the signal-to-noise ratio is variable and somewhat low for the slit edges.

If we compare these results with the radial velocity map derived for the ionised gas in NGC
7479 \citep{1996PhDT........82L} we find, from visual inspection, that the
expected velocity at the position of NGC7479\,I  ranges between  2520 and 2535
\kms\, close to an average value of our components A and B. A similar
comparison, using the H\,I map by \cite{1998MNRAS.297.1041L} which has a better
spatial and spectral resolution, yields a value between 2530 and 2540 \kms\ 
much closer to the value measured for component B. Although the latter map is
sampling neutral rather than ionised hydrogen, the former might be affected by
the complex structure we find within the region. At this stage we can only
suggest that component A shows an odd kinematic behaviour and that a detailed
study of NGC7479\,I and its surroundings is needed to clarify this issue.

\subsection{NGC7479\,II} 

\begin{figure*}
\begin{center}
\caption[A+BwingsFig]{{\sc Ngauss} fits with four Gaussian components in the NGC7479\,II
  emission line profiles, two narrow principal components and two low
  intensity components. In order, H$\alpha$ and
  [\NII]6584\AA}\label{figAB_wings} 
\includegraphics[trim=0cm 0cm 0cm 0cm,clip,angle=0,width=8cm,height=5cm]{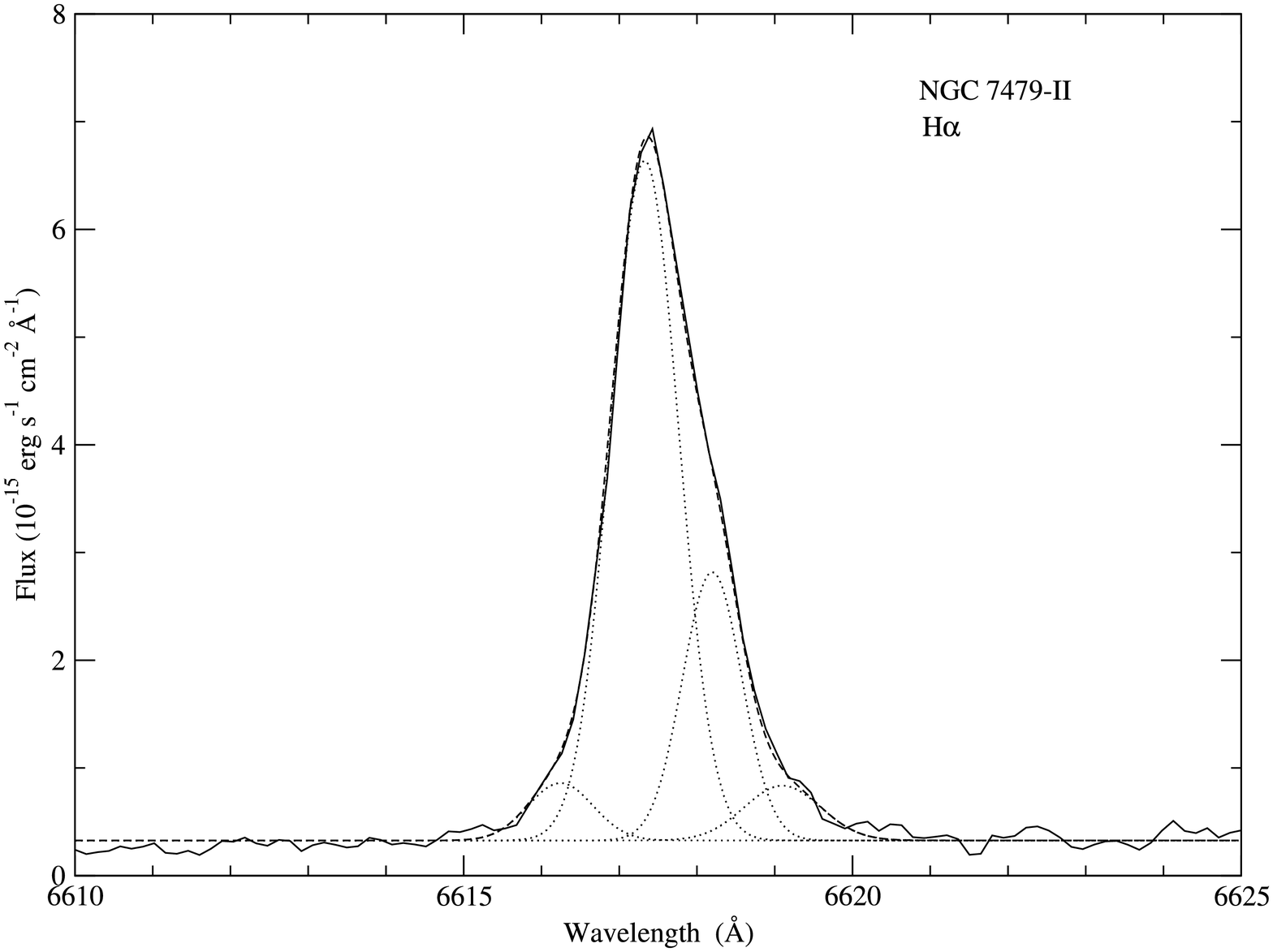}
\includegraphics[trim=0cm 0cm 0cm 0cm,clip,angle=0,width=8cm,height=5cm]{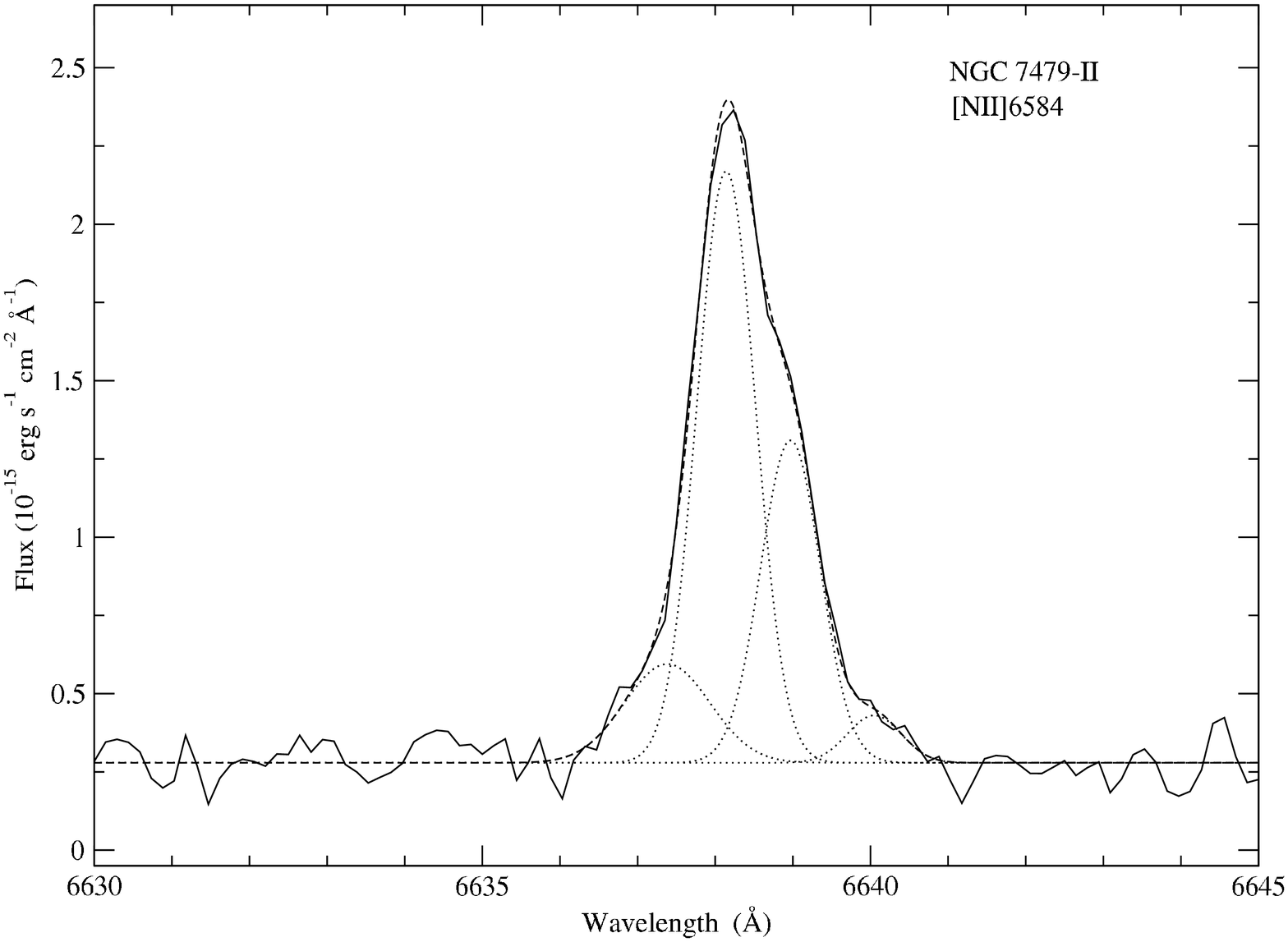}
\end{center}
\end{figure*}

A similar procedure is used to analyse the profiles of the emission lines
of NGC7479\,II. In this case, two components are identified with different radial
velocities, but as the overall signal obtained for this region is lower than for NGC7479\,I it is not 
possible to disentangle them with 2D spectra. 
 These two narrow components are found in the Gaussian profiles of
the H$\alpha$, H$\beta$, [\NII]$\lambda\lambda$6548,6584\AA\ and [\SII]$\lambda\lambda$6717,6731\AA\
lines. In this region, however, the wings of the brightest component profile could not be fitted
with one single broad component. On the other hand, we are able to fit it with two wings
only for the two lines with the strongest signal, namely H$\alpha$ and
[\NII]$\lambda$6584\AA\ (see Figure \ref{figAB_wings}). The final fitting
parameters are shown in Table \ref{tab7479_ngauss}. The overall H$\alpha$ flux, uncorrected for reddening, is found to be 1.06 x 10$^{-14}$\ergsc. The
emission lines with a low signal-to-noise ratio for which the fits do not
provide accurate results for the broad and the two wings components, are
not listed in the table.

We compare the velocity separation between the red (and blue) shifted components with the average of the central features and the H$\alpha$ luminosity of the region. The velocity separation of the red and blue shifted components in this work is in agreement with the regression found by \cite{Relano05} for their \HII\ regions sample.

We also analyse the radial velocity behaviour within the galaxy, and 
observe a similar situation to that of NGC7479\,I. There is no difference, 
within the errors, when we compare our
average value derived from the H$\alpha$ emission line with the H$\alpha$
velocities field map \citep{1996PhDT........82L}. If, however, the comparison is done against 
the H\,I rotation velocity map \citep{1998MNRAS.297.1041L}, knot B
seems to follow galactic rotation whereas knot A does not.

\subsection{NGC7479\,III} 
\label{NGC7479III}

The observed spectrum of this region shows, although weak, the
[\OIII]$\lambda\lambda$4959,5007\AA\ lines. This is probably owed
to a difference in metallicity, NGC7479\,III being slightly less metallic than
the brightest two regions discussed above, or a difference in the ionisation
structure. The latter might be due to a relatively small
difference in the hardness of the ionisation radiation field (and, what is
equivalent, in the ionising stellar population) responsible for the excitation 
observed in the spectra or a difference in the nebular geometry (see a
detailed discussion in
\citealt{2007MNRAS.382..251D,2006MNRAS.372..293H,2008MNRAS.383..209H,2010MNRAS.tmp..386P}, and references therein). Unfortunately, as we said above, since no 
temperature sensitive lines can be observed, we are not able to
derive their metallicities, and
distinguish the origin of this small difference in the excitation
present in the spectra of
these regions of NGC\,7479.

\begin{figure*}
\begin{center}
\caption[Inicial fit]{{\sc Ngauss} fits with two Gaussian components in
  the NGC7479\,III emission line profiles. In order from left to right and
  from top to bottom: H$\alpha$, [\NII]6584\AA, [\SII]6717\AA and [\SII]6731\AA.}
  \label{figsingle} 
\includegraphics[trim=0cm 0cm 0cm 0cm,clip,angle=0,width=8cm,height=5cm]{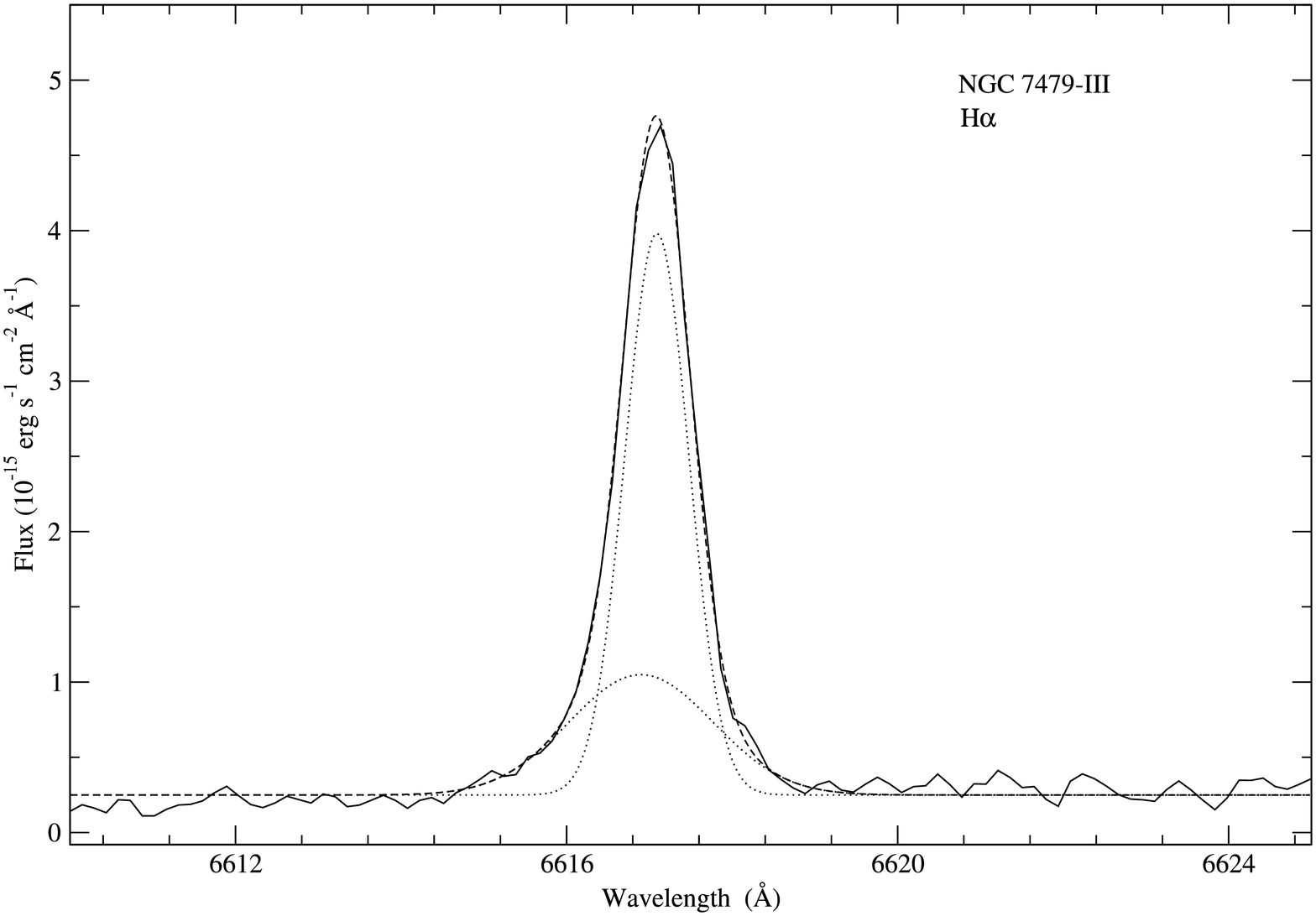}
\includegraphics[trim=0cm 0cm 0cm 0cm,clip,angle=0,width=8cm,height=5cm]{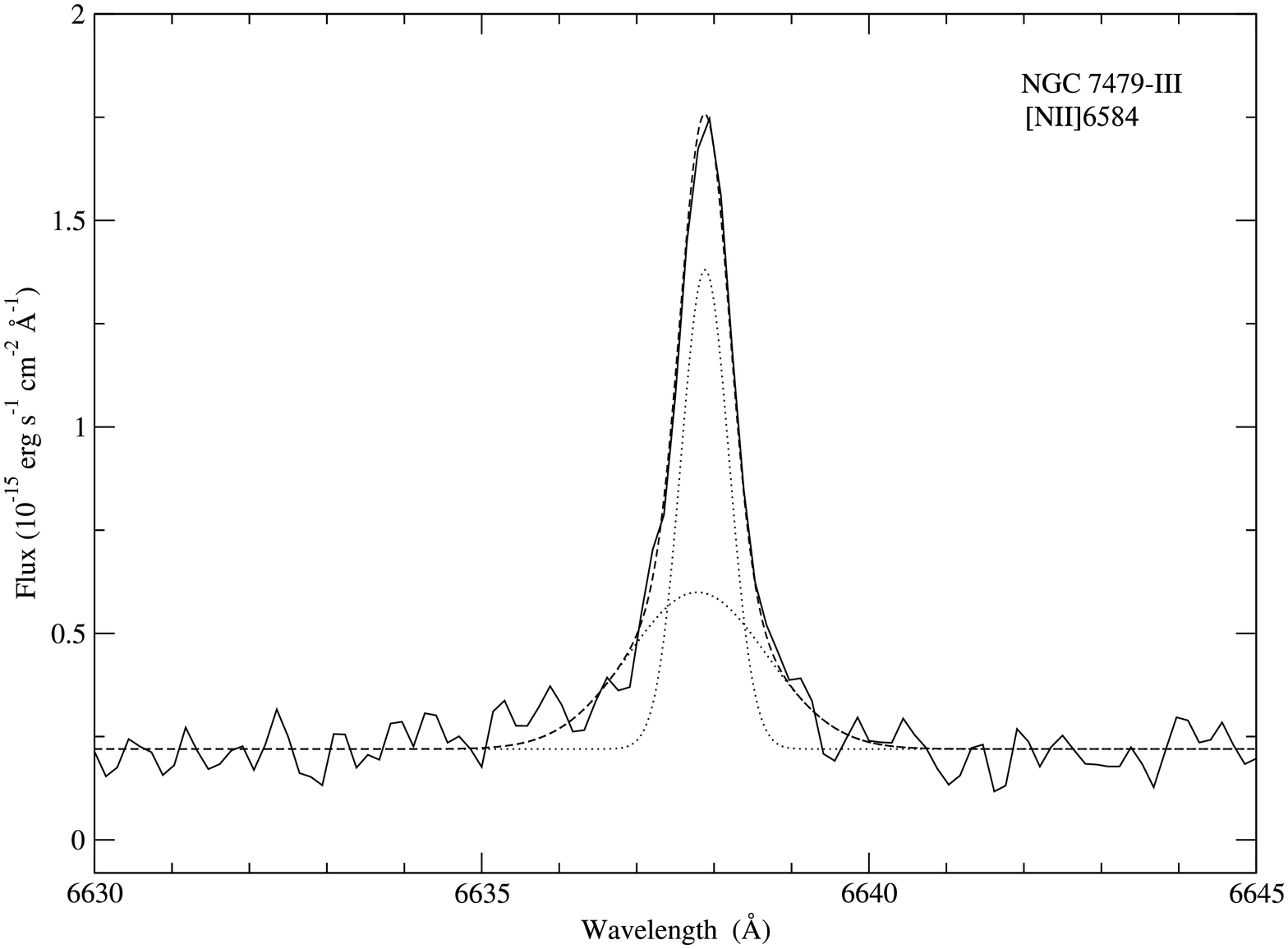}
\includegraphics[trim=0cm 0cm 0cm 0cm,clip,angle=0,width=8cm,height=5cm]{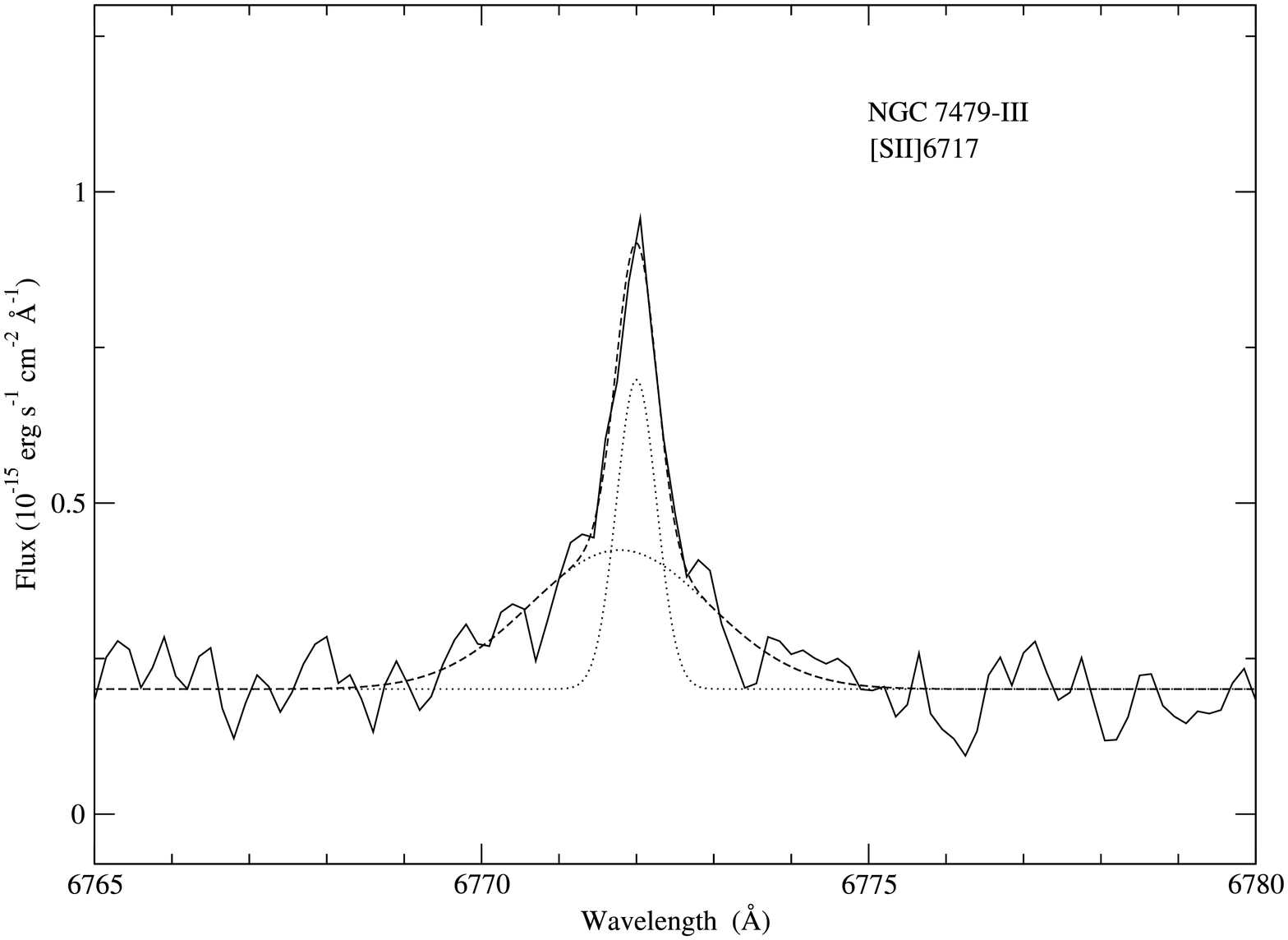}
\includegraphics[trim=0cm 0cm 0cm 0cm,clip,angle=0,width=8cm,height=5cm]{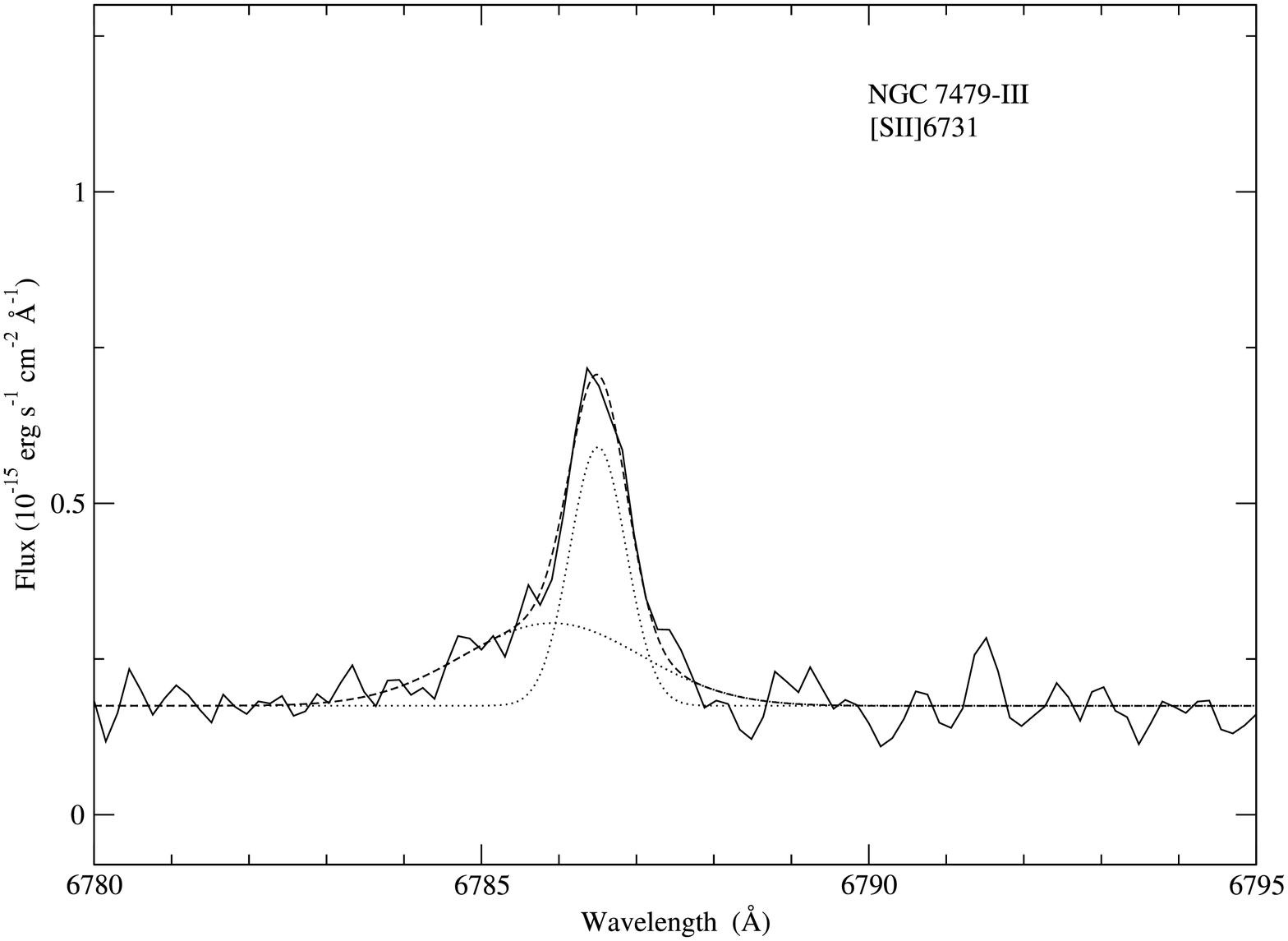}
\end{center}
\end{figure*}

We identify and fit the Gaussian profiles to the [\OIII]$\lambda$5007\AA,
H$\alpha$, [\NII]$\lambda$6584\AA\ and [\SII]$\lambda\lambda$6717,6731\AA\ 
lines in NGC7479\,III region with {\sc ngaussfit}.
In this region we have found one supersonic component, with no evidence of
multiple contributions but it is possible to detect flux excess on
the wings of the emission lines. We therefore fit a broad Gaussian component
together with the narrow component (see Figure \ref{figsingle}). 
 This procedure works well for almost all detected lines. In the case of
[\OIII]$\lambda$5007\AA , however, there are no prominent residuals when we
fit only one Gaussian component which has a dispersion of about 20\kms and we
are therefore not able to fit a broad Gaussian component. For the other listed
lines the derived radial velocities, the velocity dispersions (using
$\sigma^2_{i}$=5.4\kms\ and $T \simeq 10^{4}K$) and errors are listed in Table
\ref{tab7479_ngauss}. The overall H$\alpha$ flux, uncorrected for reddening, is found to be 5.43 x 10$^{-15}$\ergsc. It is worth mentioning that there seems to be some systematic offset
between the derived profile centres of the narrow and broad profiles which is, in average, about 10 \kms. This
happens for all lines, although it becomes more evident for [\SII]$\lambda$6731\AA.
An undiscovered blueshifted narrow component could account for the observed
behaviour, although better signal-to-noise data are needed to confirm this hypothesis.

The radial velocity average derived for the single component is in agreement with the expected radial velocity at the position of NGC7479\,III, showing only a slight offset of (10 \kms), well within the uncertainties.

\subsection{NGC6070\,I, II \& IV} 

All regions observed in NGC\,6070 show one narrow component together with the
underlying broad component. We identify and fit the Gaussian profiles to
the H$\beta$, [\OIII]$\lambda$5007\AA, H$\alpha$,
[\NII]$\lambda$6584\AA\ and [\SII]$\lambda\lambda$6717,6731\AA\ lines in
almost all regions. (see Figures \ref{dPfigngaussx2_6070I}-\ref{dPfigngaussx2_6070IV}). Table   
\ref{tab6070_ngauss} shows the derived kinematical parameters (using
$\sigma^2_{i}$ = 5.3, 5.6 and 5.3 \kms\ for regions I, II and IV
respectively; $T$ is set to $\simeq 10^{4}K$ for all regions).

\begin{table*}
\caption[NGC6070 comp]{Results of Gaussian profiles fitting to the observed emission lines in
NGC 6070. Each emission line is identified by its ion laboratory
wavelength and ion name in columns 1 and 2. According to the different
fits performed on each line, column 3 identifies each "narrow
component" (A and B, where applicable), a broad component and/or blue
and red wings (b wing and r wing respectively). Radial velocities
(V$_r$) and intrinsic velocity dispersions ($\sigma_{int}$)
together with their respective errors are expressed in \kms. The intrinsic velocity dispersions are corrected for the instrumental and thermal widths. Emission
measures (EM) are included and shown as a percentage of the component
flux relative to the total EM of the region. Upper panel: du Pont data and lover panel: Mike data.}
\label{tab6070_ngauss}
\begin{center}
\begin{tabular}{llllllllllllllllll}

\hline
\multicolumn{3}{c}{} & \multicolumn{5}{c}{\em NGC\,6070 I} &  \multicolumn{5}{c}{\em NGC\,6070 II} & \multicolumn{5}{c}{\em NGC\,6070 IV}\\
$\lambda_{0}$&ion&comp.&V$_r$&error&$\sigma_{int}$&error&EM&V$_r$&error&$\sigma_{int}$&error&EM&V$_r$&error&$\sigma_{int}$&error&EM\\
\hline
\\
\multicolumn{18}{c}{\em du Pont data} \\
\\
4860	&	H$\beta$	&	narrow	&	\ldots	&	\ldots	&	\ldots	&	\ldots	&	\ldots	&	1883.4	&	1.1	&	14.4	&	2.3	&	59	&	\ldots	&	\ldots	&	\ldots	&	\ldots	&	\ldots	\\
	&		&	broad	&	\ldots	&	\ldots	&	\ldots	&	\ldots	&	\ldots	&	1886.8	&	12.7	&	72.4	&	21.7	&	41	&	\ldots	&	\ldots	&	\ldots	&	\ldots	&	\ldots	\\
\\																																			
5007	&	[OIII]	&	narrow	&	1871.1	&	0.8	&	18.4	&	1.1	&	59	&	\ldots	&	\ldots	&	\ldots	&	\ldots	&	\ldots	&	1859.3	&	0.7	&	15.0	&	1.0	&	66	\\
	&		&	broad	&	1867.8	&	1.2	&	42.4	&	1.9	&	41	&	\ldots	&	\ldots	&	\ldots	&	\ldots	&	\ldots	&	1848.7	&	0.7	&	28.9	&	1.2	&	34	\\
\\																																			
6563	&	H$\alpha$	&	narrow	&	1868.3	&	0.6	&	17.2	&	0.8	&	69	&	1886.4	&	0.6	&	13.6	&	0.8	&	60	&	1858.0	&	0.6	&	12.2	&	0.8	&	62	\\
	&		&	broad	&	1868.3	&	0.9	&	42.9	&	2.0	&	31	&	1914.0	&	2.0	&	56.2	&	1.7	&	40	&	1848.4	&	0.7	&	31.2	&	0.9	&	38	\\
\\																																			
6584	&	[NII]	&	narrow	&	1869.7	&	0.6	&	16.6	&	1.1	&	47	&	1885.6	&	0.9	&	17.2	&	0.9	&	55	&	1863.4	&	0.8	&	13.5	&	0.9	&	54	\\
	&		&	broad	&	1872.6	&	0.6	&	34.2	&	2.4	&	53	&	1908.1	&	2.3	&	40.7	&	2.6	&	45	&	1848.4	&	0.6	&	25.5	&	2.0	&	46	\\
\\																																			
6717	&	[SII]	&	narrow	&	1865.2	&	1.7	&	21.2	&	3.2	&	55	&	\ldots	&	\ldots	&	\ldots	&	\ldots	&	\ldots	&	1857.0	&	1.4	&	16.8	&	2.6	&	49	\\
	&		&	broad	&	1868.3	&	3.3	&	37.3	&	4.5	&	45	&	\ldots	&	\ldots	&	\ldots	&	\ldots	&	\ldots	&	1844.3	&	6.1	&	33.0	&	5.0	&	51	\\
\\																																			
6731	&	[SII] 	&	narrow	&	1864.7	&	1.0	&	14.3	&	1.3	&	33	&	1883.5	&	1.3	&	23.1	&	1.9	&	79	&	\ldots	&	\ldots	&	\ldots	&	\ldots	&	\ldots	\\
	&		&	broad	&	1870.3	&	1.6	&	35.4	&	1.9	&	67	&	1942.7	&	17.4	&	44.2	&	17.1	&	21	&	\ldots	&	\ldots	&	\ldots	&	\ldots	&	\ldots	\\
\\
\multicolumn{18}{c}{\em MIKE data} \\
\\
4860	&	H$\beta$	&	narrow	&	\ldots	&	\ldots	&	\ldots	&	\ldots	&	\ldots	&	1890.9	&	0.3	&	20.0	&	0.4	&	61	&	\ldots	&	\ldots	&	\ldots	&	\ldots	&	\ldots	\\
	&		&	broad	&	\ldots	&	\ldots	&	\ldots	&	\ldots	&	\ldots	&	1902.8	&	1.1	&	60.3	&	1.6	&	39	&	\ldots	&	\ldots	&	\ldots	&	\ldots	&	\ldots	\\
\\																																			
5007	&	[OIII]	&	narrow	&	1868.4	&	0.3	&	13.6	&	0.4	&	49	&	\ldots	&	\ldots	&	\ldots	&	\ldots	&	\ldots	&	1857.3	&	0.3	&	16.7	&	0.4	&	64	\\
	&		&	broad	&	1864.7	&	0.5	&	35.4	&	0.8	&	51	&	\ldots	&	\ldots	&	\ldots	&	\ldots	&	\ldots	&	1855.9	&	0.9	&	46.0	&	2.0	&	36	\\
\\																																			
6563	&	H$\alpha$	&	narrow	&	1866.9	&	0.2	&	13.4	&	0.3	&	62	&	1890.4	&	0.2	&	18.4	&	0.3	&	62	&	1856.8	&	0.2	&	11.9	&	0.3	&	58	\\
	&		&	broad	&	1862.4	&	0.3	&	37.2	&	0.4	&	38	&	1901.1	&	0.6	&	55.2	&	0.9	&	38	&	1851.6	&	0.4	&	34.6	&	0.5	&	42	\\
\\																																			
6584	&	[NII]	&	narrow	&	1870.2	&	0.2	&	12.3	&	0.4	&	39	&	1892.6	&	0.2	&	21.3	&	0.3	&	68	&	1861.8	&	0.3	&	10.3	&	0.6	&	34	\\
	&		&	broad	&	1867.3	&	0.4	&	29.5	&	0.6	&	61	&	1902.7	&	1.1	&	62.5	&	1.1	&	32	&	1855.6	&	0.6	&	27.0	&	0.9	&	66	\\
\\																																			
6717	&	[SII]	&	narrow	&	1868.7	&	0.2	&	13.3	&	0.3	&	38	&	\ldots	&	\ldots	&	\ldots	&	\ldots	&	\ldots	&	1860.4	&	0.3	&	12.0	&	0.5	&	33	\\
	&		&	broad	&	1865.2	&	0.4	&	31.6	&	0.5	&	62	&	\ldots	&	\ldots	&	\ldots	&	\ldots	&	\ldots	&	1854.2	&	0.4	&	29.3	&	0.6	&	67	\\
\\																																			
6731	&	[SII] 	&	narrow	&	1868.6	&	0.5	&	11.3	&	1.0	&	26	&	1889.1	&	0.3	&	23.0	&	0.5	&	70	&	\ldots	&	\ldots	&	\ldots	&	\ldots	&	\ldots	\\
	&		&	broad	&	1866.4	&	0.7	&	27.1	&	0.8	&	74	&	1896.7	&	2.5	&	49.8	&	4.4	&	30	&	\ldots	&	\ldots	&	\ldots	&	\ldots	&	\ldots	\\
\hline 
\end{tabular}
\end{center}
\end{table*}

For these three regions, we are also able to compare the du Pont data with
higher resolution spectra data obtained with the double echelle spectrograph
MIKE in July 2004. The agreement in the kinematical parameters can be seen in Table
\ref{tab6070_ngauss}, which shows the reliability of the data here
presented. The overall H$\alpha$ flux, uncorrected for reddening, for du Pont data are 1.68 x 10$^{-13}$, 6.26 x 10$^{-14}$ and 2.15 x 10$^{-13}$\ergsc\ for regions I, II and IV respectively. And for MIKE data are 1.74 x 10$^{-13}$, 1.29 x 10$^{-13}$ and 1.88 x 10$^{-13}$\ergsc. Only a slight bias towards lower velocity dispersion can be seen
for NGC6070\,I, which can be probably attributed to differences in the
centering of the slit during acquisition of the target. There seems to be some systematic offset
between the derived profile centres of the narrow and broad profiles. Theses values are, in average, about -25\kms\ for du Pont data and -10\kms\ for MIKE data in the case of NGC\,6070 II, and about 10\kms\ and 5\kms\ for NGC\,6070 IV.

\begin{figure*}
\begin{center}
\caption[Ajuste 6070\,I]{{\sc Ngauss} fits with two Gaussian components in the NGC6070\,I emission line profiles. du Pont data on the left side and MIKE data on the right side. In order from top to bottom panels: [\OIII]5007\AA, H$\alpha$, [\NII]6584\AA and
  [\SII]6717\AA.} \label{dPfigngaussx2_6070I}
\includegraphics[trim=0cm 0cm 0cm 0cm,clip,angle=0,width=8cm,height=5cm]{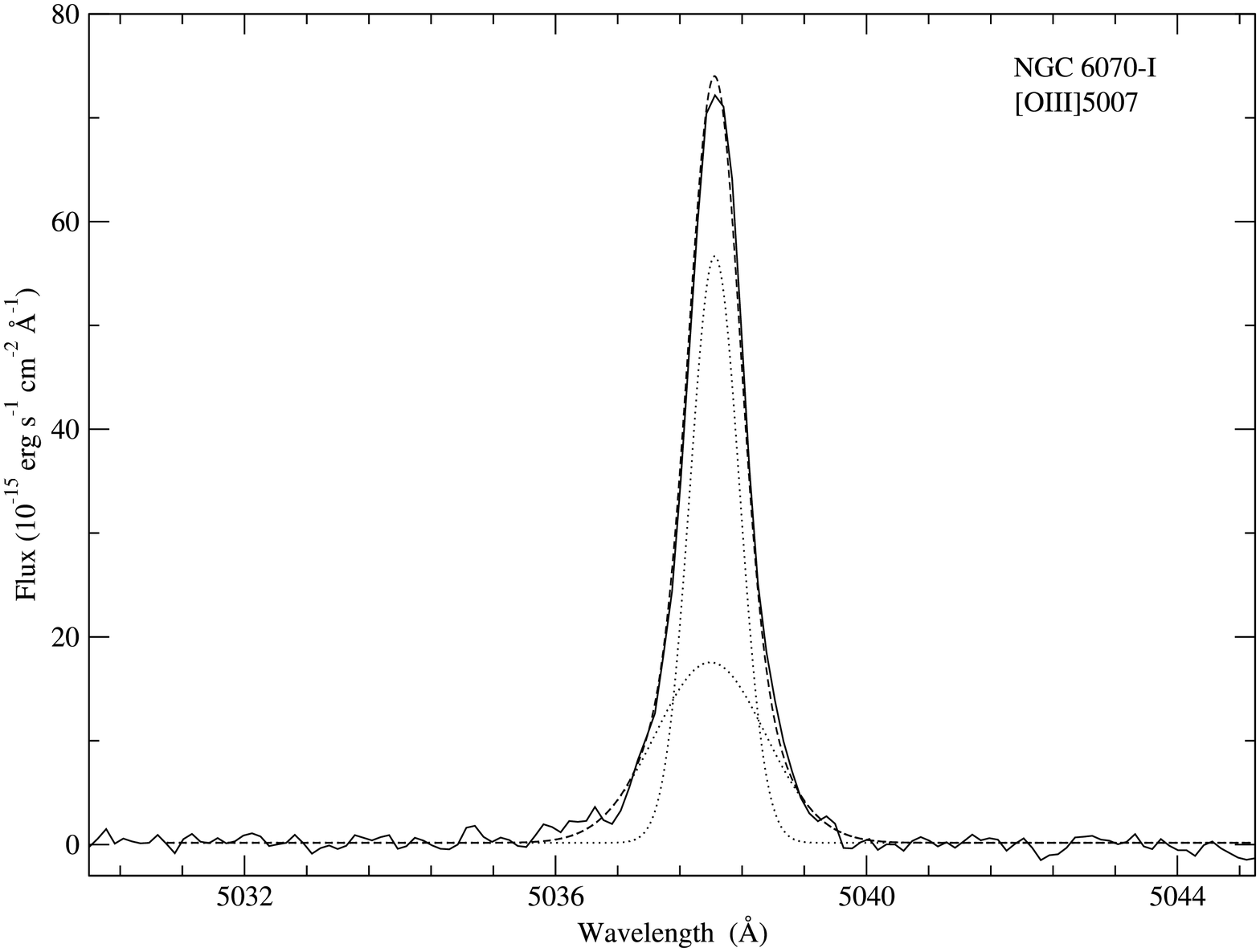}
\includegraphics[trim=0cm 0cm 0cm 0cm,clip,angle=0,width=8cm,height=5cm]{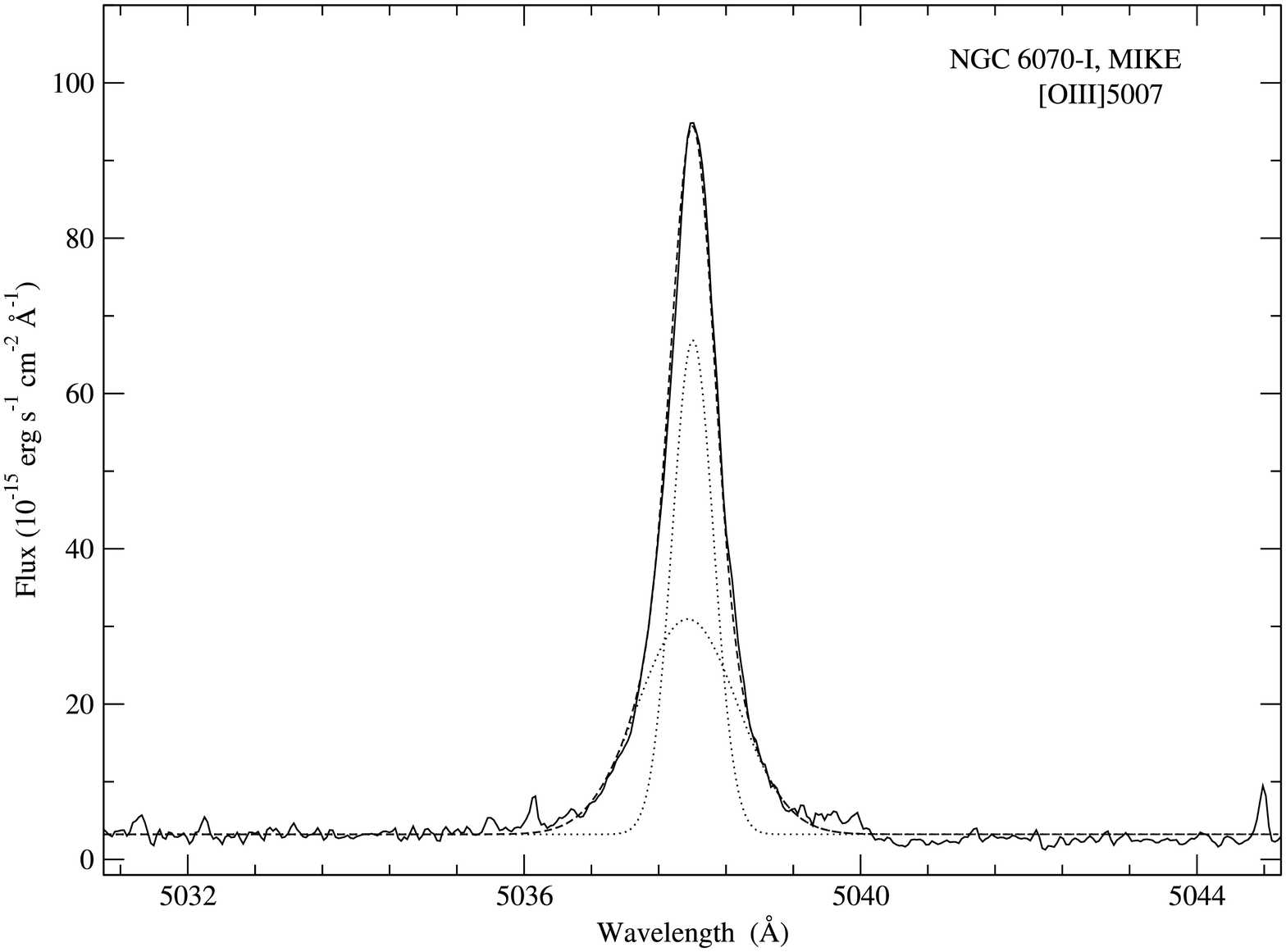}
\includegraphics[trim=0cm 0cm 0cm 0cm,clip,angle=0,width=8cm,height=5cm]{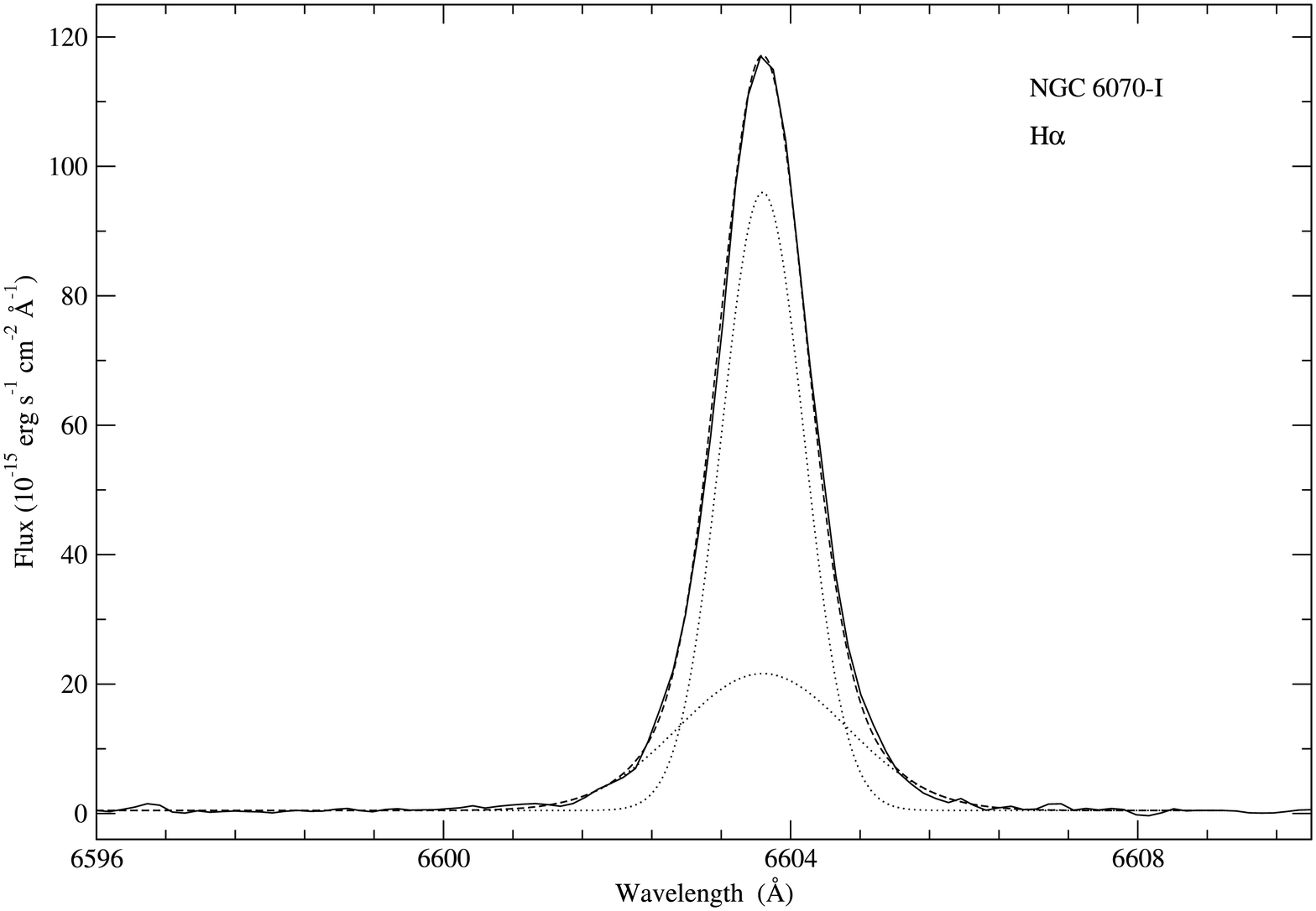}
\includegraphics[trim=0cm 0cm 0cm 0cm,clip,angle=0,width=8cm,height=5cm]{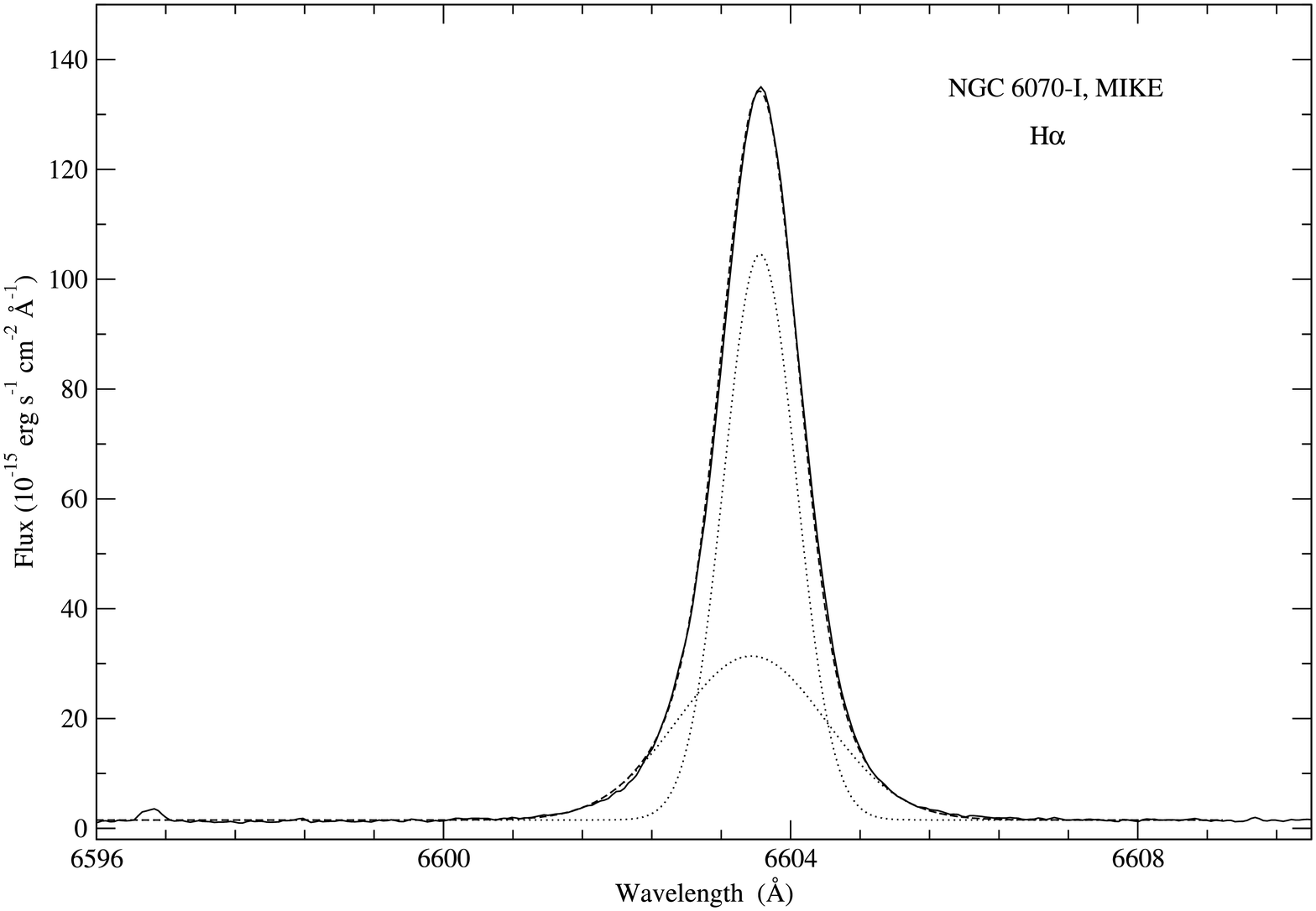}
\includegraphics[trim=0cm 0cm 0cm 0cm,clip,angle=0,width=8cm,height=5cm]{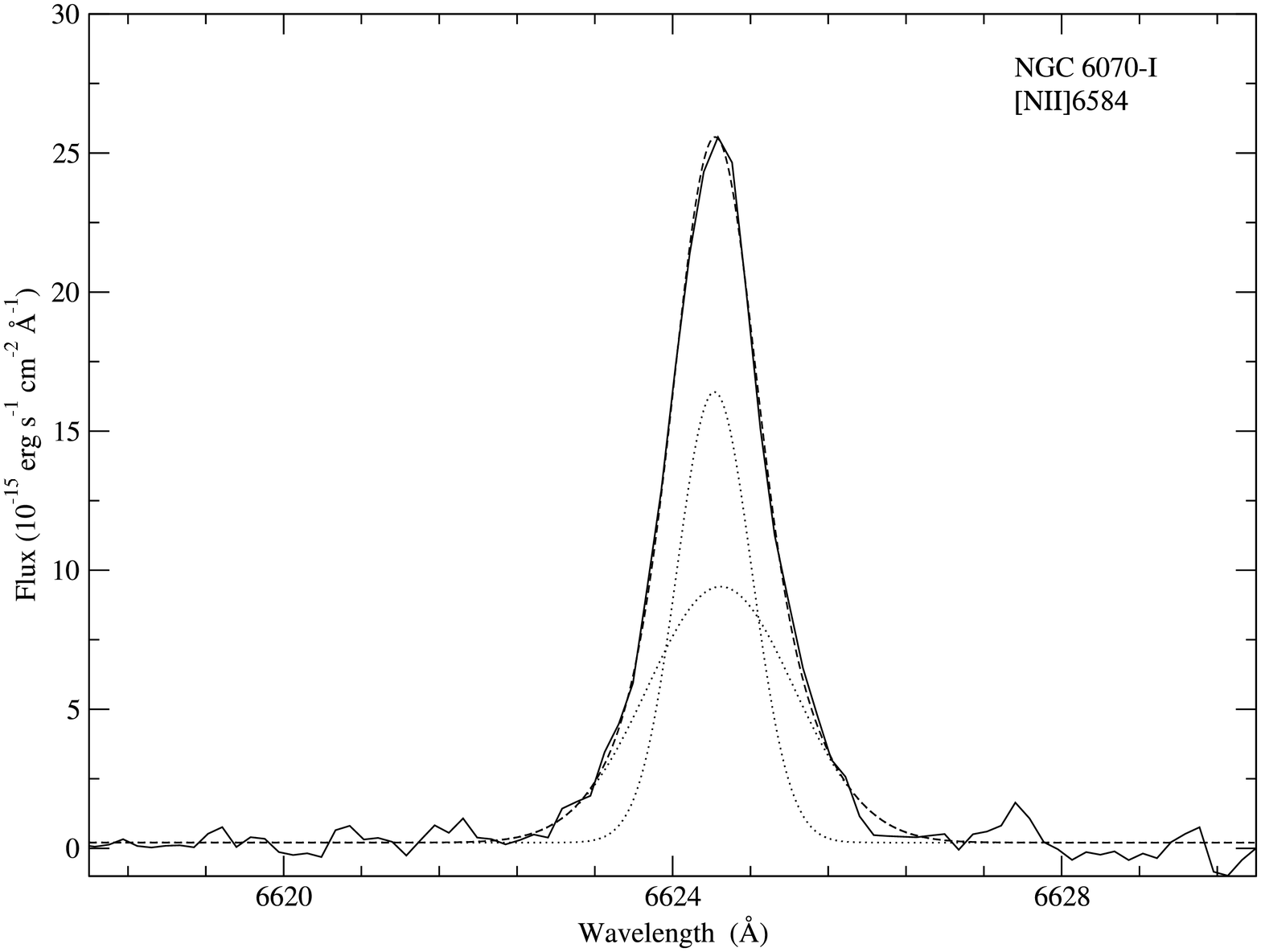}
\includegraphics[trim=0cm 0cm 0cm 0cm,clip,angle=0,width=8cm,height=5cm]{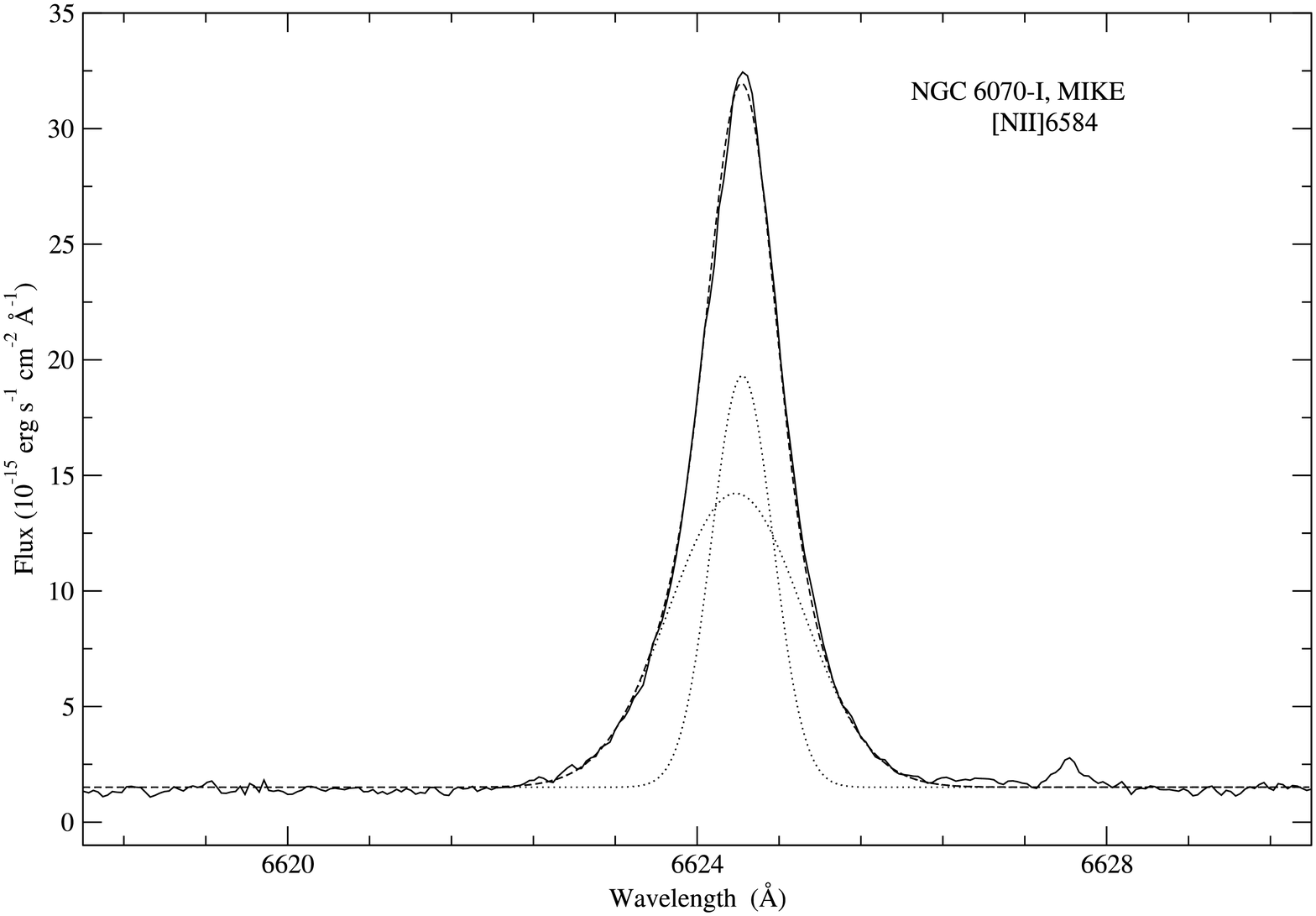}
\includegraphics[trim=0cm 0cm 0cm 0cm,clip,angle=0,width=8cm,height=5cm]{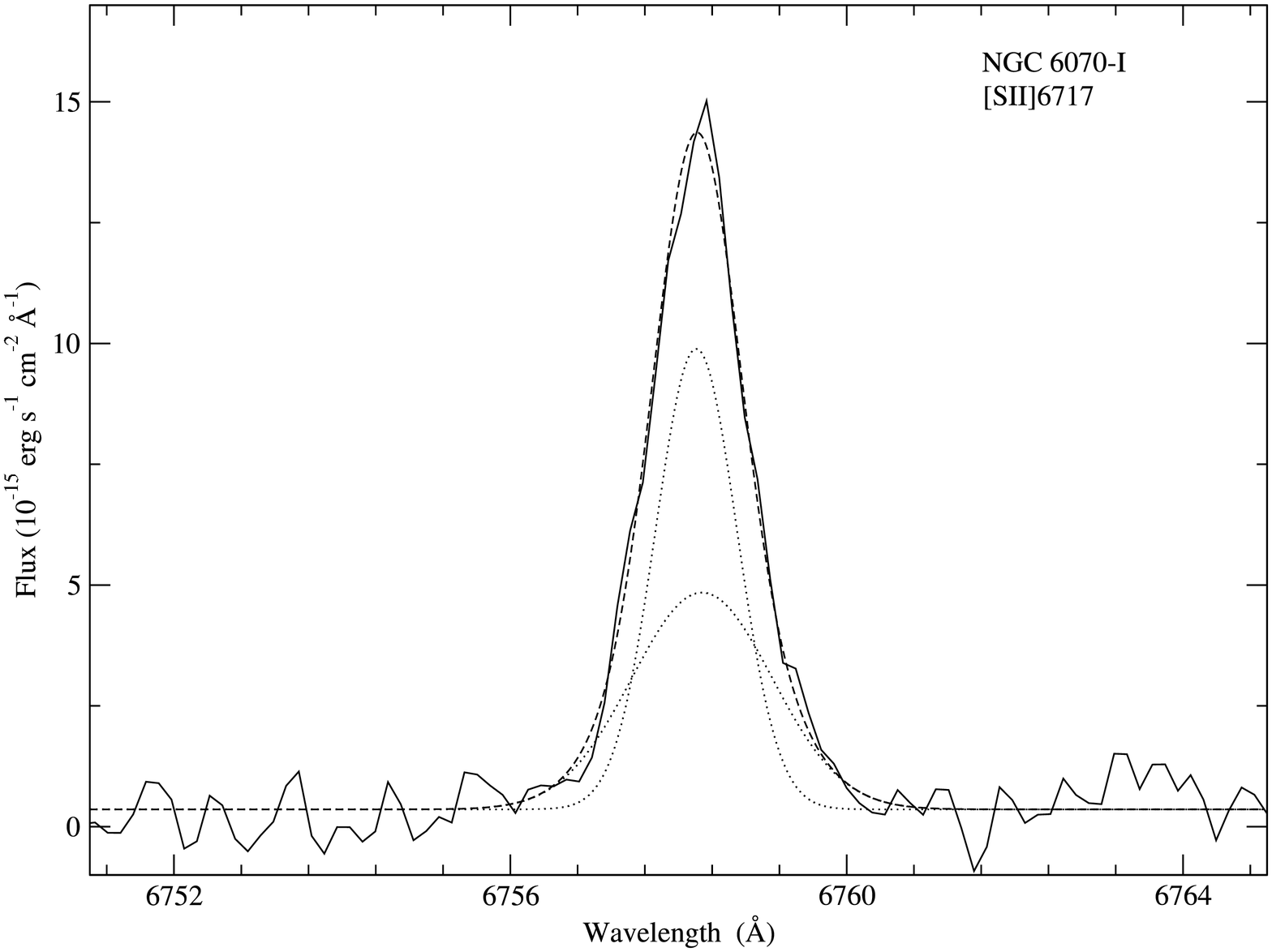}
\includegraphics[trim=0cm 0cm 0cm 0cm,clip,angle=0,width=8cm,height=5cm]{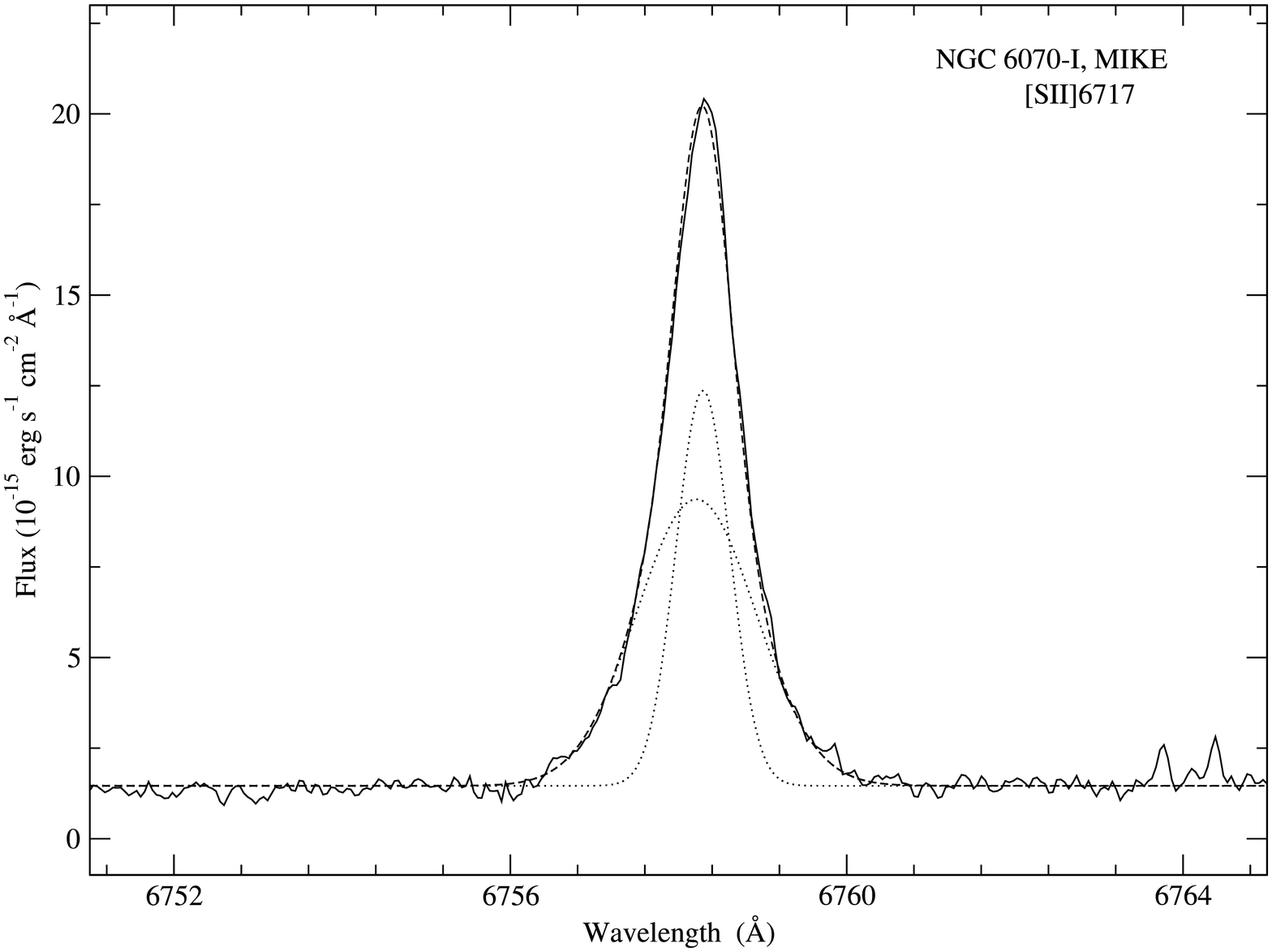}
\end{center}
\end{figure*}

\setcounter{figure}{7}
\begin{figure*}
\begin{center}
\caption[Ajuste 6070\,I]{({\it continued})} 
\includegraphics[trim=0cm 0cm 0cm 0cm,clip,angle=0,width=8cm,height=5cm]{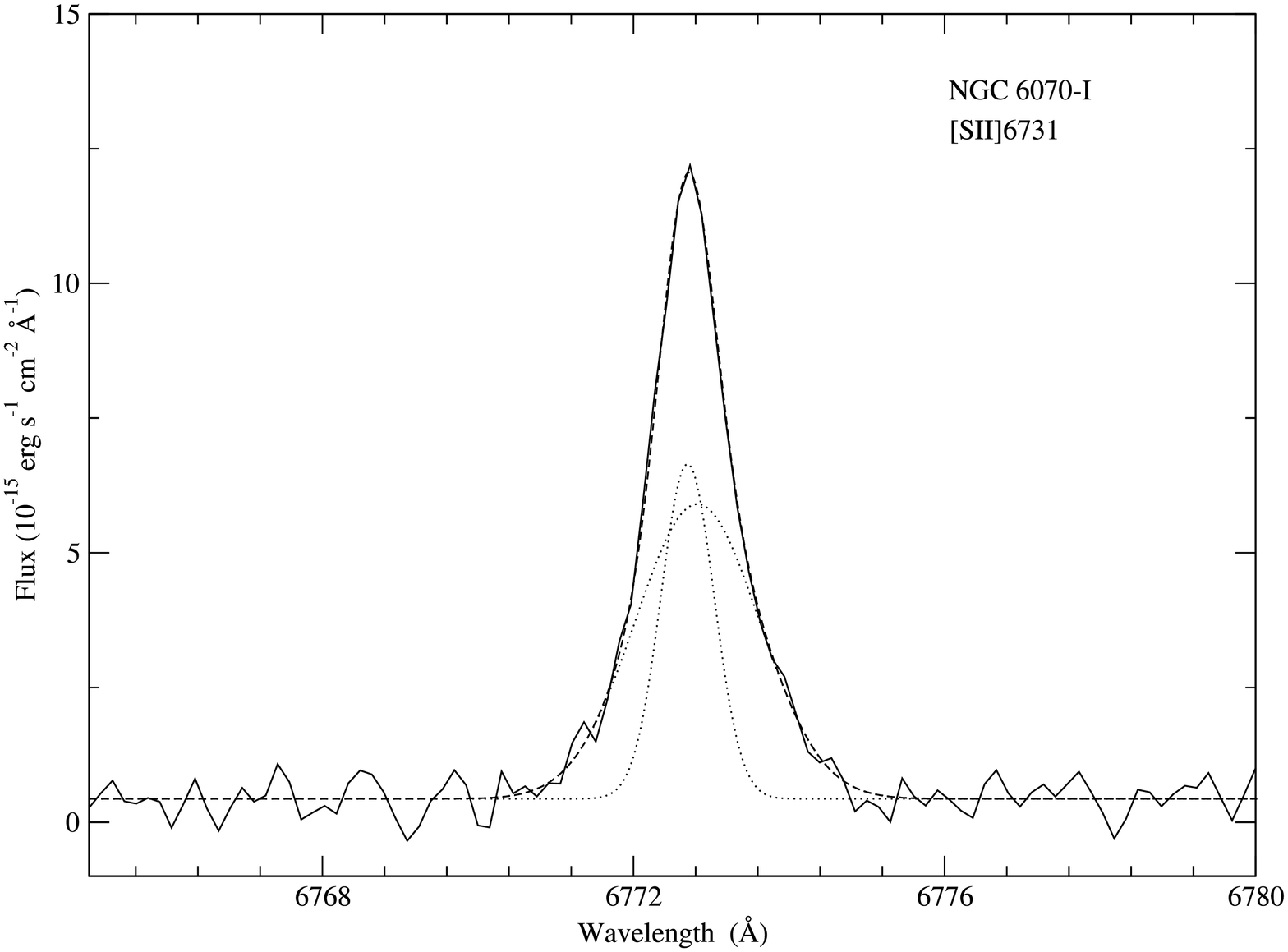}
\includegraphics[trim=0cm 0cm 0cm 0cm,clip,angle=0,width=8cm,height=5cm]{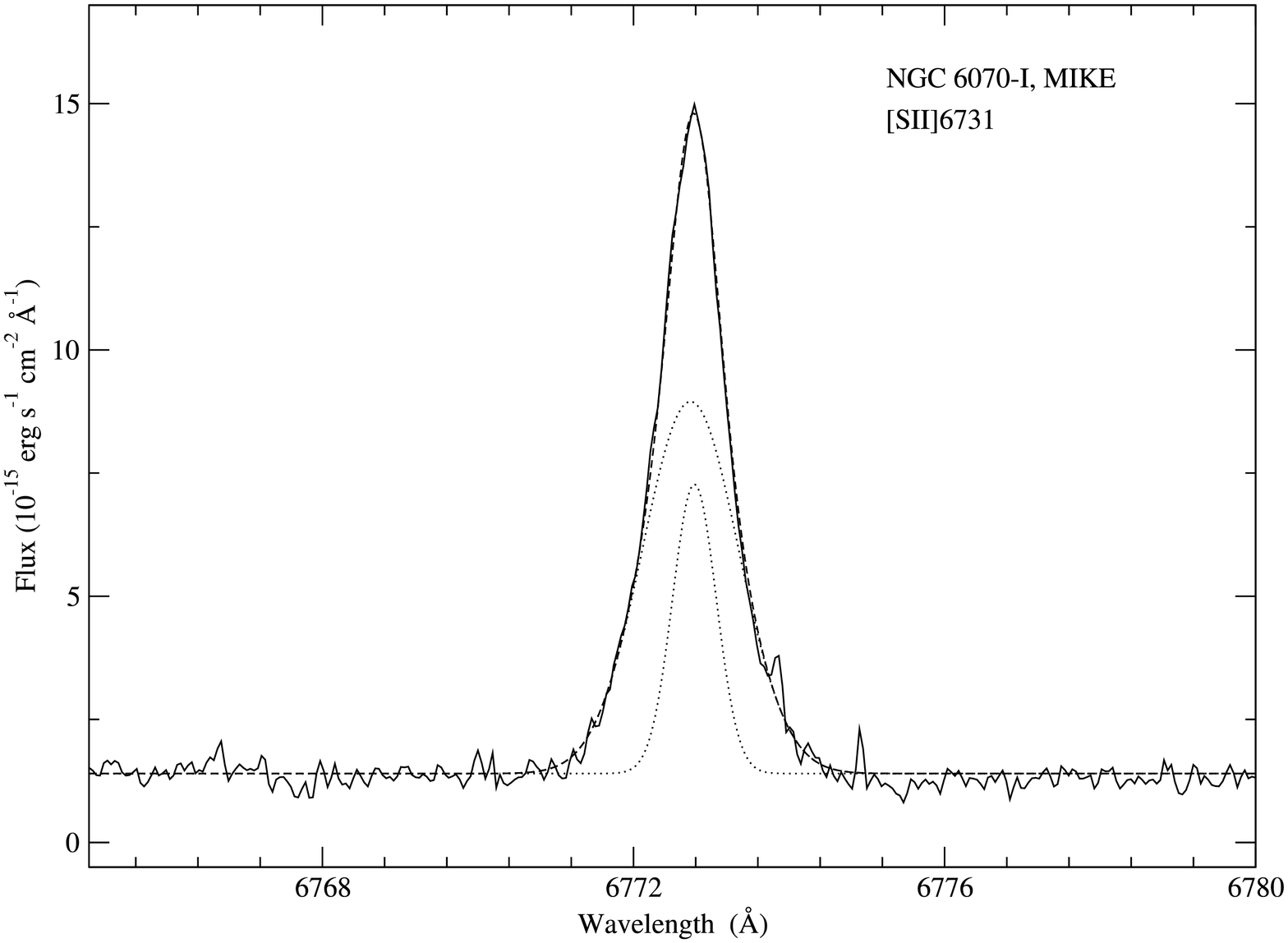}
\end{center}
\end{figure*}

\begin{figure*}
\begin{center}
\caption[Ajuste 6070\,II]{{\sc Ngauss} fits with two Gaussian components in the
  NGC6070\,II emission line profiles. du Pont data on the left side and MIKE data on the right side. In order from top to bottom panels: H$\beta$, H$\alpha$,
  [\NII]6584\AA, and [\SII]6731\AA.} \label{dPfigngaussx2_6070II}
\includegraphics[trim=0cm 0cm 0cm 0cm,clip,angle=0,width=8cm,height=5cm]{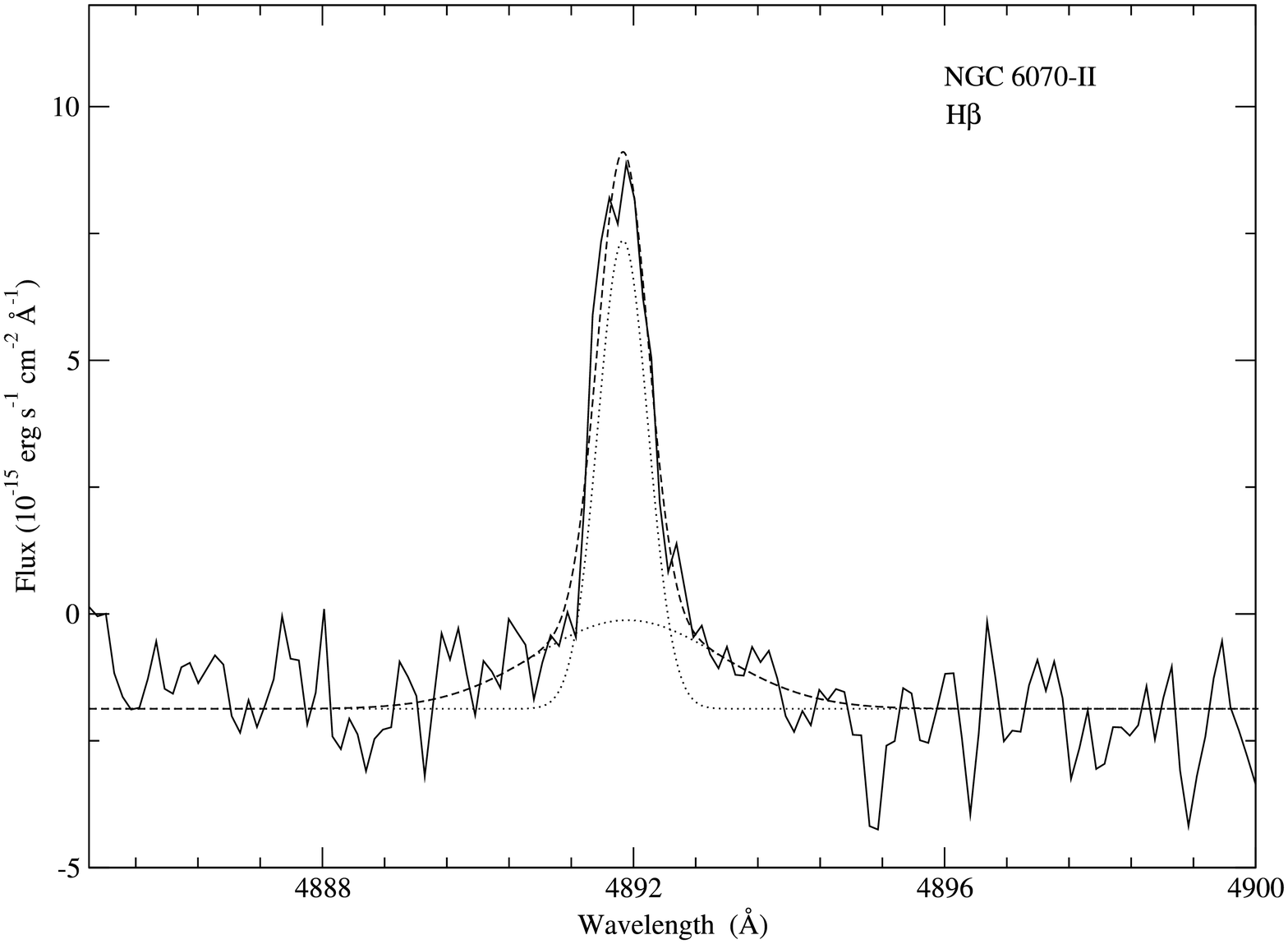}
\includegraphics[trim=0cm 0cm 0cm 0cm,clip,angle=0,width=8cm,height=5cm]{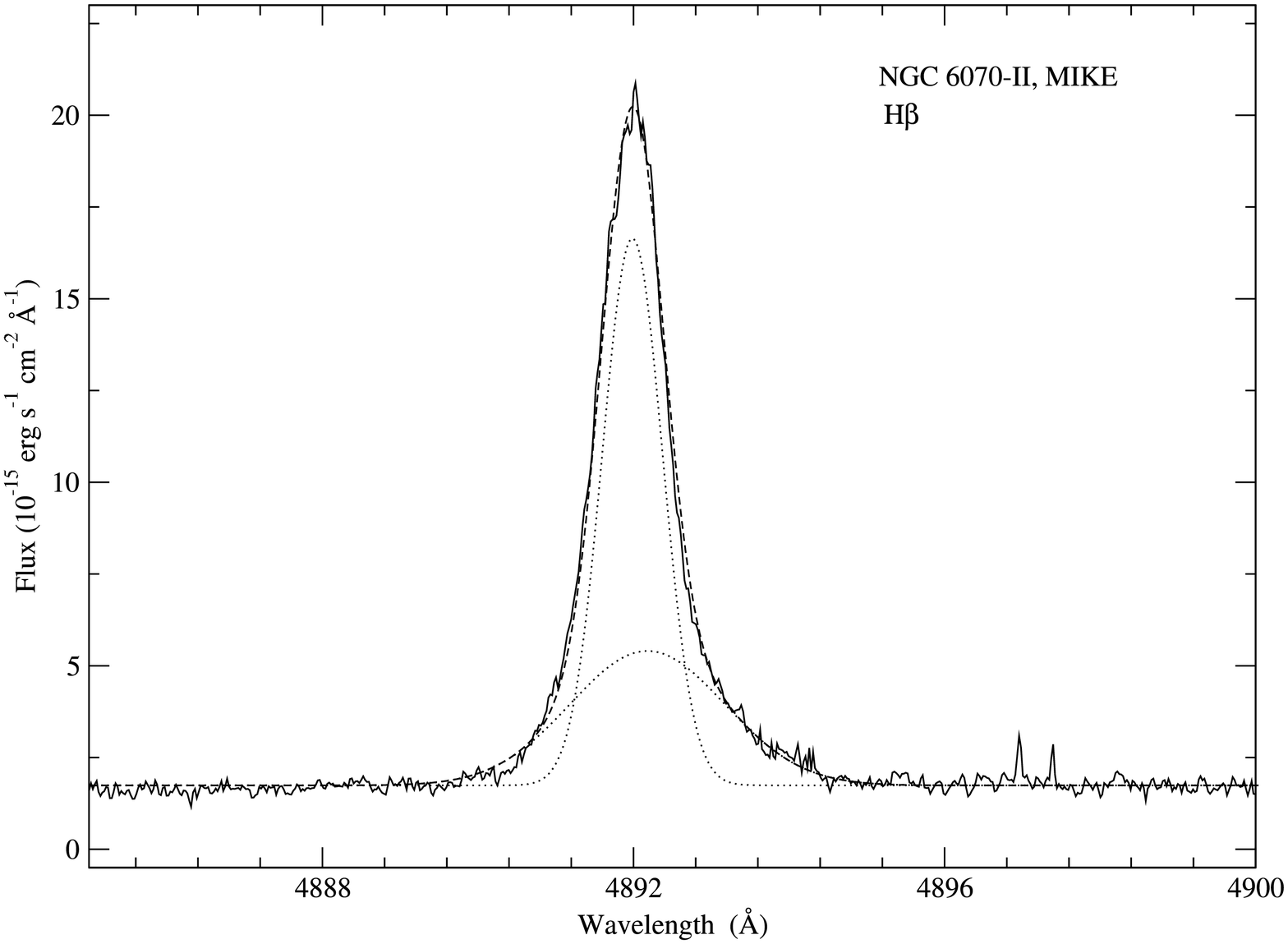}
\includegraphics[trim=0cm 0cm 0cm 0cm,clip,angle=0,width=8cm,height=5cm]{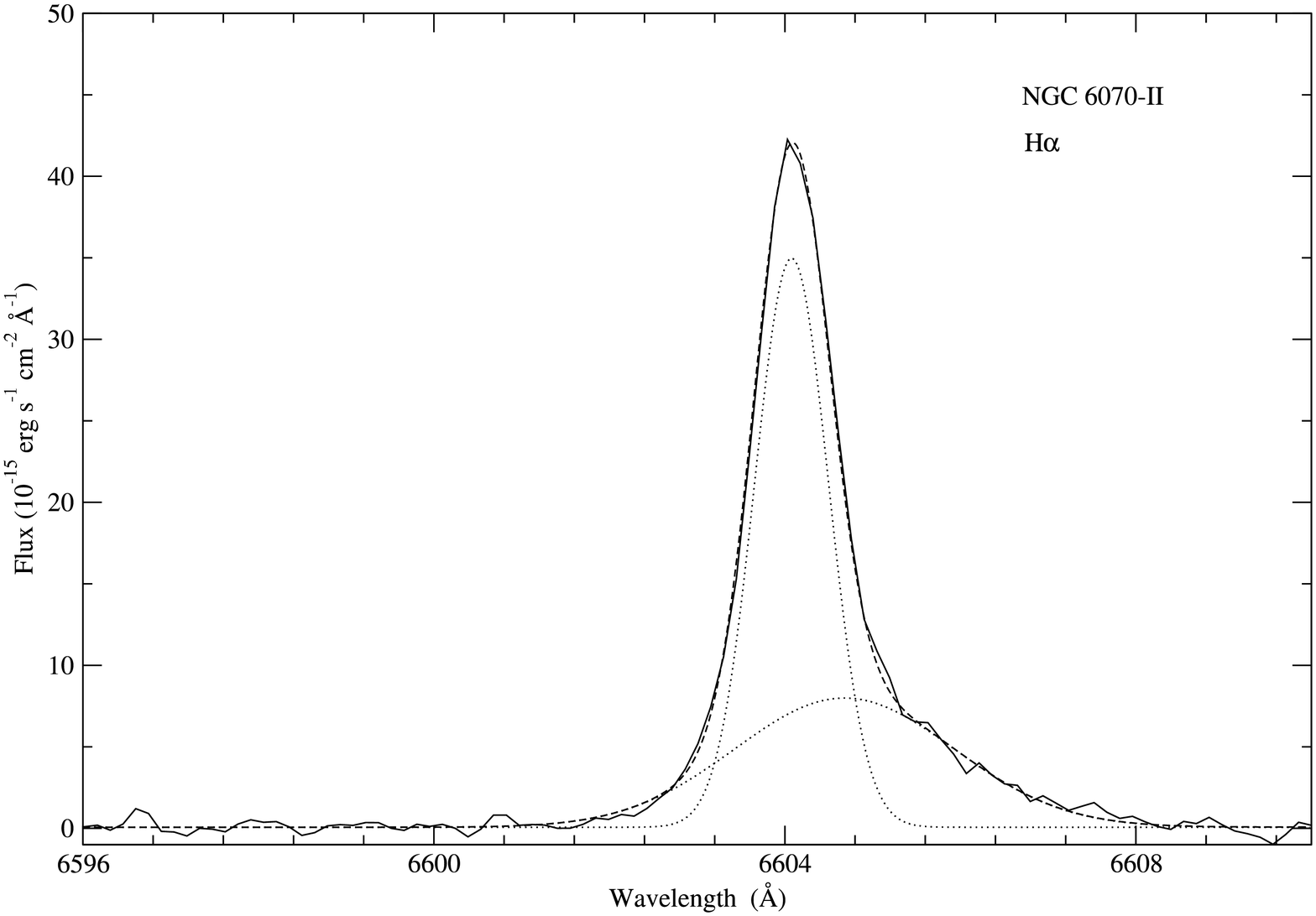}
\includegraphics[trim=0cm 0cm 0cm 0cm,clip,angle=0,width=8cm,height=5cm]{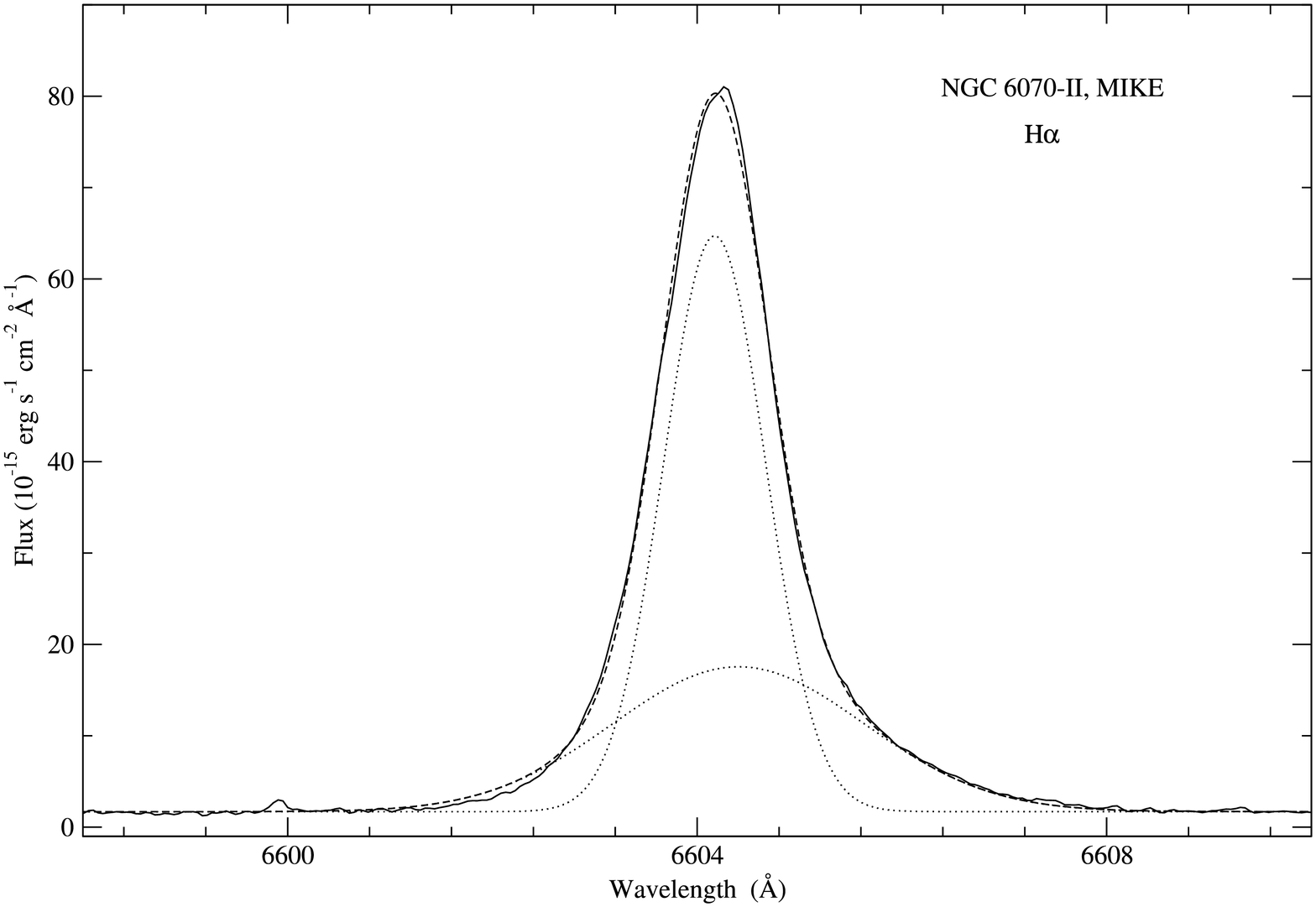}
\includegraphics[trim=0cm 0cm 0cm 0cm,clip,angle=0,width=8cm,height=5cm]{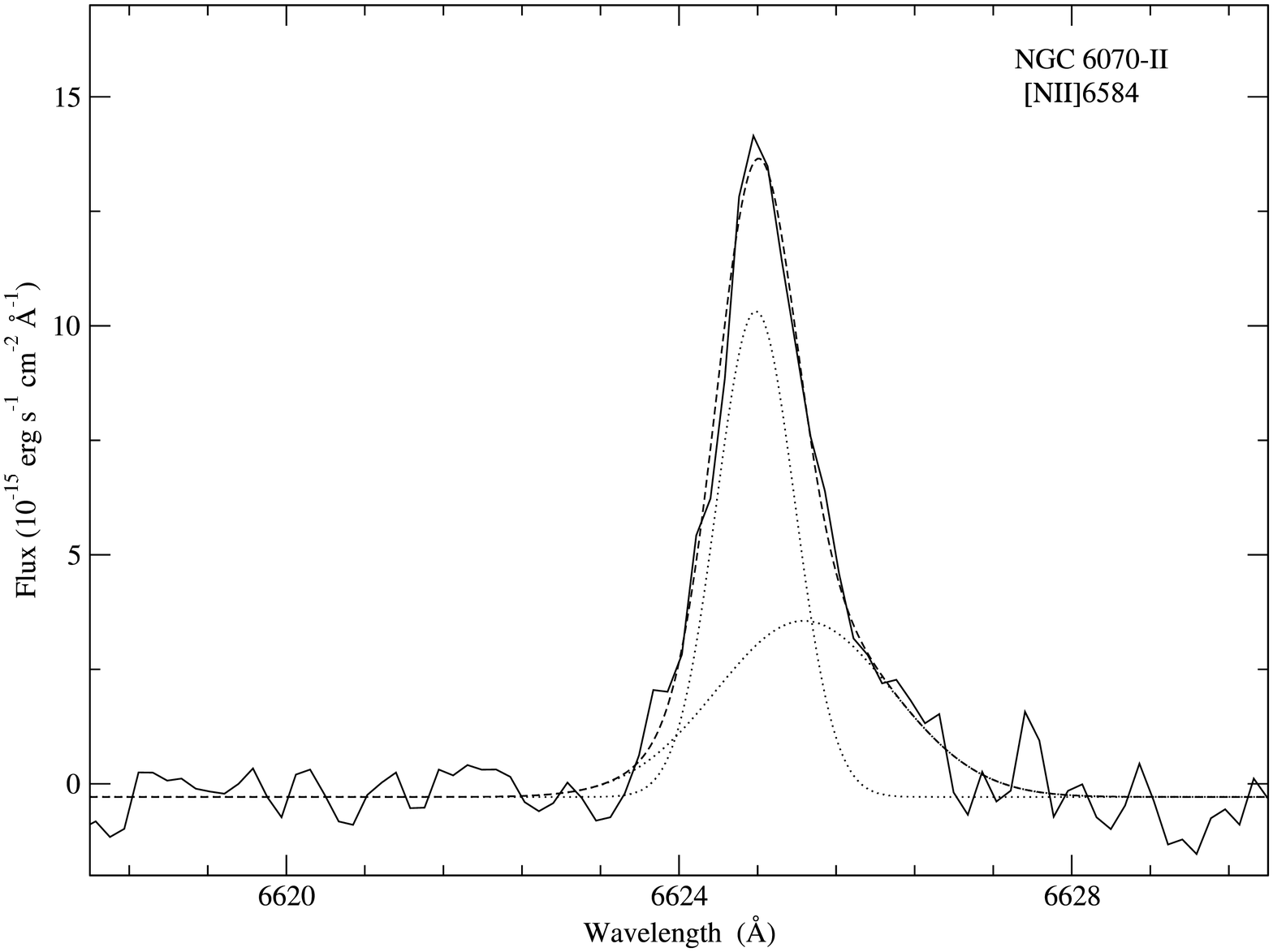}
\includegraphics[trim=0cm 0cm 0cm 0cm,clip,angle=0,width=8cm,height=5cm]{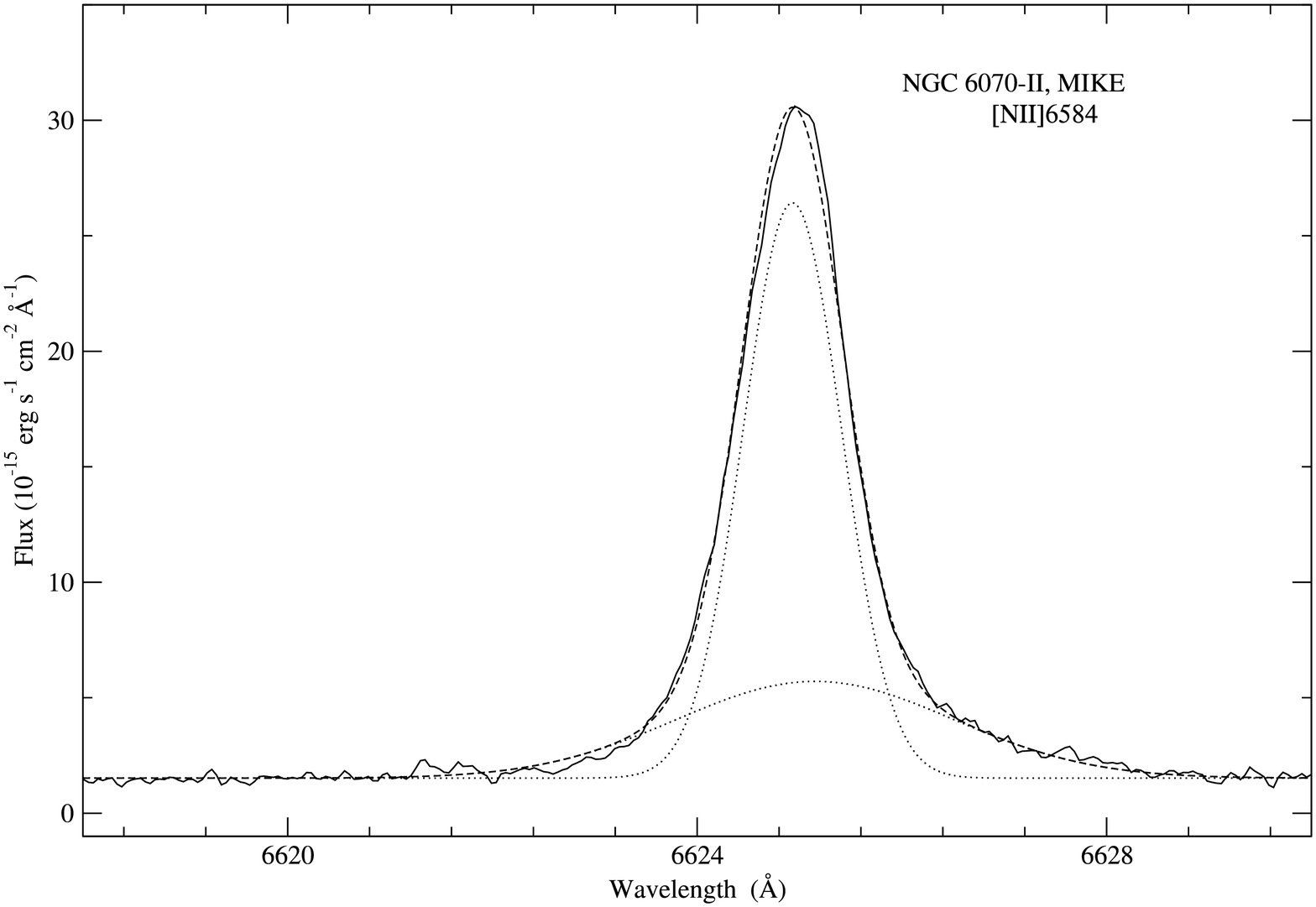}
\includegraphics[trim=0cm 0cm 0cm 0cm,clip,angle=0,width=8cm,height=5cm]{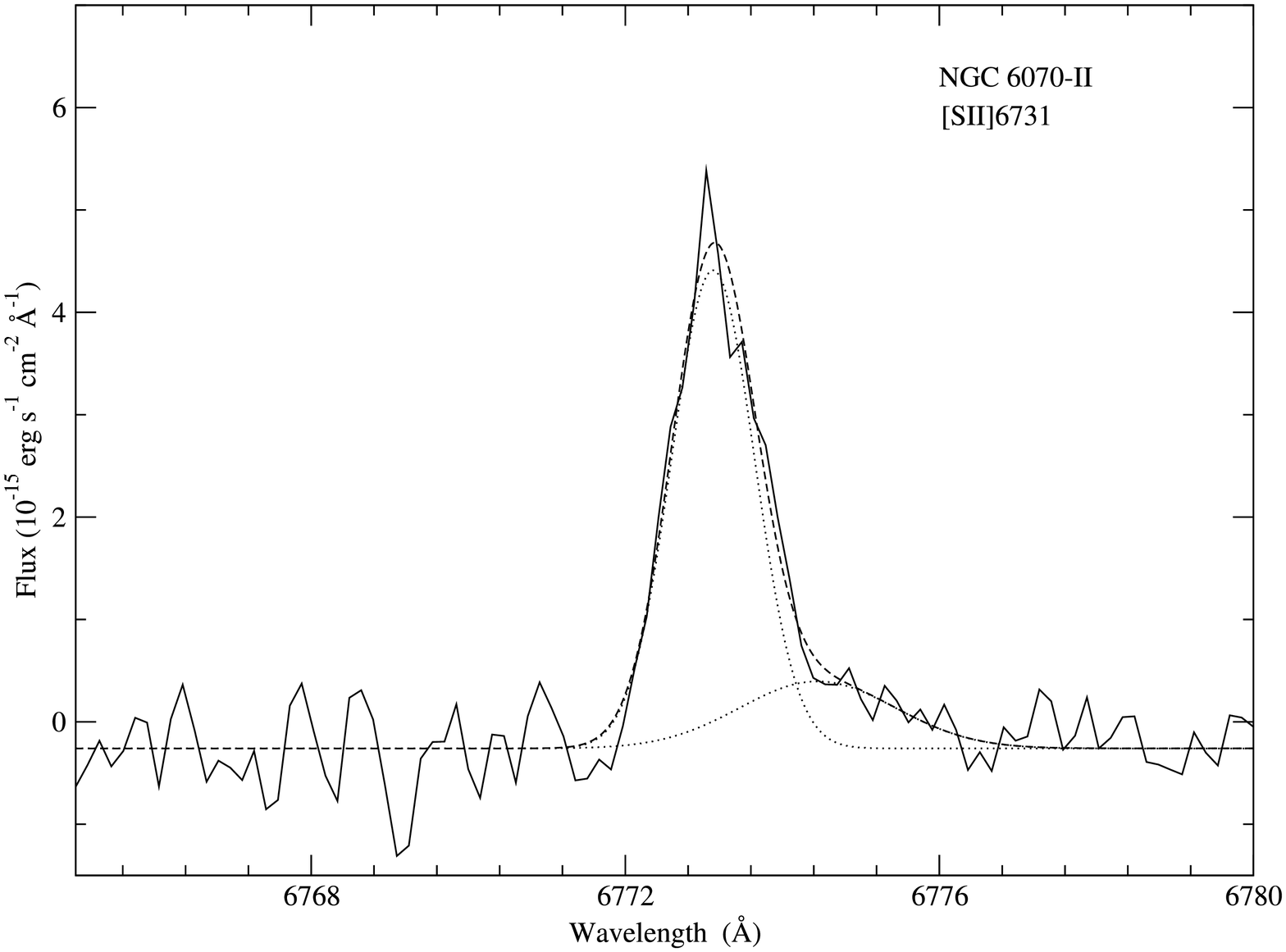}
\includegraphics[trim=0cm 0cm 0cm 0cm,clip,angle=0,width=8cm,height=5cm]{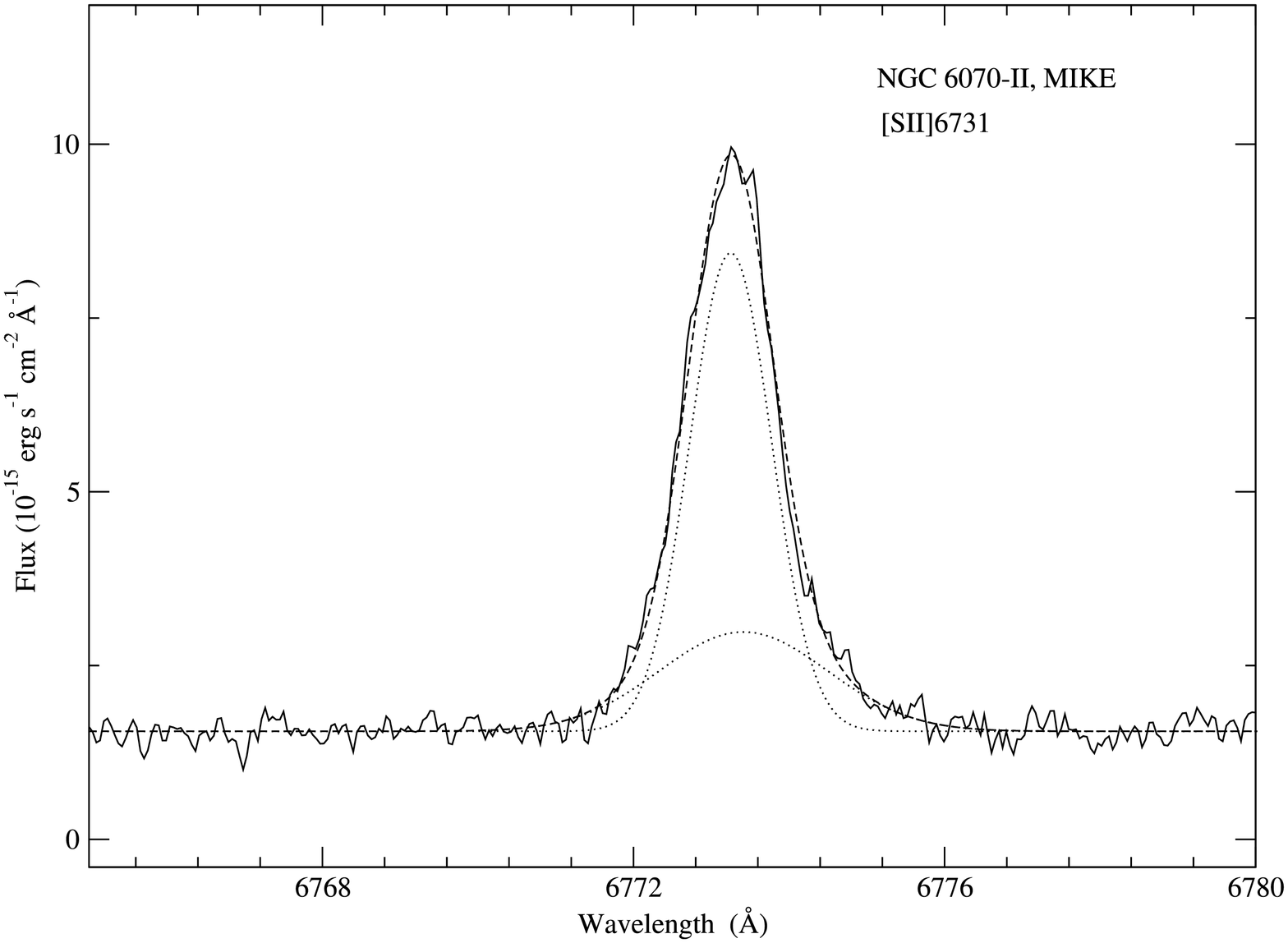}
\end{center}
\end{figure*}

\begin{figure*}
\begin{center}
\caption[Ajuste 6070\,IV]{{\sc Ngauss} fits with two Gaussian components in the
  NGC6070\,IV emission line profiles. du Pont data on the left side and MIKE data on the right side. In order from top to bottom panels: [\OIII]5007\AA, H$\alpha$,
  [\NII]6584\AA and [\SII]6717\AA.} \label{dPfigngaussx2_6070IV}
\includegraphics[trim=0cm 0cm 0cm 0cm,clip,angle=0,width=8cm,height=5cm]{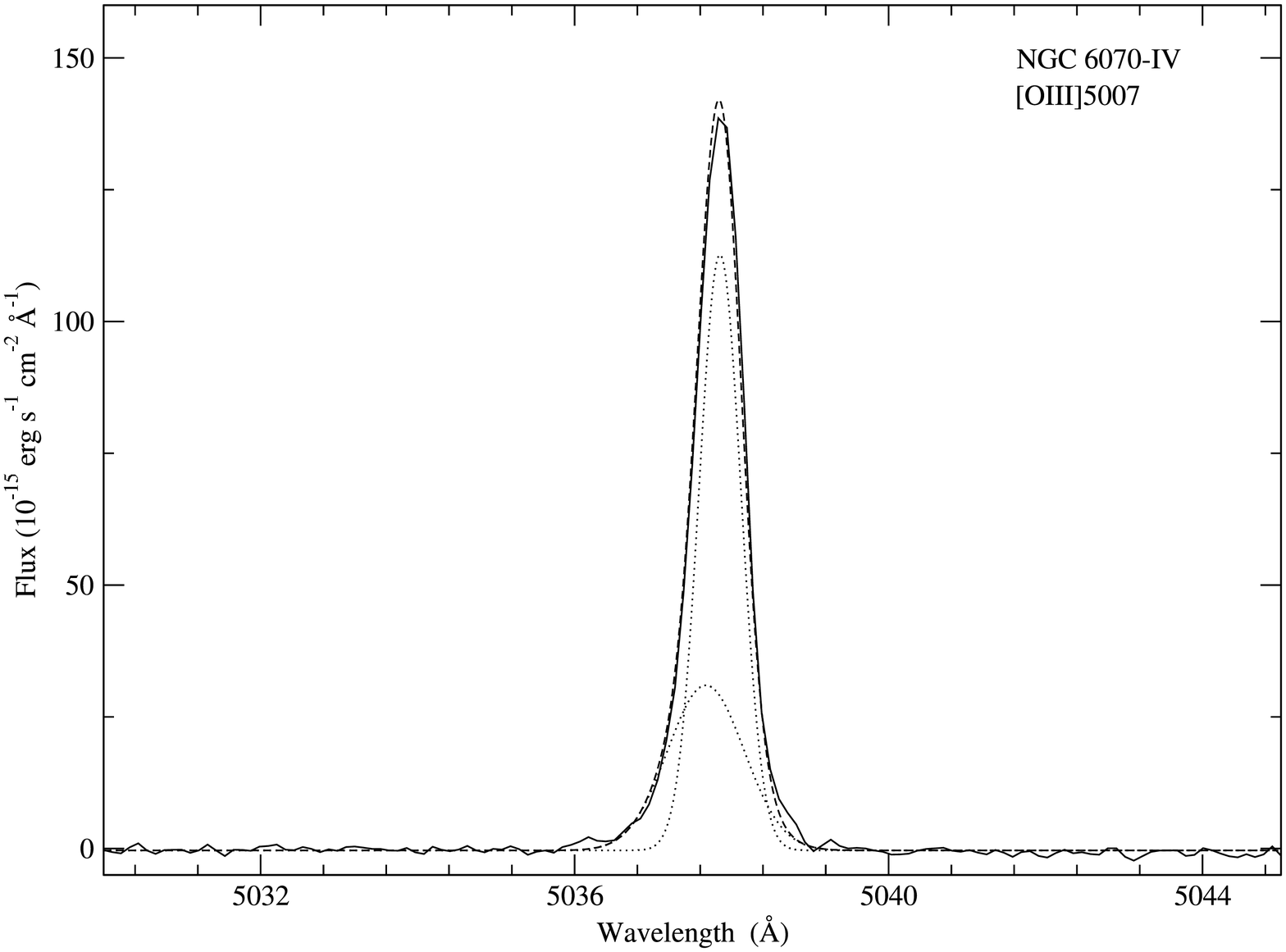}
\includegraphics[trim=0cm 0cm 0cm 0cm,clip,angle=0,width=8cm,height=5cm]{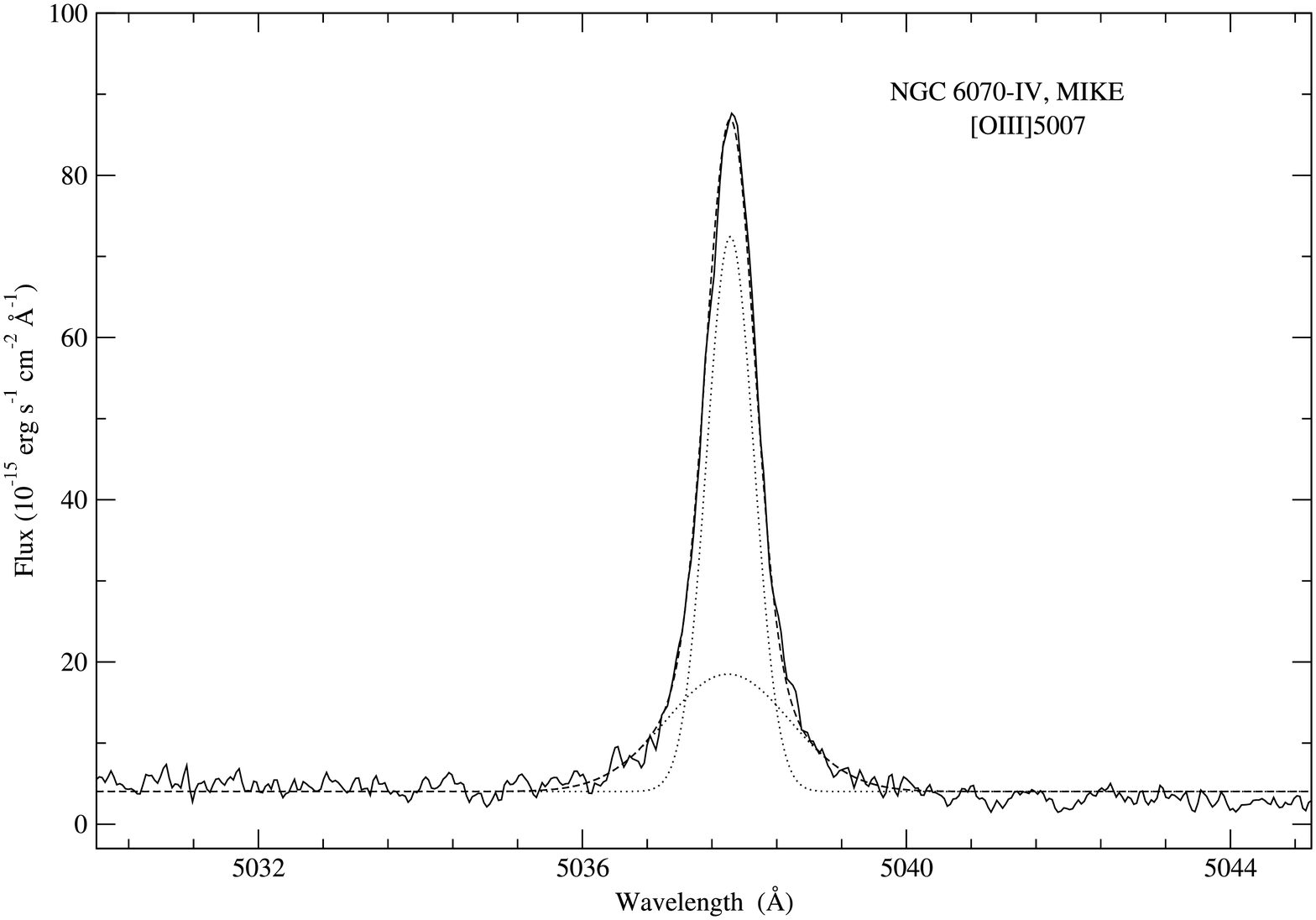}
\includegraphics[trim=0cm 0cm 0cm 0cm,clip,angle=0,width=8cm,height=5cm]{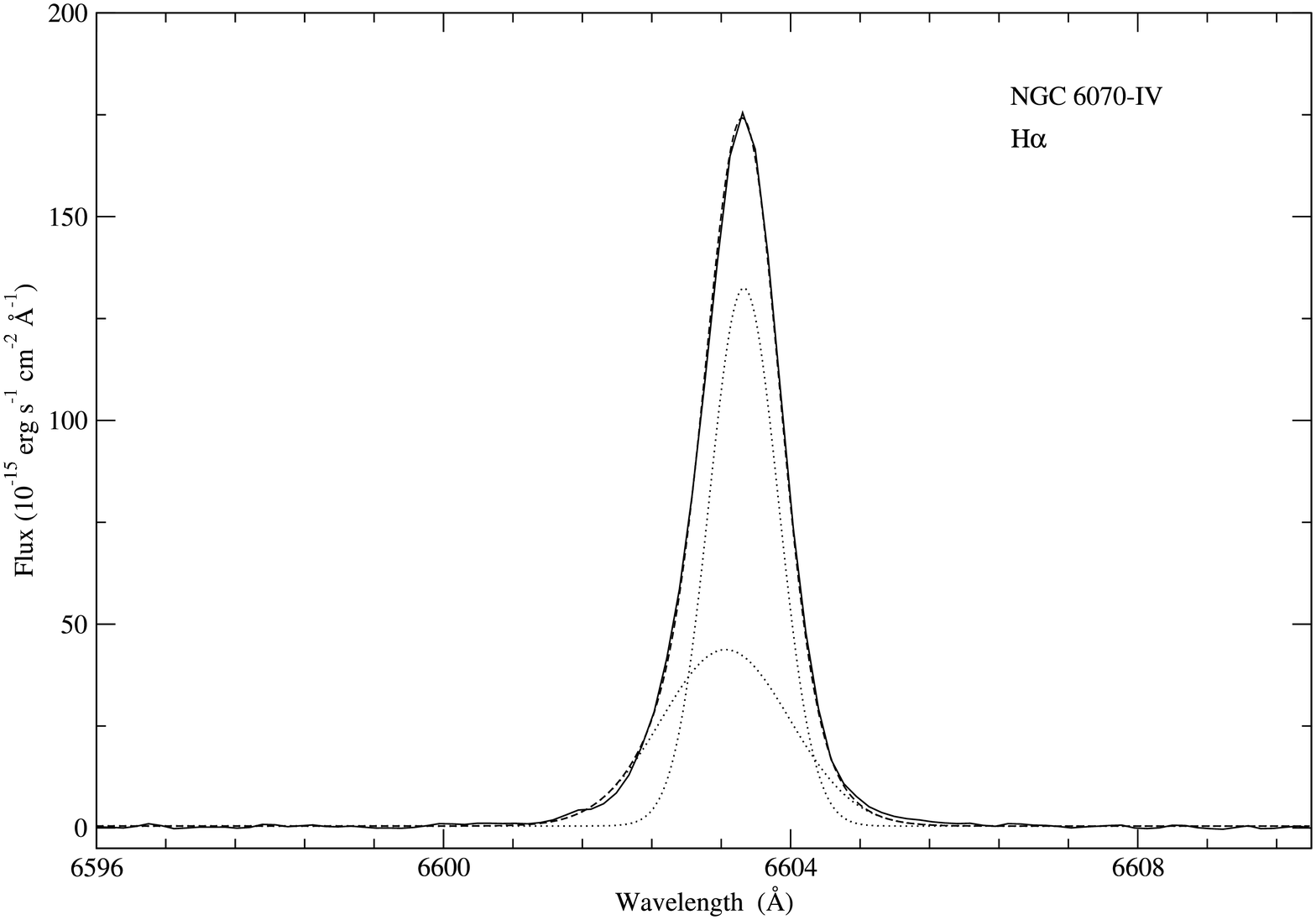}
\includegraphics[trim=0cm 0cm 0cm 0cm,clip,angle=0,width=8cm,height=5cm]{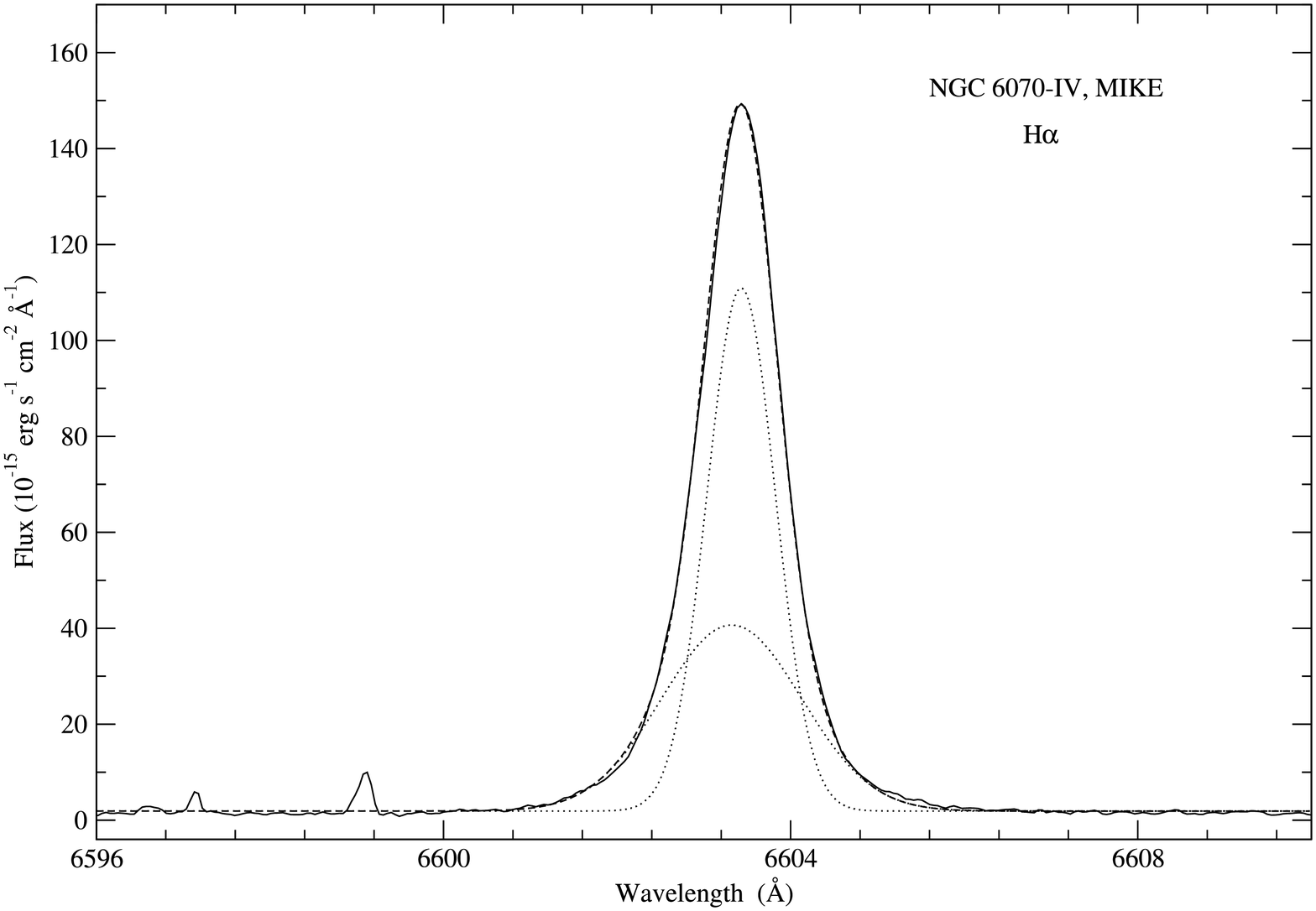}
\includegraphics[trim=0cm 0cm 0cm 0cm,clip,angle=0,width=8cm,height=5cm]{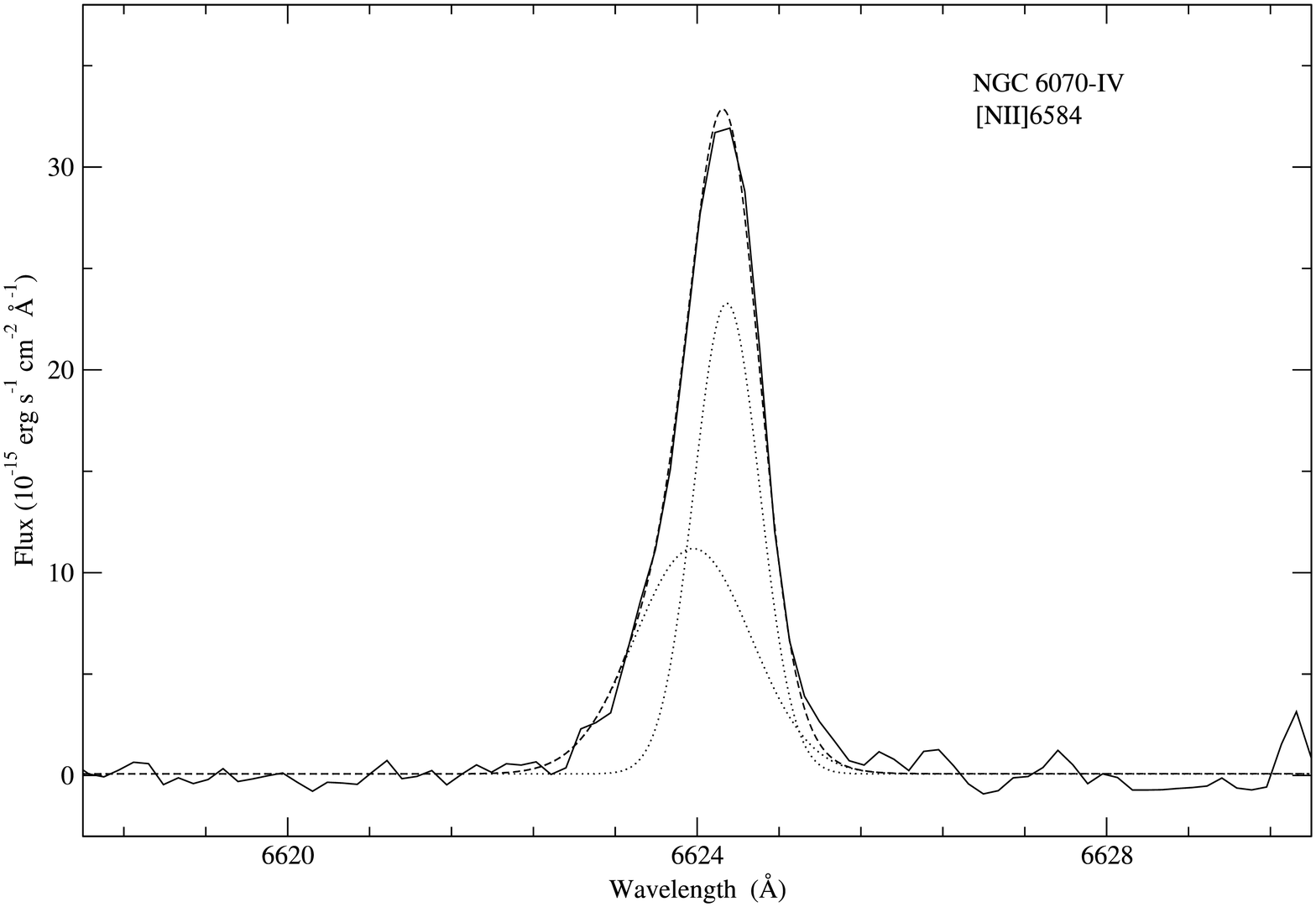}
\includegraphics[trim=0cm 0cm 0cm 0cm,clip,angle=0,width=8cm,height=5cm]{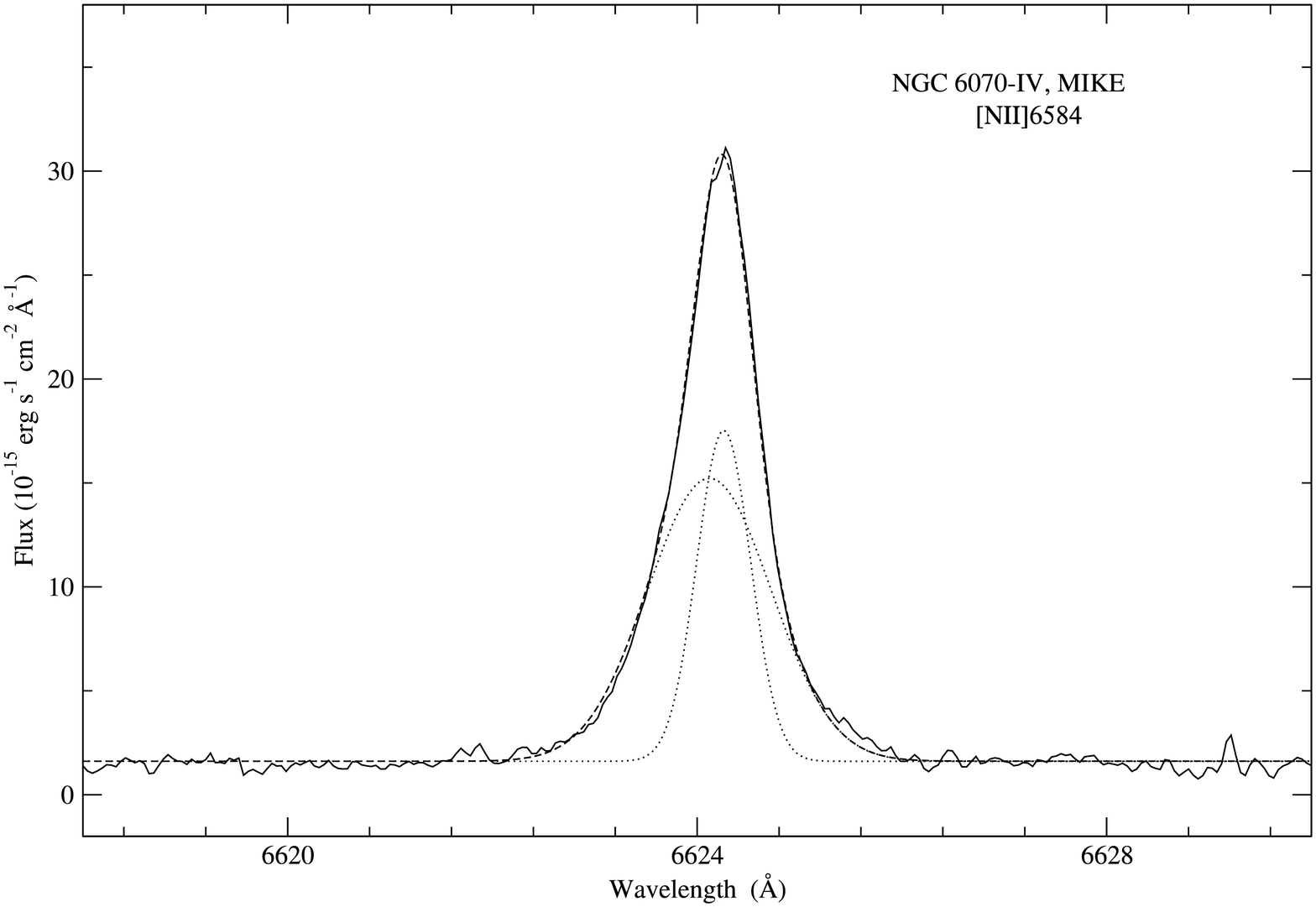}
\includegraphics[trim=0cm 0cm 0cm 0cm,clip,angle=0,width=8cm,height=5cm]{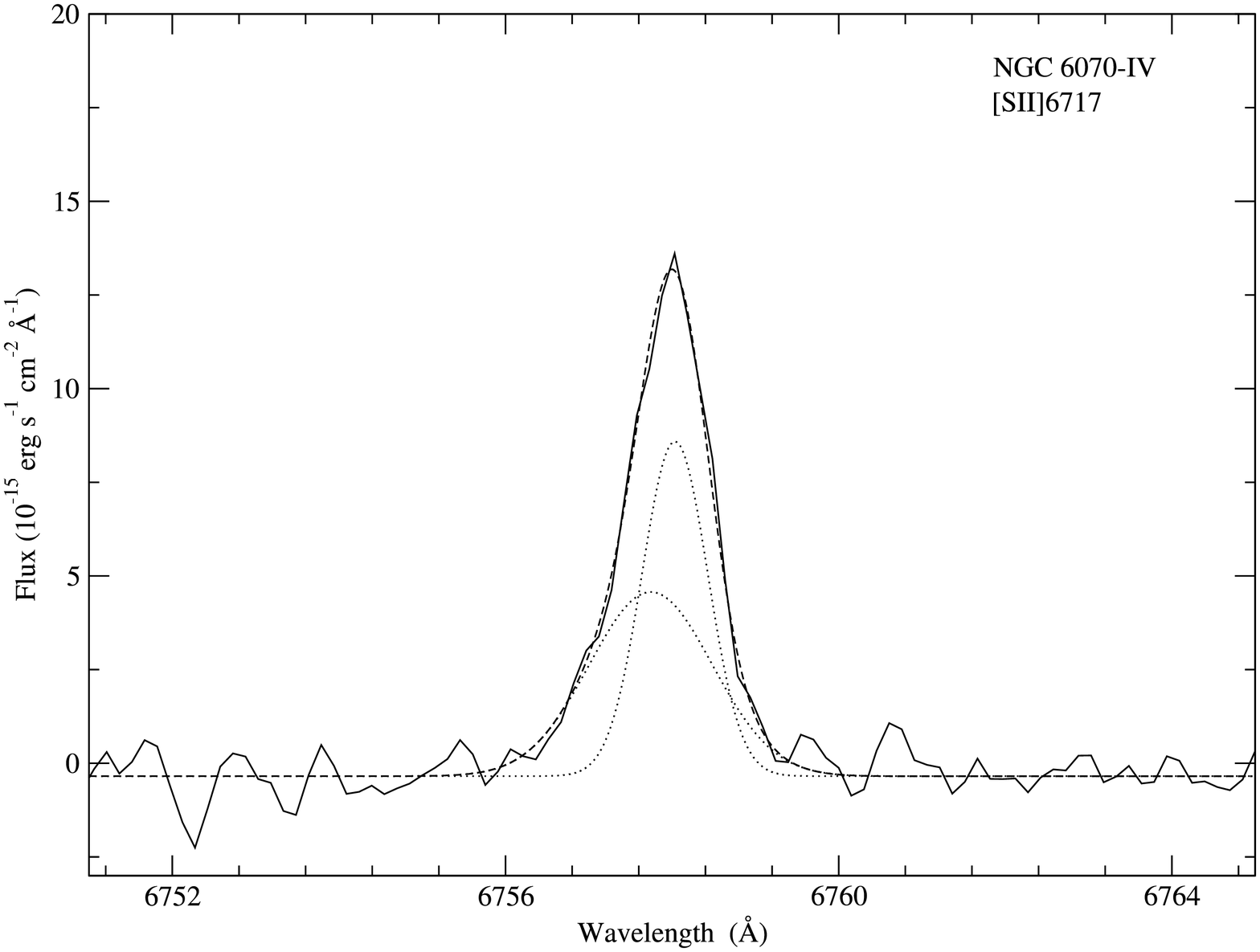}
\includegraphics[trim=0cm 0cm 0cm 0cm,clip,angle=0,width=8cm,height=5cm]{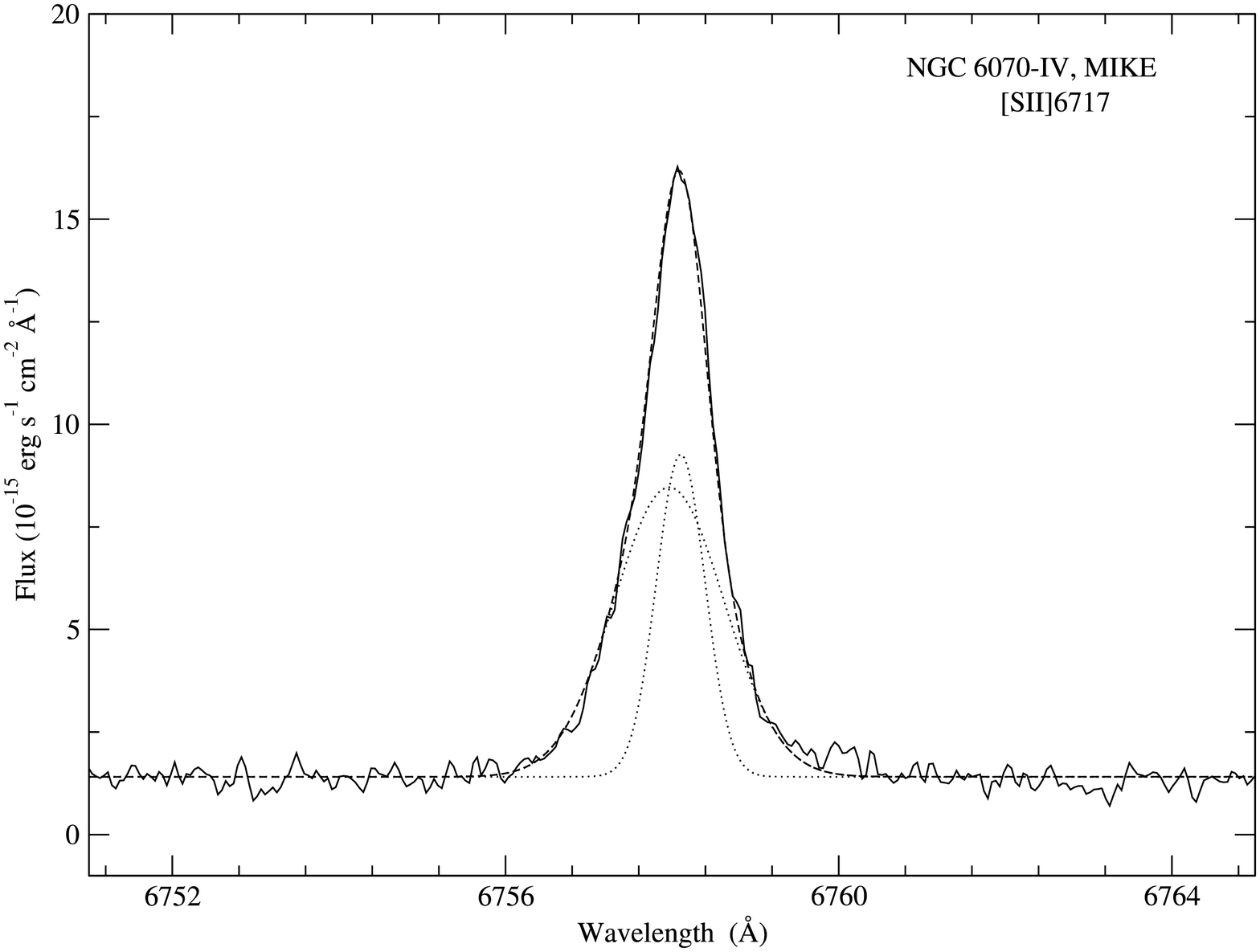}
\end{center}
\end{figure*}

Regarding radial velocities of the three observed regions in NGC\,6070, there is no detailed velocity map in the
 literature. We are able to compare them with rotation curves derived from
 long slit observations made by
\cite{2002A&A...393..389M}. Although the orientation of the slits does not match exactly the location of all our regions, the velocities are in fair agreement considering the angular distance of the H{\sc ii} regions to the centre of NGC\,6070.

\subsection{Relation between H$\alpha$ Luminosities and Velocity Dispersion} 
\label{L_Sigma}

\begin{figure}
\begin{center}
\caption[LvsSigma]{$\log(L) - \log(\sigma)$ relation for our HII regions. Luminosities and velocity dispersions are derived from our spectrophotometric data. The plot includes results from individual components labelled A and B (where applicable) together with velocity dispersions for the overall profile (labelled as ``global'') which were derived from a single Gaussian fitting to the line profile (color solid error bars). The narrow components in NGC\,6070 were labelled A, for which we only plot the MIKE measured data. A few Giant H{\sc ii} regions from \cite{B02} (blue dashed error bars) together with their linear fit to their ``young'' Giant H{\sc ii} regions are plotted as a reference value. The luminosities are not corrected for extinction.}
\label{Fig:elesigma}
\includegraphics[angle=0,width=9cm,height=7cm]{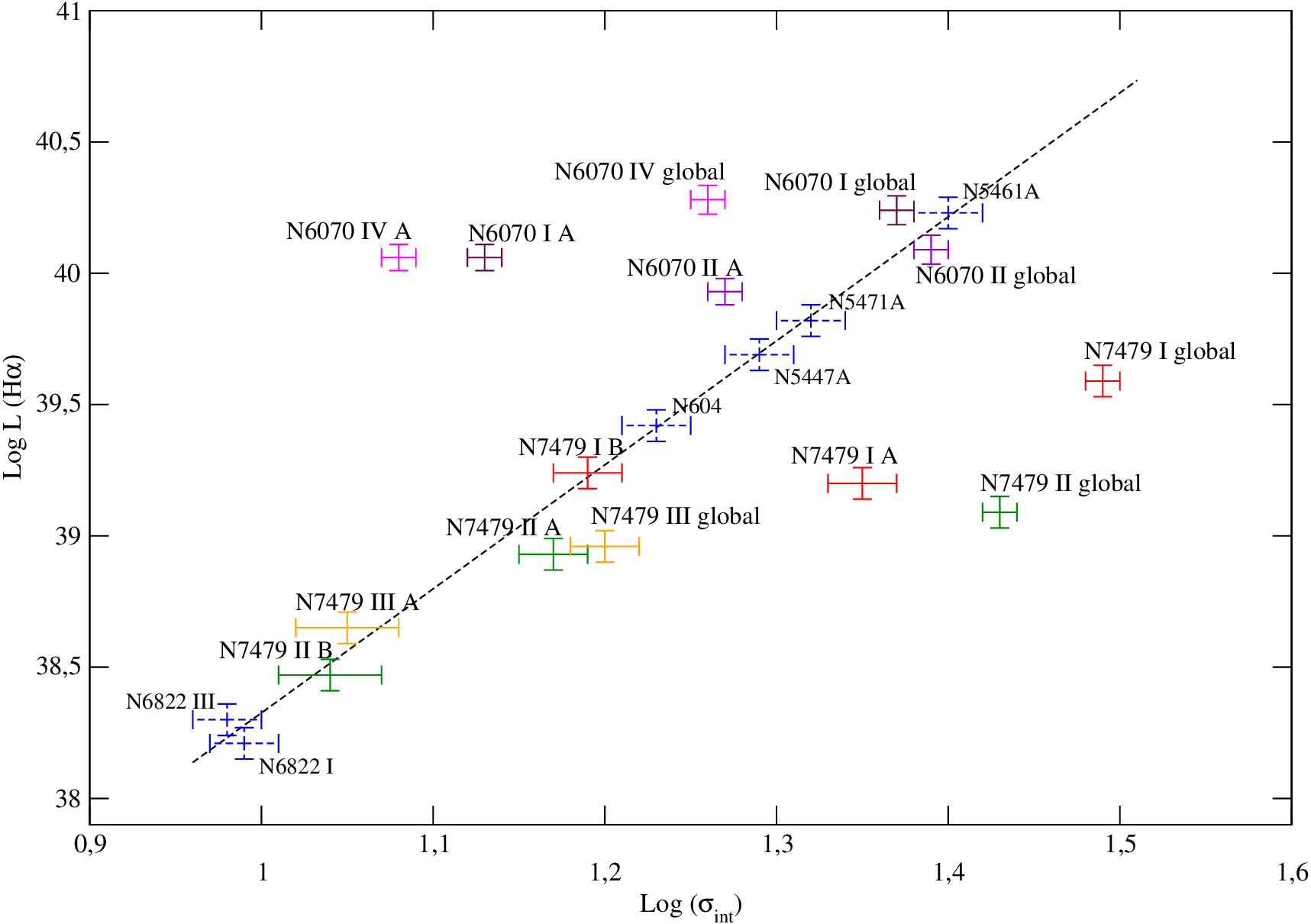}
\end{center}
\end{figure}

We have also analysed the impact that the presence of multiple
components has on the location of
H{\sc ii} regions in the $\log(L) - \log(\sigma)$ plane. To evaluate
this we have plotted in Figure \ref{Fig:elesigma}
the luminosities and velocity dispersions derived directly from our
spectrophotometric data. Luminosities were derived
from the fluxes measured directly from the
component fitting to our echelle spectra (uncorrected by reddening), and using distances as
published by \cite{1999A&AS..135..145R} for NGC\,7479 and by \cite{2002A&A...389...68G} for NGC\,6070. We are aware that slit spectra might not provide the
most realistic sample of the overall flux of the ionised nebula, but
the accuracy of acquisition and guiding is not good enough as to allow
us to perform a proper identification with published luminosity values
in fields relatively ``crowded'' with H{\sc ii} regions by \cite{F97}
and \cite{2006A&A...455..539R}. Velocity dispersion values and its errors are taken directly from Tables \ref{tab7479_ngauss} and \ref{tab6070_ngauss}. The luminosity errors have been estimated taking into account the errors in the amplitude (A) and the FWHM in the component fitting (F=1.0645*A*FWHM), and the distance errors. As a reference
value we have plotted a few Giant H{\sc ii} regions from
\cite{B02} together with their linear fit to their ``young''
GH{\sc ii} regions.

Inspection Figure \ref{Fig:elesigma} reveals that the distribution of
points in the $\log(L) - \log(\sigma_{int})$ plane are strongly dependent on
the components derived from the profile fitting. Individual
components, labelled A and B (when applicable) as in their respective
tables, and global values are those derived from a single Gaussian
fitting to the line profile. As expected, individual components have
smaller fluxes and velocity dispersions than the global profile and
points are therefore shifted in the diagram. The incidence of this
analysis has, however, a different outcome for regions in NGC\,6070
and NGC\,7479. In NGC\,6070 the values derived from the global profile
seem to lie, within observational errors, in the expected location if
they follow the relation expected for virialised systems. When the
individual components are plotted in the same diagram, only
NGC\,6070\,II lies close to the regression, and NGC\,6070\,I and IV
seem to be too luminous for their measured velocity dispersion. On the
other hand, two out of three regions in NGC\,7479 lie well away this
regression if parameters are derived from a single Gaussian, but four
out of five individual components show a very good agreement when
plotted in the same diagram.

The interpretation of this behaviour is not straightforward.
Definitely, the presence of more than one Gaussian component rules the 
final position of the H{\sc ii} regions in the $\log(L) - \log(\sigma)$
plane. Our high resolution spectroscopic data has allowed us to
disentangle multiple kinematic components, which reflects in a big
improvement in the x axis of the plot. But limitations in spatial
resolution prevent us from identifying multiple components, if
present, which do not become evident in the spectral signature. This
has been a problem in previous attempts to derive relations in the
$\log(L) - \log(\sigma)$ as discussed by \cite{B02} when they
were able to obtain photometry for individual knots within Giant H{\sc
ii} regions in M101.

 \section{Summary and Conclusions}

From new high resolution spectra of the H{\sc ii} regions  NGC7479\,I,
NGC7479\,II, NGC7479\,III, NGC6070\,I, NGC6070\,II and NGC6070\,IV obtained at
the 100-inch du Pont Telescope, LCO, we have confirmed the giant nature of all
these regions. We have also found that all of them show evidence
of wing broadening evident mainly in the H$\alpha$ line and confirmed in other
emission lines. In five of them we are able to fit a broad component which explains the integral profile
wings. Only in NGC7479\,II can we fit the integrated profile wings
with two narrow components symmetrically shifted in
velocity with respect to the average of the two intense components.

NGC7479\,I and NGC7479\,II reveal the presence of two separated kinematic
components with relatively narrow profiles in all analysed emission lines. For NGC7479\,I region we can also
generate 2D velocity images in H$\alpha$, [\NII]$\lambda$6584\AA\ and
[\SII]$\lambda$6717\AA\ lines. These images show that the two narrow
components are spatially resolved and the kinematical information helps to split the knot in the velocity-distance plane. There is also
evidence of the presence of a broad component in all the narrower 
extracted sections, at least in the H$\alpha$ emission line. 
For NGC7479\,II region we can also fit two relatively narrow Gaussian
components. In both regions the
radial velocity average value of our components A and B derived from the H$\alpha$ emission line matches the global velocity in
the galaxy when we project our value over the isovelocity contours in
H$\alpha$ velocity field map. A similar comparison, using the H\,I map yields
a value much closer to the value measured for component B. Then, we could only
suggest that component A shows an odd kinematic behaviour. New observations
with better spatial and spectral resolution are needed to clarify this point.

In the rest of the studied H{\sc ii} regions, we find
one narrow Gaussian component together with an underlying broad 
component,
with no evidence of multiple narrow contributions. The radial velocity average derived for the narrower component corresponds
to the one expected from galaxy rotation curves, within the observational errors, in all of these cases.

In most cases, the velocity dispersions of the broad components for these HII regions derived by optimal Gaussian fit are in agreement with \cite{2009MNRAS.396.2295H,Hagele+10} for circumnuclear star-forming regions with $\sigma_{int}$ in H$\beta$ line $\sim$34 to 65\kms, and those derived by \cite{CHK94} and \cite{M99} for 30 Doradus nebula with $\sigma_{int}$ in H$\alpha$ line $\sim$45\kms.

The estimated offsets between the narrow and broad components in NGC\,7479 III, NGC\,6070 II, and NGC\,6070 IV are in complete agreement with the values found in NGC\,2903 and NGC\,3310 by \cite{2009MNRAS.396.2295H,Hagele+10} respectively, which are between -25 and 35 \kms.

The distribution of the regions in the $\log(L) - \log(\sigma)$ plane are strongly dependent on the components derived from the profile fitting. Individual components have
smaller fluxes and velocity dispersions than the global profile and
points are therefore shifted in the diagram. In NGC\,6070 the values derived from the global profile
seem to lie, within observational errors, in the expected location if
they follow the relation expected for virialised systems. When the
individual components are plotted in the same diagram, only
NGC\,6070\,II lies close to the regression, and NGC\,6070\,I and IV
seem to be too luminous for their measured velocity dispersion. On the
other hand, two out of three regions in NGC\,7479 lie well away this
regression if parameters are derived from a single Gaussian, but four
out of five individual components show a very good agreement when
plotted in the same diagram.
Definitely, the presence of more than one Gaussian component rules the
final position of the H{\sc ii} regions in the $\log(L) - \log(\sigma)$
plane.

\section*{Acknowledgements} 

We are grateful to R.\ Terlevich for useful discussions and comments
about the 2D velocity 
images that greatly improved the data analysis. We thank to C.\ Feinstein for
providing us with the original data and images used for his published work and we are grateful to the director and staff of LCO for technical assistance and warm hospitality. Finally, we appreciate the comments and suggestions by the referee which significantly improved this paper.
This research has made use of the NASA/IPAC Extragalactic Database (NED) which
is operated by the Jet Propulsion Laboratory, California Institute of
Technology, under contract with the National Aeronautics and Space
Administration.  

Support from the Spanish MEC through grant AYA2007-67965-C03-03 and
from the Comunidad de Madrid under grant S-0505/ESP/000237 (ASTROCAM) is
acknowledged by GH. VF and GB thank the Universidad
Aut\'onoma de Madrid, specially to \'Angeles D\'{\i}az, for their hospitality.

\clearpage 

\label{biblio} 
\bibliography{biblos} 
\bibliographystyle{mn2e} 

\end{document}